\documentclass[10pt,journal,compsoc]{IEEEtran}

\usepackage{cite} 
\usepackage{lipsum}
\usepackage{stfloats}
\usepackage{graphicx,subfigure,amsmath,comment}
\usepackage{url}
\usepackage{epstopdf}

\newcommand{\fref}[1]{Fig.~\ref{#1}}
\newcommand{\sref}[1]{Section~\ref{#1}}
\newcommand{\tref}[1]{Table~\ref{#1}}

\newcommand{\pipe}{Mechanical Room}
\newcommand{\ups}{UPS Room}
\newcommand{\ring}{Particle Accelerator Ring}
\newcommand{\hall}{Experimental Hall}
\newcommand{\tunnel}{Tunnel}
\newcommand{\side}{Side Tunnel}

\usepackage{todonotes} 

\begin{document}

\title{A Channel Measurement Campaign for mmWave Communication in Industrial Settings}

\author{
    \IEEEauthorblockN{Adrian Loch\IEEEauthorrefmark{1}, Cristina Cano\IEEEauthorrefmark{2}, Gek Hong (Allyson) Sim\IEEEauthorrefmark{3}, Arash Asadi\IEEEauthorrefmark{3}, Xavier Vilajosana\IEEEauthorrefmark{2}}\\
   \IEEEauthorblockA{\IEEEauthorrefmark{1}IMDEA Networks Institute\\
	}
    \IEEEauthorblockA{\IEEEauthorrefmark{2}Universitat Oberta de Catalunya (UOC)\\
	}
    \IEEEauthorblockA{\IEEEauthorrefmark{3}Secure Mobile Networking lab (SEEMOO), Technische Universit\"{a}t Darmstadt
	}
}
\IEEEtitleabstractindextext{%
Industry 4.0 relies heavily on wireless technologies. Energy efficiency and device cost have played a significant role in the initial design of such wireless systems for industry automation. However, high reliability, high throughput, and low latency are also key for certain sectors such as the manufacturing industry. In this sense, existing wireless solutions for industrial settings are limited. Emerging technologies such as millimeter-wave (mmWave) communication are highly promising to address this bottleneck. Still, the propagation characteristics at such high frequencies in harsh industrial settings are not well understood. Related work in this area is limited to isolated measurements in specific scenarios. In this work, we carry out an extensive measurement campaign in highly representative industrial environments. Most importantly, we derive the statistical distributions of the channel parameters of widely accepted mmWave channel models that fit these environments. This is a highly valuable contribution, since researchers in this field can use our empirical model to understand the performance of their mmWave systems in typical industrial settings. Beyond analyzing and discussing our insights, with this paper we also share our extensive dataset with the research community.

}

\maketitle

\IEEEdisplaynontitleabstractindextext

\IEEEpeerreviewmaketitle

\IEEEraisesectionheading{\section{Introduction}\label{sec:introduction}}


The upcoming industry 4.0 revolution with the integration of wireless technologies in the industrial setting seems unquestionable today. 
The common design goal in the industrial wireless solutions available today (such as WirelessHART and 6TiSCH) is to trade off performance for gains in energy efficiency and device cost.
However, some sectors, such as the manufacturing industry, require high performance communications in terms of reliability and latency and are not so concerned with the cost and energy consumption of the wireless devices \cite{martinez2018square}. 
 We believe this can be a barrier for wireless adoption in the manufacturing industry and that mmWave communication is the way forward as it better fits industrial requirements\cite{Engelhard2018sigcom, Cheffena2016commag}.
The large available bandwidth in mmWave bands ($30$-$300$ GHz) allows for supporting Gigabit-per-second (Gbps) links while maintaining low latency. In addition, the short communication range in mmWave is suitable for industrial applications, where communication is mostly local. Furthermore, the reliance of mmWave communication on directional beamforming can significantly reduce interference. 

{\bf Motivation.} Understanding the behavior of the wireless medium is key prior to the deployment of new wireless technologies in a given environment, in particular when operating at very high frequencies. For example, related work in the area of mmWave includes numerous studies and measurement campaigns that analyze channel characteristics for mmWave communication in urban scenarios~\cite{geng2009millimeter,soma2000modeling, elrefaie1997propagation, samimi20163, maccartney2016millimeter, maccartney2017rural, meinila2009winner, Hur:jv, Fan:2016dg, zhang2017millimeter, zhao2017channel}. These studies were a crucial step in the process of standardizing the use of mmWave bands in 5G cellular networking. The outcome of such studies is critical to evaluate whether such bands are suitable for the intended use cases. Also, the results are the foundation of the design of algorithms, protocols, and deployment plans (e.g., networking planning). Unfortunately, work on characterizing wireless channels in industrial settings is limited~\cite{rappaport1991statistical, zaaimia201660}. This is to a great extend due to the difficulty in accessing industrial facilities, thus making such measurement datasets of high value. 

Before deploying mmWave in industrial scenarios, it is crucial to build an understanding of channel propagation characteristics in such peculiar environments. Urban and industrial scenarios differ significantly due to: $( i )$ the dominant material used for construction (e.g., concrete) and for the machinery (e.g., metal) in industrial setups; and $( ii )$ space planning, which is aimed at fitting the maximum number of machinery with minimal regards to beauty or other factors that are relevant in urban planning. Since accessing industrial facilities is often tedious due to regulatory reasons, it is important to pursue such measurements campaigns in order to formulate an empirical model that can be widely used within the community. 

{\bf Contribution.} In this work, we perform the first extensive industrial mmWave measurement using both specialized measurement hardware and commercial off-the-shelf (COTS) 60 GHz hardware. Our measurements took place at the ALBA Synchrotron, which is a particle accelerator facility in the city of Barcelona. We chose this facility because it provides a unique set of representative industrial environments: $( i )$ server rooms representing data centers, $( ii )$ the experimental hall resembling large production plants, $ (iii) $ underground tunnels along with the particle accelerator ring which resemble the environments within a constrained space such as subway tunnels, and $(iv)$ the cooling facility, which is a widespread environment in many industries, including large pipes and reflecting surfaces. 

Our measurement campaign consists of high accuracy measurements across a wide range of locations within the above scenarios. Specifically, the outcome of our campaign is 70 gigabytes of data, which includes both raw physical layer IEEE 802.11ad traces as well as upper layer metrics such as throughput and packet error rates. While the latter provides a broad view on the potential performance of mmWave networks in industrial settings, the former allows us to obtain detailed insights into the operation of the physical layer. We use the above raw traces to fit the parameters of a widely accepted mmWave channel model for the particular case of an industrial environment. We obtain the empirical distributions of each parameter for each of the aforementioned scenarios. This is a highly valuable contribution to the community, since it enables researchers in this field to generate arbitrary channels that are representative of industrial scenarios. Our work opens the door to an accurate understanding of the mmWave channel in such scenarios and enables first-of-its-kind mmWave systems for industry 4.0. In particular, our contributions are as follows:

\begin{itemize}
\item We collect a large dataset of mmWave traces at the ALBA Synchrotron in the city of Barcelona using both COTS IEEE 802.11ad devices as well as high accuracy measurement hardware.
\item While related work focuses on individual mmWave performance measurements in particular scenarios, we generalize our practical insights to obtain a highly accurate channel model. Such a channel model is of much higher relevance to the community than individual traces.
\item We share both our dataset and our model with the community, enabling other researchers to build on our measurements in order to use them as a foundation for the evaluation of arbitrary mmWave systems for industrial scenarios.
\item We carry out our measurement campaign at a unique location which encompasses a wide variety of scenarios that fit many typical industrial settings.
\item We find statistical distributions that fit each model parameter across all of our scenarios, thus obtaining a broad and universal model. 
\end{itemize}

The remainder of this paper is structured as follows. In Section~\ref{sec:related_work} we survey related work in the area of mmWave channel studies. After that, we present our measurement methodology in Section~\ref{sec:methodology}, which includes a detailed description of our hardware setup. We then describe each of the scenarios that we analyze at the ALBA Synchrotron in Section~\ref{sec:setup}. In Section~\ref{sec:model} we introduce the channel model that we use as a basis for our study. Next, we present our measurements and fit the parameters of the model in Section~\ref{sec:results}. Finally, we conclude the paper in Section~\ref{sec:conc}.


%


\section{Related Work}
\label{sec:related_work}


Many works aim at characterizing and modeling the propagation characteristics of mmWave channels empirically. Most of them are related to this article concerning their methodology and measurement hardware. To this aim, we first provide an overview on seminal works on mmWave channel characteristics and their methodology~\cite{geng2009millimeter,soma2000modeling, elrefaie1997propagation, samimi20163,maccartney2016millimeter,maccartney2017rural,meinila2009winner,Hur:jv, Fan:2016dg,zhang2017millimeter,zhao2017channel} and then we focus on the related literature characterizing wireless propagations in industrial environments~\cite{rappaport1991statistical, zaaimia201660}.

\subsection{mmWave Channel Characterization} 
Although mmWave communication has drawn much attention within the past few years, it was initially studied for LOS communication in indoor~\cite{geng2009millimeter} and outdoor~\cite{soma2000modeling, elrefaie1997propagation} environments. In the following, we summarize the latest developments in mmWave channel modeling. 

In~\cite{samimi20163}, Samimi {\it et al.} leverage the data from the measurement campaigns of the NYU WIRELESS \cite{nyuwireless} team within the past five years~\cite{maccartney2016millimeter,maccartney2017rural} to derive a statistical 3D channel model for mmWave links. Their measurements are performed in the 28, 38, and 78 GHz bands, which are the candidate frequencies for mmWave cellular networking. The authors provide a detailed description of the methodology they use and the assumptions they make in order to identify path clusters in time and space. In particular, time-delayed versions of the transmitted signal should be partitioned into {\it time clusters} which help identifying the effective multi-path components and compute the delay spread. The same applies to the spatial lobes which characterize the angle of arrival and angle of departure (i.e., angular spread). This characterization of the signal in time and space plays a significant role in the accuracy of the channel model. Indeed, Samimi {\it et al.} elaborate on the fact that the bin size for each cluster should be decided based on the environment. 

In parallel, further work in this area has modeled the mmWave channel leveraging a slightly different methodology. The main differences are: $ ( i ) $ combining ray-tracing simulations with actual measurement to reduce the overhead of measurement, and $ ( ii ) $ simplifying the parameterization of the models. This work includes the mmWave spatial channel models proposed in 3GPP~\cite{3GPP25.996} and WINNER II~\cite{meinila2009winner}. Both models characterize the channel based on the delay spread, azimuth spread, shadow fading, and spatial autocorrelation. However, each of the models defines some of these parameters in a slightly different manner. Also, 5G Public Private Partnership (5GPPP) projects such as mmMAGIC \cite{mmmagic} have studied channel models for the mmWave band. Specifically, mmMAGIC covered the frequency range from 6 GHz to 100 GHz and focused on scenarios such as street canyons, open squares, indoor offices, shopping malls, airports, stadiums, and subway stations. Our work stands apart from this model since we consider industrial settings and focus on the characteristics of the 60 GHz band.

In \cite{Hur:jv}, the authors focus on deriving radio propagation parameters at 28 GHz from measurements in urban environments such as New York, USA and Daejeon, Korea. Specifically, they obtain the value of the delay spread and the angular spread using both practical measurements and ray tracing techniques. In~\cite{Fan:2016dg}, Fan {\it et al.} perform indoor channel measurements in 2-4 GHz, 14-16 GHz, and 28-30 GHz. Their study focuses on characterizing the angular and delay spreads for the aforementioned frequency bands. Their measurement results confirm that the number of multipath components decreases at higher frequencies. A similar study in~\cite{zhang2017millimeter} characterizes the mmWave channel at 25.5 GHz, 28 GHz, 37.5 Ghz, and 39.5 GHz in indoor scenarios. The authors claim that their measurements are more accurate than earlier work in the field because they consider a number of different elevations. 
In~\cite{zhao2017channel}, the authors perform outdoor channel measurements at 32 GHz. They compare their findings with the values reported by NYU WIRELESS, mmMAGIC, and 3GPP. Their results match the models of NYU WIRELESS and mmMAGIC but deviate from the results reported by 3GPP. The authors explain that this difference is due to the fact that the 3GPP model was derived from limited measurements on a small subset of carrier frequencies. 


\subsection{Industrial Channel Characterization} 

The characteristics of the wireless channel differ significantly in industrial environments compared to typical indoor scenarios such as office or home environments. The main differences are structural (e.g., ceiling height) and environmental (e.g., wall/floor material and 
metallic surfaces)~\cite{zaaimia201660,rappaport1991statistical,shuangchannel}. As a result, channel measurement and modeling in industrial environments is significantly less explored than urban outdoor/indoor scenarios. The limited related work in this field includes the seminal paper by Rappaport {\it et al.} \cite{rappaport1991statistical}, which models the wireless channel for factory settings but is limited to a carrier frequency of 1.3 GHz. Similarly, \cite{hawbaker1990indoor} also considers an industrial scenario but only for communication up to 4 GHz. Recent work extends this analysis to the mmWave band. For instance, the authors of \cite{zaaimia201660} focus on modeling mmWave channels at 60 GHz within a data center. The result of their measurement in terms of path loss and delay spread indicates that the wireless channel in data centers does not match other well-known scenarios due to the aforementioned differences. While we consider data centers a type of industrial environment, our work stands apart from the above study because we consider a much broader set of scenarios that are representative for factories. To the best of our knowledge, no further earlier work considers such scenarios. This highlights the relevance of our contribution in this paper.

\section{Methodology}
\label{sec:methodology}

To measure the performance of mmWave communications in the industrial scenarios described in Section~\ref{sec:setup}, we use both a commercial off-the-shelf (COTS) setup as well as specialized measurement hardware. The former reveals how commodity devices would perform in such environments, whereas the latter allows us to gain deep insights into propagation characteristics. For the bulk of our experiments, we use both setups such that we can relate the performance of COTS devices to the actual channel conditions. In the following, we describe each of the setups in detail.

\subsection{Commercial Off-The-Shelf Setup}
\label{subsec:cots_setup}

\begin{figure}[t!]
\centering
 \includegraphics[width=\columnwidth]{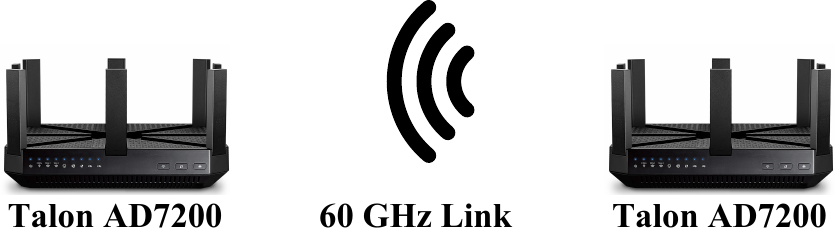}
 \caption{Commercial Off-The-Shelf Setup.}
 \label{fig:cots_setup}
\end{figure}

In our COTS setup, we use commodity hardware that implements the IEEE 802.11ad amendment. This amendment enables WiFi networks to operate in the mmWave band. Specifically, IEEE 802.11ad focuses on the 60 GHz band. At the time of writing, IEEE 802.11ad is the most promising candidate for mmWave indoor communications, which motivates our choice of the corresponding COTS hardware. In particular, we use TP-Link Talon AD7200 routers, which incorporate a tri-band Qualcomm QCA9500 chip that supports IEEE 802.11ad. Figure~\ref{fig:cots_setup} depicts our setup, in which we configure one router as a 60 GHz access point (AP) and a second router as a 60 GHz client. The above chip consists of a baseband module and an antenna module. Both modules are physically separated to allow the router manufacturer to place the antenna at a convenient location within the device case. The antenna module consists of an electronically steerable phased antenna array with 32 radiating elements. This allows the module to use analog beam-forming, which is crucial to overcome the high path loss in the mmWave band.

We install a customized LEDE/OpenWRT \cite{lede} system on the routers that enables us to use the ``Talon Tools'' \cite{talon-tools:project}. These tools are a framework for practical IEEE 802.11ad research based on the TP-Link Talon routers. The framework gives us full access to the network interfaces of the router and enables us to configure it both as an access point as well as a client. Moreover, the framework uses open-source drivers for the QCA9500 chip which allow us to gain detailed information about its operation in the 60 GHz band.

Our experiment methodology with the above COTS hardware is as follows. We place one device as an access point at a designated location in each of our scenarios, as discussed in Section~\ref{sec:setup}. We then place a second device as a client at a number of different positions within the scenario. At each position, we establish a 60 GHz connection and generate traffic on the link using iperf.
During this data communication, we record a number of metrics both at the access point and at the client at regular intervals. Specifically, we record the TCP throughput, the identifiers of the beam-patterns chosen at each side of the link, the Modulation and Coding Scheme (MCS), the number of transmitted packets, the Packet Error Rate (PER), and a Signal Quality Indicator (SQI) which reflects changes in the Signal-to-Noise Ratio (SNR). Our measurements allow us to depict the above metrics for each measurement location in each scenario, which provides crucial insights regarding how a particular environment influences the performance of communications in the mmWave band.

\subsection{Measurement Hardware Setup}
\label{subsec:measurement_setup}

While the above COTS setup allows us to understand the performance of mmWave networks in industrial settings at the link layer, it barely reveals any information about the operation of the physical layer. The latter is key to characterize such industrial settings and develop a representative channel model. Next, we present the details of the specialized measurement hardware that we use to extract physical layer channel parameters and formulate such a model (c.f. Section~\ref{sec:model}).

\begin{figure}[t!]
\centering
 \includegraphics[width=\columnwidth]{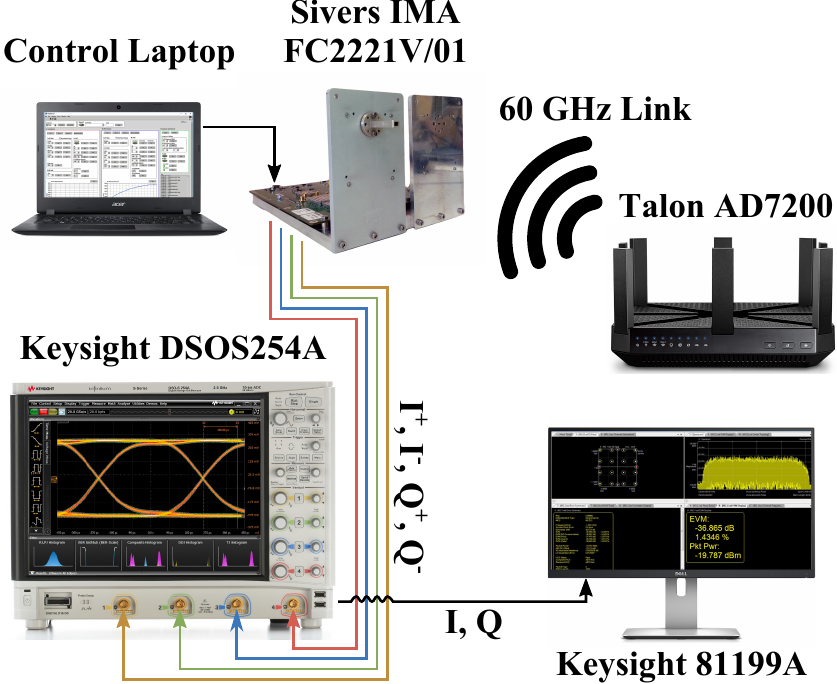}
 \caption{Measurement Hardware Setup.}
 \label{fig:hw_setup}
\end{figure}

As discussed in Section~\ref{subsec:cots_setup}, the IEEE 802.11ad standard is a promising candidate for indoor mmWave communication. Thus, we focus our physical layer analysis on the specifics of full-bandwidth 802.11ad channels. To this end, we transmit frames using the Talon devices introduced in Section~\ref{subsec:cots_setup} and analyze them using our measurement hardware. Figure~\ref{fig:hw_setup} depicts our measurement setup. We configure a Talon device as an access point and place it at a designated location in each scenario (c.f. Section~\ref{sec:setup}). Since we do not associate any client to this access point, the device simply transmits periodic beacon frames to announce its presence. We capture those beacons at each location at which we placed a client device in Section~\ref{subsec:cots_setup}. Specifically, we use a Sivers IMA FC2221V/01 V-band down-converter attached to a horn antenna to receive the signal in the 60 GHz band and convert it to a baseband signal. We then capture the output of the down-converter using a Keysight DSOS254A oscilloscope. Since this oscilloscope has a bandwidth of 2.5 GHz, we can record the full signal. We then feed the recording to the Keysight Wideband Waveform Center, which is a software that implements a full IEEE 802.11ad decoder. This software detects, demodulates, and decodes each of the beacons, providing full insight into the physical layer parameters of the channel.

Our experiment methodology with the above measurement hardware setup is as follows. At each client location, we capture a full beacon sequence of the access point. The access point transmits such sequences at a fixed interval of 102.4 milliseconds \cite{Nitsche:2015vq} to announce its presence. For the case of the Talon router, each sequence contains 32 individual beacons since the access point implements 32 different beam-patterns. Each of the beam-patterns covers a certain azimuthal region. The periodicity of this burst of beacons allows us to easily synchronize the oscilloscope to the signal and record stable traces. In order to obtain averages of each channel parameter in our later analysis, we capture a number of beacon bursts at each client location. Each individual beacon results in different Channel State Information (CSI) since the transmit beam-pattern is different. Thus, after capturing the trace with the oscilloscope, we split it into individual beacons using traditional signal processing techniques in Matlab. We then decode each beacon of each trace individually using the Keysight Wideband Waveform Center. From the decoded data, we extract the beam-pattern identifier which is embedded in each beacon. Finally, at each location we average the channel parameters corresponding to each identifier separately.

In addition to the beam-pattern identifier, we obtain full CSI in terms of amplitude and phase. This includes the channel impulse response, the channel frequency response, the SNR, and the Error Vector Magnitude (EVM) for each beacon. To build our channel model, we extract the amplitude and delay of each channel tap, along with the number of taps in the channel as well as the potential clustering of the taps due to the scarce multi-path propagation in the mmWave band. For a detailed description of our channel model, we refer the reader to Section~\ref{sec:model}.
\section{Experimental Setup}
\label{sec:setup}

In the following, we describe the layout of the physical environments in which we collect measurement data. We consider five distinct environments: a \pipe{}, an \hall{}, a \ring{}, a \tunnel{}, and a \ups{}. We select these areas because they effectively represent the scenarios found in a typical industrial site. For instance, the \pipe{} contains a large number of pipes and other reflective infrastructure that is present in most industrial scenarios. The \ring{} resembles a production line and the \tunnel{} resembles a hallway interconnecting different production areas. Last but not least, the \ups{} could represent a data center or any environment with large machinery. For each environment, we collect data both using our COTS setup as well as our setup based on measurement hardware (c.f. Section~\ref{sec:methodology}).


\subsection{\pipe{}}
\begin{figure}[t!]
\centering
 \includegraphics[width=\columnwidth]{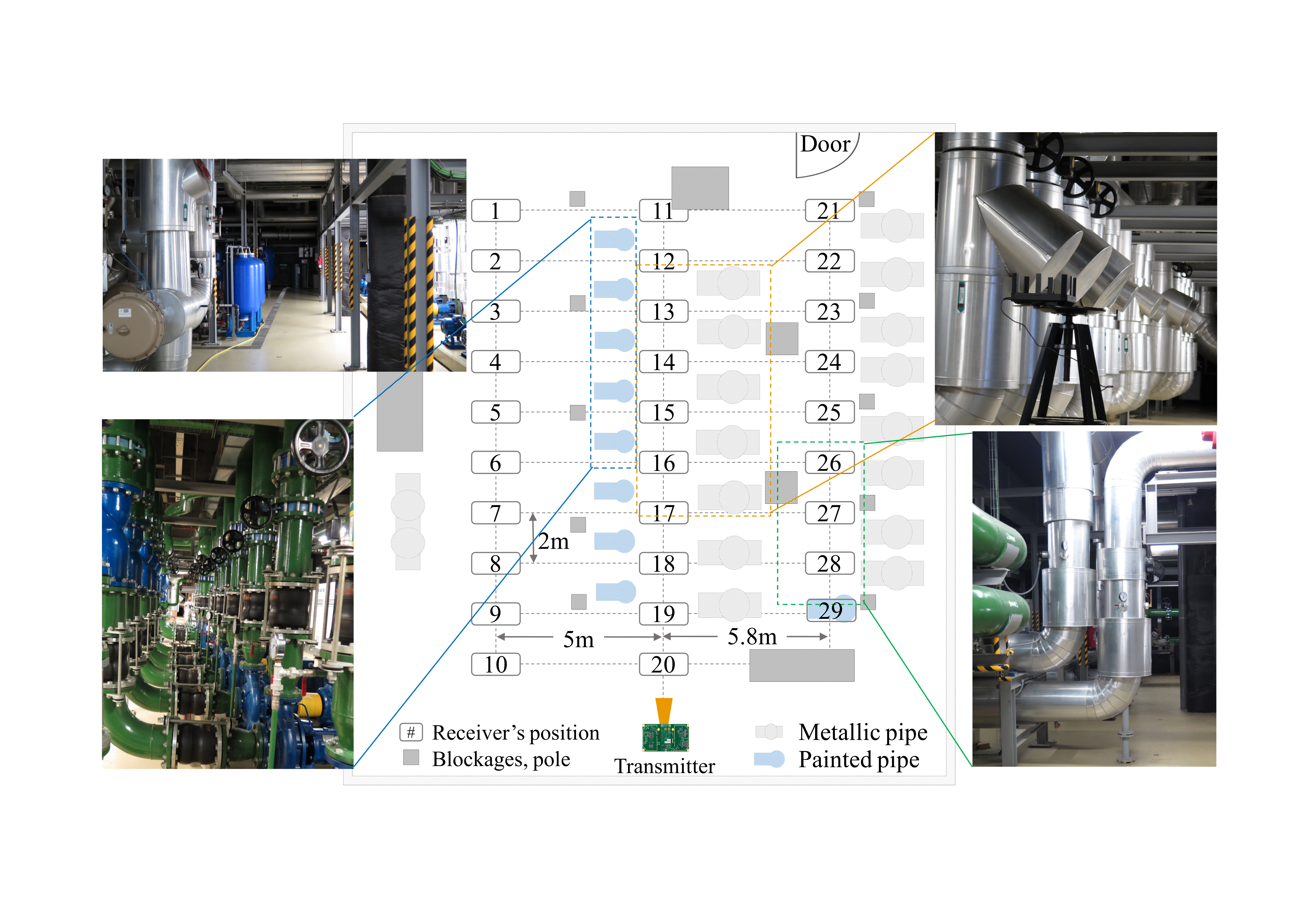}
 \caption{Layout of the \pipe{}.}
 \label{fig:pipe}
\end{figure}

\fref{fig:pipe} shows the layout of the \pipe{}. This room consists of the following main structures: (i) metallic and painted circular pipes with variable sizes and (ii) large and small poles. These structures have a strong impact on several radio propagation effects such as diffraction, reflection, and multi-path transmission.  
In addition, they also act as blockages for some paths between the transceivers. However, in contrast to the \ups{}, blockage is typically not full but partial. In order to evaluate the effects of these structures on the communication, we place the receiver at 28 different positions as shown in \fref{fig:pipe}. Still, communication is not viable at locations 11 and 29 since permanent infrastructure fully blocks the link. The \pipe{} also provides environments with both sparse and dense infrastructure on the left and right side of the layout, respectively.      

\subsection{\hall{}}
\label{subsec:hall_setup}
\begin{figure}[tbh!]
\centering
 \includegraphics[width=\columnwidth]{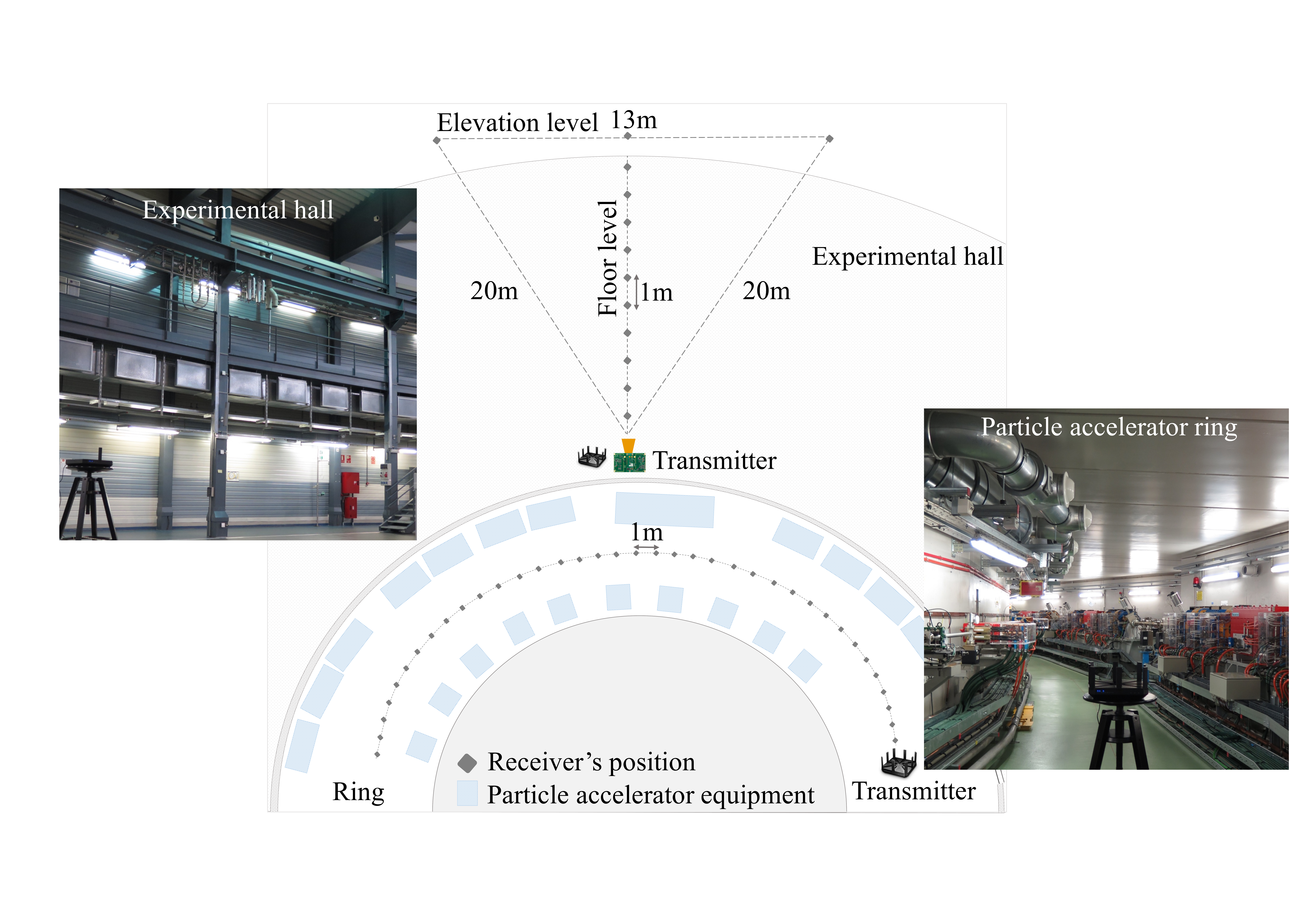}
 \caption{\hall{} and \ring{}.}
 \label{fig:ring}
\end{figure}

\fref{fig:ring} depicts the layout of both the \ring{} as well as the \hall{}. The \hall{} is a large double-height empty circular space that encloses the \ring{} as well as surrounding beamlines that receive the synchrotron light. In \fref{fig:ring}, we show the upper half of the ring as well as a fraction of the hall. The ring is located at the lower level of the hall. The upper level of the hall consists of a walkway that encircles the entire space and provides access to other areas within the particle accelerator plant. It features metal staircases, multiple metal platforms and doors, as well as concrete walls. We collect measurements both in the open space within the \hall{} as well as at the second level. This allows us to measure links which are even with the ground as well as links with non-zero elevation that connect both levels.
At the ground level, we take measurements at intervals of $1$m up to a total link length of $10$m to characterize signal propagation for increasing distances. For links connecting the upper and lower levels, we consider an elevation angle of about 12$^{\circ}$ and a link distance of $20$m as depicted in \fref{fig:ring}. This measurement allows us to study the performance of industrial mmWave links for cases where the transmitter and the receiver are placed at significantly different heights. Unlike residential or office indoor environments, such height differences are common in industrial settings.

\subsection{\ring{}}
The layout of the \ring{} is depicted in the lower part of \fref{fig:ring}. The walkway inside the ring is surrounded with particle accelerator equipment. This equipment typically features a reflective metal structure as well as a protective polycarbonate plastic plate in some cases. Within the ring, we focus on measuring the range of a mmWave transmission. Since the above metal structures reflect the mmWave signal, we analyze whether transceivers can still communicate even when they are in non-line-of-sight (NLOS) of each other due to the curvature of the ring.

\begin{figure}[tb!]
\centering
 \includegraphics[width=\columnwidth]{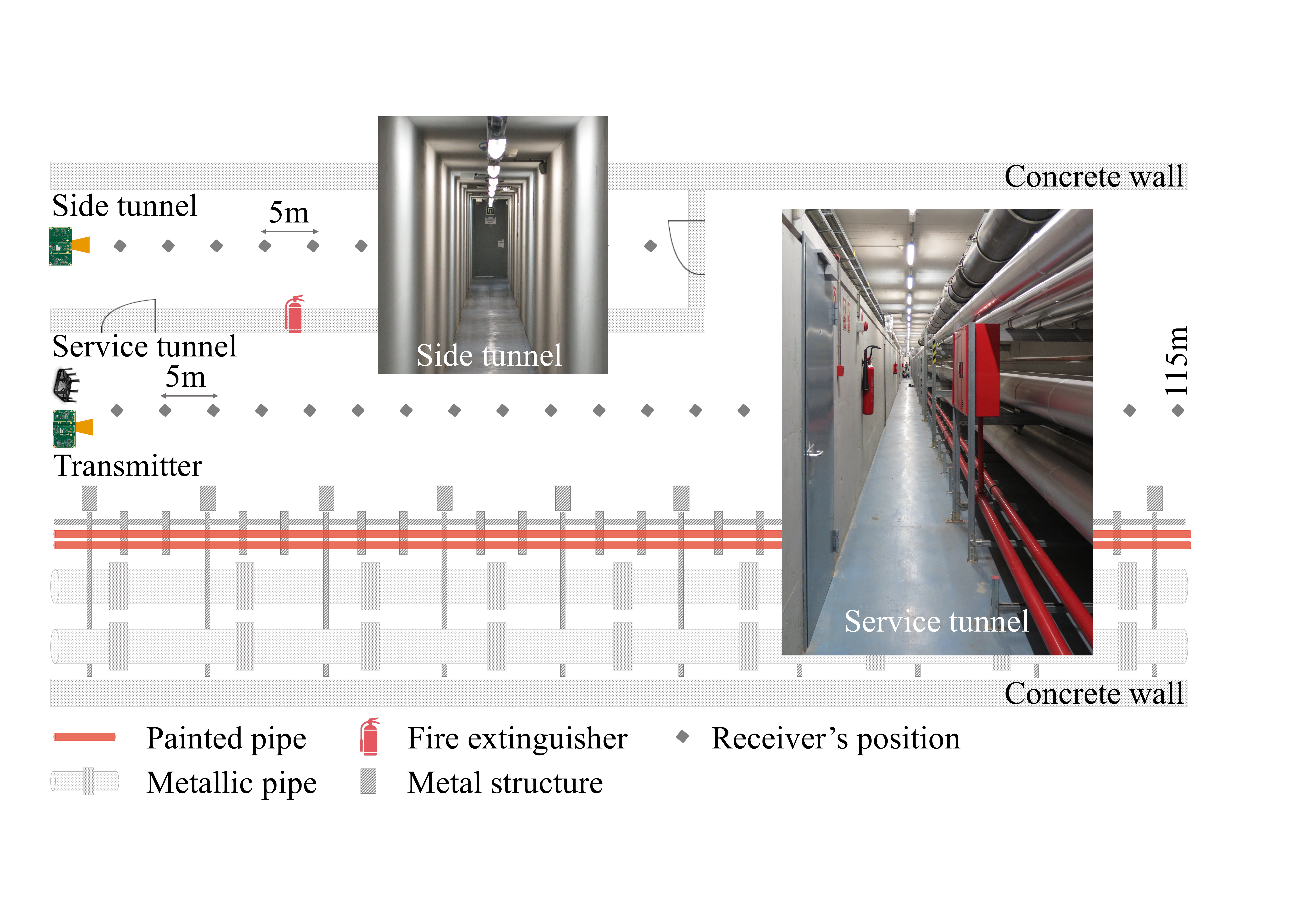}
 \caption{\side{} and Service Tunnel.}
 \label{fig:tunnel}
\end{figure}

\subsection{\tunnel{}}

We collect measurements from two tunnels at the particle accelerator plant. One is a narrow concrete \emph{side} tunnel, whereas the other one is a longer and wider \emph{service} tunnel.

\subsubsection{Side Tunnel} 
\label{subsec:side_tunnel}
The side tunnel has concrete walls and a length of 80 meters, as shown at the upper part of \fref{fig:tunnel}. The width of the tunnel is one meter. It is empty except for light fixtures attached at regular intervals to the ceiling. One side of the tunnel is a plain concrete wall, whereas the other features a metallic door at about half of the length of the tunnel. We place one transceiver at one end of the tunnel and perform measurements for increasing link lengths. In particular, we collect data at five meter intervals. The lack of metallic elements in the side tunnel allows us to understand signal propagation in a narrow enclosed space with very low reflectiveness.

\subsubsection{Service Tunnel} 
\label{subsec:serv_tunnel}
The service tunnel is similar to the side tunnel but features a number of metallic objects which increase the likelihood of reflections. In particular, one side of the tunnel is covered with pipes of different sizes attached to a metal support structure that runs along the whole length of the tunnel (200 meters). Further, fire extinguishers are placed at regular intervals along the tunnel. \fref{fig:tunnel} depicts the layout of the service tunnel. Note that the side tunnel described in Section~\ref{subsec:side_tunnel} lies within the service tunnel but only covers a fraction of its length. We again collect measurements at a regular interval of five meters.

\subsection{\ups{}}

\begin{figure}[tbh!]
\centering
 \includegraphics[width=\columnwidth]{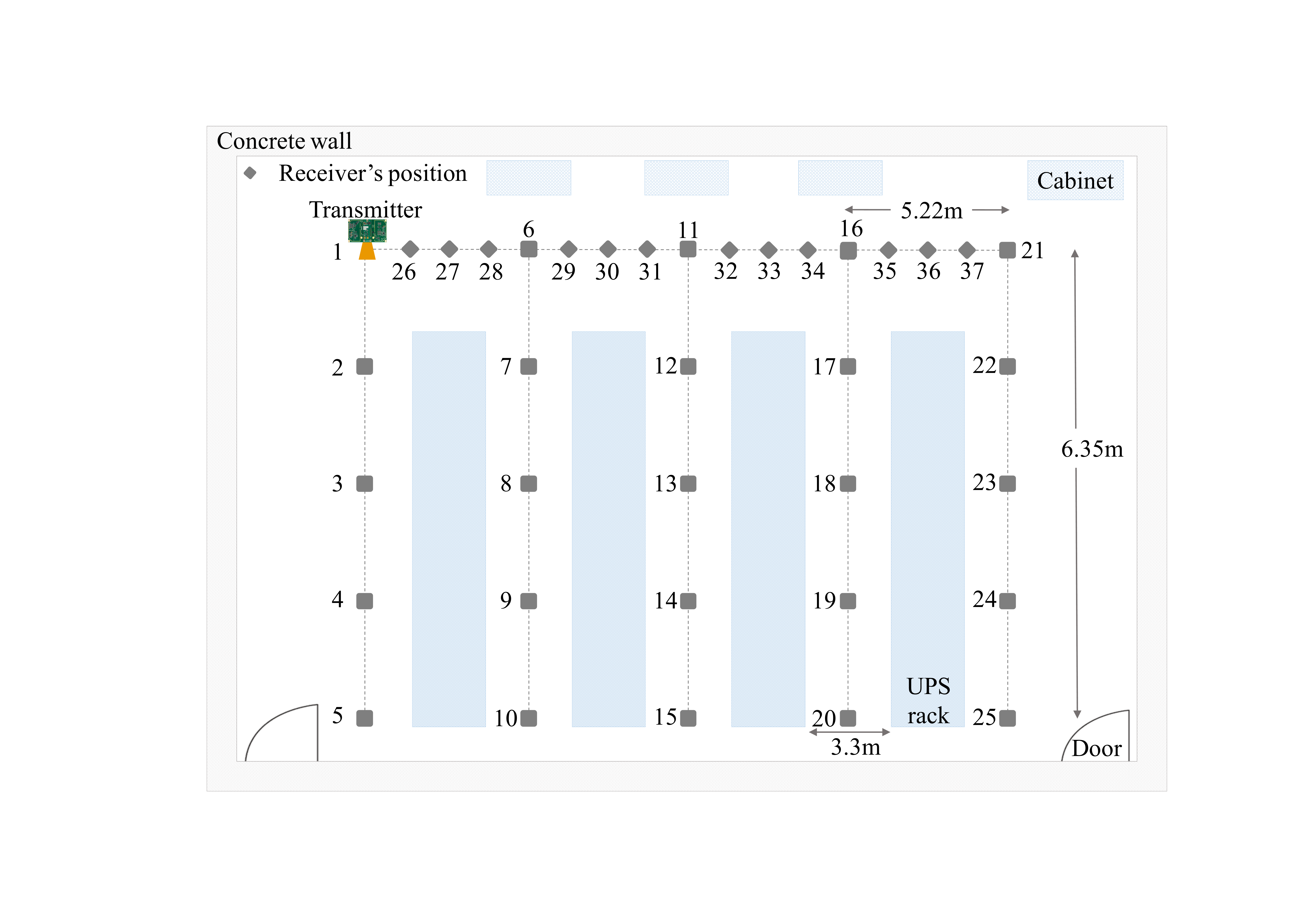}
 \caption{Measurement locations in the \ups{}.}
 \label{fig:ups}
\end{figure}

As a last scenario, we consider a \ups{} with multiple rows of racks that hold the UPS devices. The racks are massive metallic enclosures that fully block any signal in the mmWave band. The goal of the measurement at this location is to study mmWave coverage via reflections in NLOS industrial scenarios. Beyond a \ups{}, similar environments are present in data centers or machinery rooms. As shown in \fref{fig:ups}, we place the transmitter at a fixed location in the upper left corner of the room and move the receiver along the grid of locations depicted in the diagram.

\section{Channel Model}
\label{sec:model}


In the following, we introduce the channel model that we use to characterize the propagation environment in industrial settings. We base our analysis on the channel model defined as part of IEEE 802.11ad \cite{intel_model} because we focus our study on this widely used standard (c.f. Section~\ref{sec:methodology}). Specifically, our model is a well-known and widely accepted geometrical channel model with some extensions that account for the particularities of mmWave propagation.

\subsection{IEEE 802.11ad Model}
\label{subsec:ad_model}
We first introduce the channel model as defined in \cite{intel_model}. Equation~\ref{equ:overall_cir} shows the overall Channel Impulse Response (CIR), which is the sum of $i$ path clusters $C^{(i)}$ with amplitude $A^{(i)}$. Each cluster $C^{(i)}$ is defined as in Equation~\ref{equ:cluster_cir}.

\begin{figure*}[b]ç
\begin{equation}
\begin{aligned}
\label{equ:overall_cir}
h\left( t, \phi_{tx}, \theta_{tx}, \phi_{rx}, \theta_{rx} \right) = \displaystyle\sum_{i} A^{(i)} C^{(i)} \left(t-T^{(i)}, \phi_{tx}-\Phi_{tx}^{(i)}, \theta_{tx}-\Theta_{tx}^{(i)}, \phi_{rx}-\Phi_{rx}^{(i)}, \theta_{rx}-\Theta_{rx}^{(i)} \right)
\end{aligned}
\end{equation}
    
\begin{equation}
\begin{aligned}
\label{equ:cluster_cir}
C^{(i)}\left(t, \phi_{tx}, \theta_{tx}, \phi_{rx}, \theta_{rx} \right) = \displaystyle\sum_{k} \alpha^{(i,k)} \delta\left( t - \tau^{(i,k)} \right)
\delta\left( \phi_{tx} - \phi_{tx}^{(i,k)} \right) \delta\left( \theta_{tx} - \theta_{tx}^{(i,k)} \right)
\delta\left( \phi_{rx} - \phi_{rx}^{(i,k)} \right)
\delta\left( \theta_{rx} - \theta_{rx}^{(i,k)} \right)
\end{aligned}
\end{equation}
\end{figure*}

%

A cluster is the sum of $k$ paths of amplitude $\alpha^{(i,k)}$ and spaced in time $\tau^{(i,k)}$ seconds. The remaining deltas in Equation~\ref{equ:cluster_cir} account for the azimuth and elevation of the path. Basically, a path is only part of the cluster if the antenna points towards the corresponding angle. Otherwise, the deltas in Equation~\ref{equ:cluster_cir} cancel out and the path is not part of $C^{(i)}$. Both the angle-dependency as well as the the cluster structure account for the particularities of mmWave propagation. The former is a direct result of the use of directional antenna beam-patterns. The latter is due to the scarce multi-path propagation environment in the mmWave band. Due to the high path loss and material absorption, most reflections fall below the noise floor at the receiver. In most cases, only the Line-of-Sight (LOS) path and first-order reflections are detectable at the receiver. While related work shows that second-order reflections are feasible as well \cite{Nitsche:2015vq}, they typically occur only in environments with excellent reflectors such as glass. Path clusters are a result of reflections on uneven surfaces such as rugged concrete. The incoming ray scatters on the surface, resulting in one main reflected ray and multiple secondary reflected rays with similar travel direction but with less power each. At the receiver, the main reflection appears as a strong tap in the CIR followed by a number of weaker taps that result of the secondary rays. This bundle of rays forms a path cluster. While similar effects occur also at lower frequencies, the rich multi-path environment at such frequencies masks the clusters in the CIR.

\subsection{Angle-Agnostic Model}
\label{sec:angle_agnostic_model}

The above channel model is strongly dependent on the steering of the particular beam-pattern in use. This ties the channel model to a specific antenna model. To avoid this limitation, we generalize the model in Section~\ref{subsec:ad_model} to be angle-agnostic. As a result, the model in Equations~\ref{equ:overall_cir} and \ref{equ:cluster_cir} are generalized as depicted in Equations~\ref{equ:overall_cir_noangles} and \ref{equ:cluster_cir_noangles}.

\begin{equation}
\label{equ:overall_cir_noangles}
h\left( t \right) = \displaystyle\sum_{i} A^{(i)} C^{(i)} \left(t-T^{(i)} \right)
\end{equation}

\begin{equation}
\label{equ:cluster_cir_noangles}
C^{(i)}\left( t \right) = \displaystyle\sum_{k} \alpha^{(i,k)} \delta\left( t - \tau^{(i,k)} \right)
\end{equation}

This angle-agnostic model still includes all of the cluster and path information, but does not eliminate paths based on antenna steering. Since we aim at computing a statistical channel model which is not limited to a specific location or antenna model, this is a suitable approach. To characterize the channels in our industrial setting, we need the distribution of the following parameters:

\begin{itemize}

	\item The number of clusters in the channel $i$
	\item The amplitude of each cluster $A^{(i)}$
	\item The delay of each cluster $T^{(i)}$
	\item The number of paths within a cluster $k$
	\item The amplitude of each individual path in a cluster $\alpha^{(i,k)}$
	\item The delay of each individual path in a cluster $\tau^{(i,k)}$

\end{itemize}

In Section~\ref{sec:results}, we compute the empirical CDFs of each of the above parameters for each of our scenarios. To this end, we use the data of our exhaustive channel measurements that we carry out as described in Section~\ref{subsec:measurement_setup}. Based on the above statistical distributions, we obtain analytical expressions that describe the behavior of each channel parameter for each of our scenarios.
\section{Results}
\label{sec:results}

In this section, we present the results of our extensive measurement campaign with both the COTS setup described in \sref{subsec:cots_setup} and the measurement hardware setup discussed in \sref{subsec:measurement_setup}. We collect about 70 gigabytes of raw channel traces from which we derive statistical channel parameter distributions. We then correlate our lower layer insights with the performance in terms of throughput that we observe at higher layers. Our analysis reveals that reflectors play a fundamental role---our \side{} scenario, which is a unique location built entirely out of concrete and thus lacking any reflective surfaces, clearly shows a different behavior compared to all of our scenarios. We conclude that the particular characteristics of industrial scenarios are not a hurdle for millimeter-wave communication but rather a benefit in most cases. The vast amplitude of the \hall{}, the reflective behavior of the pipes along the \tunnel{}, and the curvature of the \ring{} are beneficial for signal propagation and thus result in increased coverage compared to a home or office scenario.

\subsection{Parameter Fitting}
\label{subsec:param_fitting}

First, we present an experimental evaluation of the different parameters of the model in \sref{sec:model} for the scenarios described in \sref{sec:setup}. We have obtained these results using the measurement hardware setup described in \sref{subsec:cots_setup}.
Due to safety restrictions in the \ring{} and the \ups{}, we could not perform this analysis in those scenarios. We thus exclude them in the foregoing evaluation. However, we still present upper-layer throughput results for those scenarios in \sref{subsec:throughput}.
For each parameter of the model, we compute its empirical Cumulative Distribution Function (CDF) and perform a goodness of fit analysis. 
The parameterized fitted CDFs are used to draw conclusions on the observed data and can later be exploited to generate synthetic industrial mmWave channels. The research community can use such synthetic channels to evaluate, for instance, the performance of physical and medium access techniques designed for industrial mmWave. Our channel model in \sref{sec:model} is not limited to a specific antenna, which means that it suits the data collected with any of the horn antennas that we use along with our measurement equipment described in \sref{subsec:measurement_setup}. Instead of modeling the decay of the amplitude of each path as transceivers move out of the boresight of each other, we provide the overall picture given a certain location and steering. This approach is the one that fits our goal best, since it allows us to obtain a statistical model that aggregates our measurement insights across locations and steerings for a given antenna beamwidth. Modelling the specific beamwidth of the antenna would be suitable to capture the channel at a particular location. However, that approach would not be suitable to capture the general propagation environment in industrial scenarios.

\subsubsection{Number of Clusters}
\label{subsubsec:num_clusters}

\begin{figure*}[hhht!] 
\centering
\subfigure[\tunnel{}]{\includegraphics[width=0.5\columnwidth]{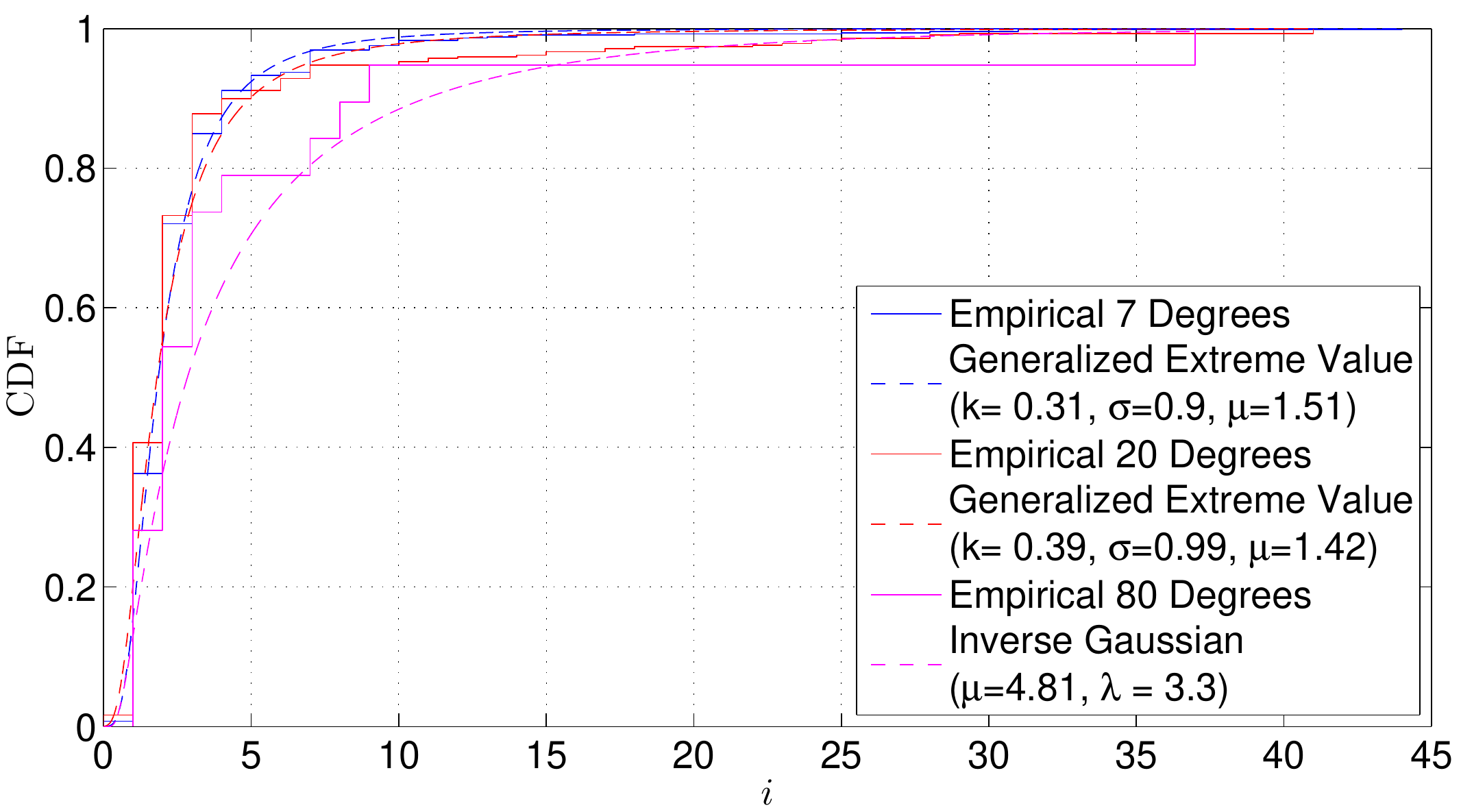}}
\subfigure[\hall{}]{\includegraphics[width=0.5\columnwidth]{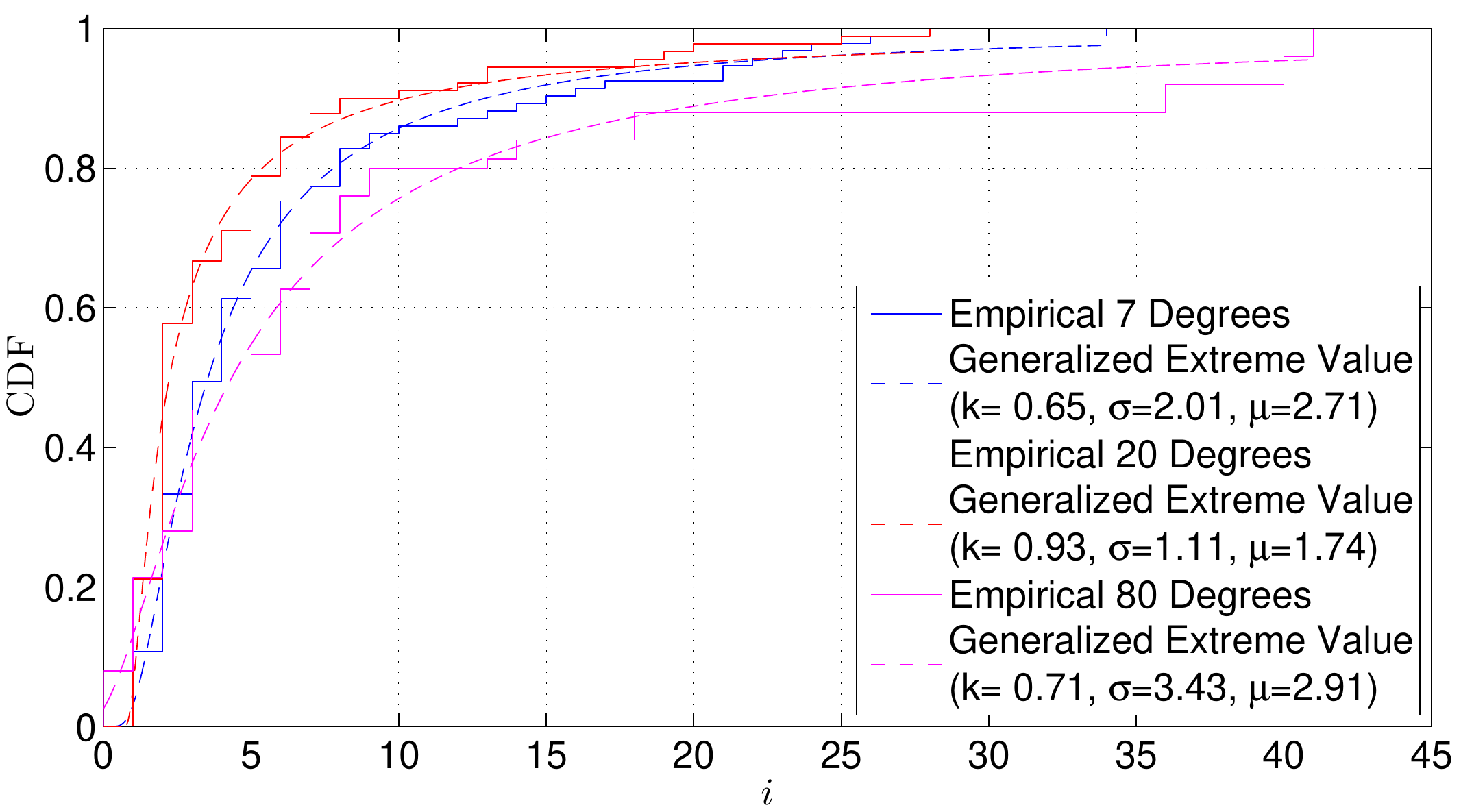}\label{fig:npaths_hall}}
\subfigure[\pipe{}]{\includegraphics[width=0.5\columnwidth]{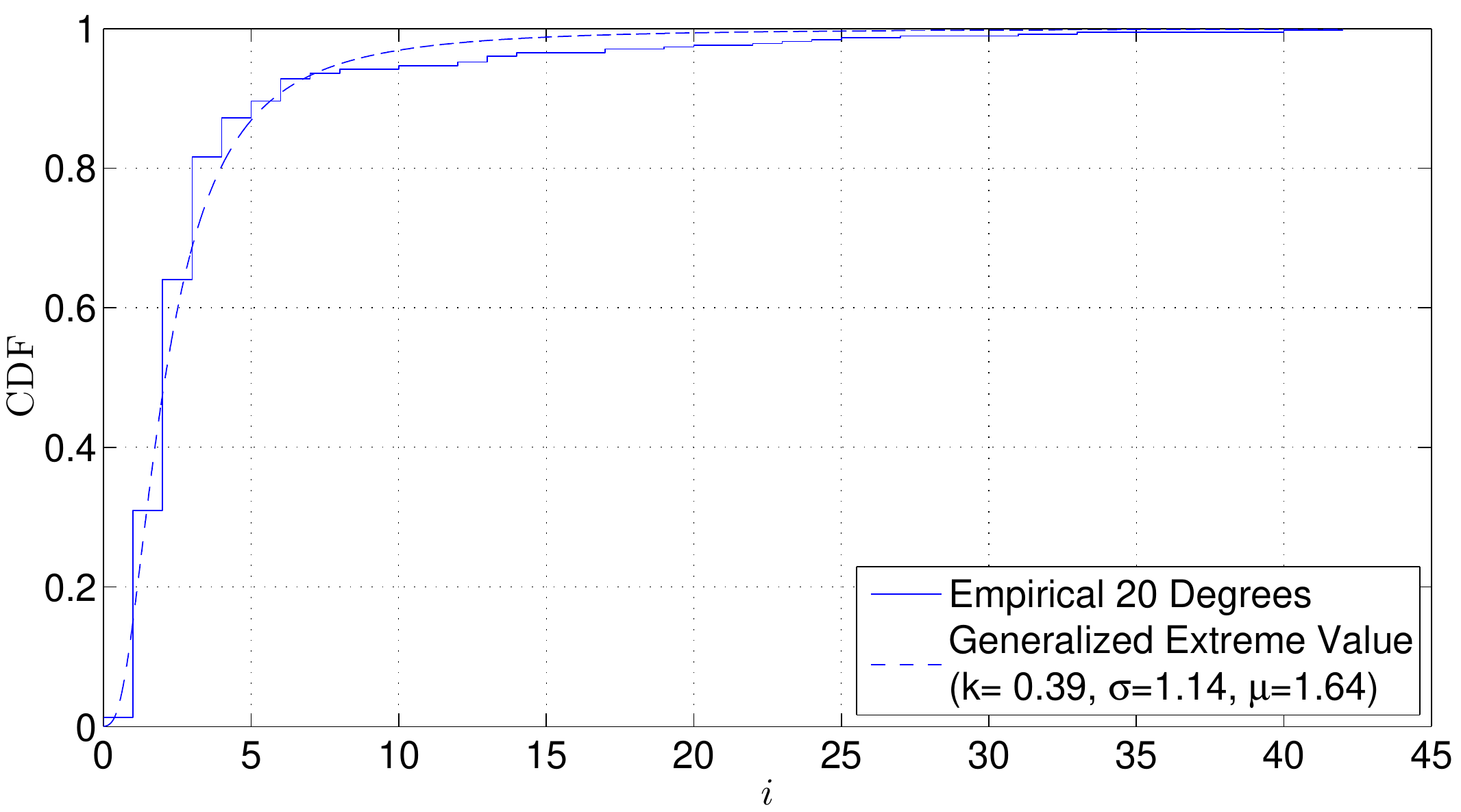}}
\subfigure[\side{}]{\includegraphics[width=0.5\columnwidth]{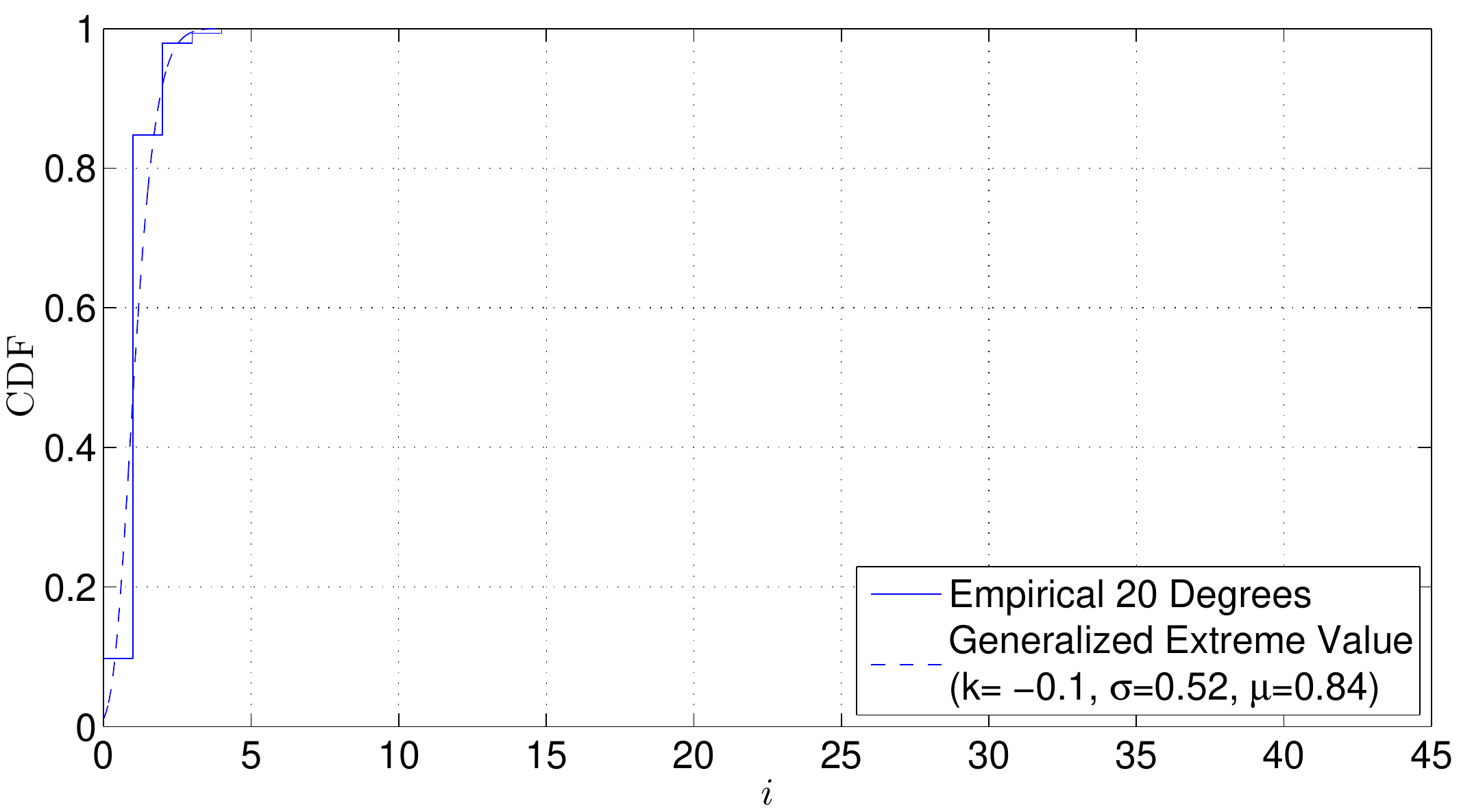}\label{fig:npaths_side}}
\caption{Empirical and best goodness of fit CDFs of the number of clusters ($i$).}
\label{fig:npaths}
\end{figure*}

\fref{fig:npaths} shows the CDFs of the number of clusters ($i$) for the \tunnel{}, \hall{}, \pipe{}, and \side{}. As discussed in \sref{sec:model}, each cluster contains the scattered rays that propagate along one geometrical path. The cluster of the LOS path typically consists of a single ray since it does not reflect on any surface, but clusters from reflected paths often include a bundle of rays. \fref{fig:npaths} depicts the aggregated data at the different positions described in \sref{sec:setup}. We aggregate the data for each of the available beam widths (i.e., $7$, $20$ and $80$ degrees) across all locations to generalize our results. As expected, \fref{fig:npaths} shows that the amount of observable clusters is larger the wider the beam width since wider beams capture more propagation paths. For the \pipe{} and the \side{}, we only collect measurements with the $20$-degree antenna because the $7$- and $80$-degree antennas do not reveal further paths. In the case of the \pipe{}, this is due to the large amount of obstacles. When using the $20$-degree antenna, communication is not limited due to range but due to blockage. Thus, switching to the narrow $7$-degree antenna does not provide additional path information. While the $80$-degree antenna could capture additional paths, its gain is too limited at most locations in the \pipe{}. In the \side{}, the geometry of the environment allows us to receive a strong signal even for large link lengths. The $7$-degree antenna does not provide further information. The gain of the $80$-degree antenna is again too limited.

%


\begin{table}[tb!]
\centering
    \caption{Quartiles of the number of clusters ($i$).\label{tbl:npaths}\vspace{-3.5mm}}{%
\begin{tabular}{|c|c|c|c|c|} \hline
Scenario & Degree & 1st Quartile & Median & 3rd Quartile\\ \hline
Tunnel & 7 & 2 & 2 & 2 \\ \hline
Tunnel & 20 & 1 & 2 & 2 \\ \hline
Tunnel & 80 & 1 & 2 & 3\\ \hline
Exp. Hall & 7 & 2 & 4 & 6 \\ \hline
Exp. Hall & 20 & 2 & 2 & 3 \\ \hline
Exp. Hall & 80 & 2 & 5 & 8 \\ \hline
Pipe Room & 20 & 2 & 2 & 3 \\ \hline
Side Tunnel & 20 & 1 & 1 & 1 \\ \hline
\end{tabular}}
\end{table}

In \fref{fig:npaths}, we observe that the \hall{} is the scenario with the highest number of clusters. Specifically, among $20$-$40\%$ of the samples show more than $5$ clusters and among $10$-$20\%$ of the samples show more than $10$ clusters. That is, we observe a large number of reflective propagation paths. While the \hall{} is large, the beamlines and other machinery located next to the \ring{} offer a large number of reflective surfaces that result in the high number of clusters in \fref{fig:npaths_hall}.

In the \tunnel{} and \pipe{} instead, we find that among $5$-$20\%$ of the samples show more than $5$ paths and less than $10\%$ show 10 or more clusters. In the \pipe{}, this is due to the large number of blockages. While the machinery in the room allows for a large number of reflections, during our measurements we observed that most of them were shadowed by the machinery itself. This is an interesting difference to the \hall{}---while both environments are highly reflective, the contrast in terms of size and thus likelihood of blockage results in very different channels. In the case of the \tunnel{}, the limited number of clusters is due to the elongated geometry of the environment. While the pipes on one side of the \tunnel{} (c.f. \sref{subsec:serv_tunnel}) allow for reflections, geometrically we only receive a handful of them at each measurement location. This is in contrast to the \hall{} or the \pipe{}, where reflectors are distributed homogeneously. This effect is further exacerbated in the \side{}, since it does not feature reflective pipes but only concrete walls. Hence, the \side{} is the scenario with the smallest number of paths. As depicted in \fref{fig:npaths_side}, we have not found any instance of more than five paths. \tref{tbl:npaths} depicts numerical values of the 1st quartile, median, and 3rd quartile of the CDFs in \fref{fig:npaths} to enable computational use of the data. 

In \fref{fig:npaths} we observe that, with the only exception of the $80$-degree antenna in the \tunnel{}, all empirical distributions of the number of clusters can be well described by a Generalized Extreme Value distribution with parameters $0.31 \leq k \leq 0.93$, $0.9 \leq \sigma \leq 3.43$, $1.42 \leq \mu \leq 2.91$. 
The data also suggests that, with the exception of the \side{}, the distributions for this parameter are quite similar.
As discussed above, the propagation environment of the \side{} is indeed peculiar and does not resemble the characteristics of the other rooms.
Thus, the dissimilarity in the data from the \side{} compared to the other rooms is inherit to the specific propagation environments.
 
\subsubsection{Inter-cluster Delays}
\label{subsubsec:inter-cluser-delays}
 
\begin{figure*}[hhh!] 
\centering
\subfigure[\tunnel{}]{\includegraphics[width=0.5\columnwidth]{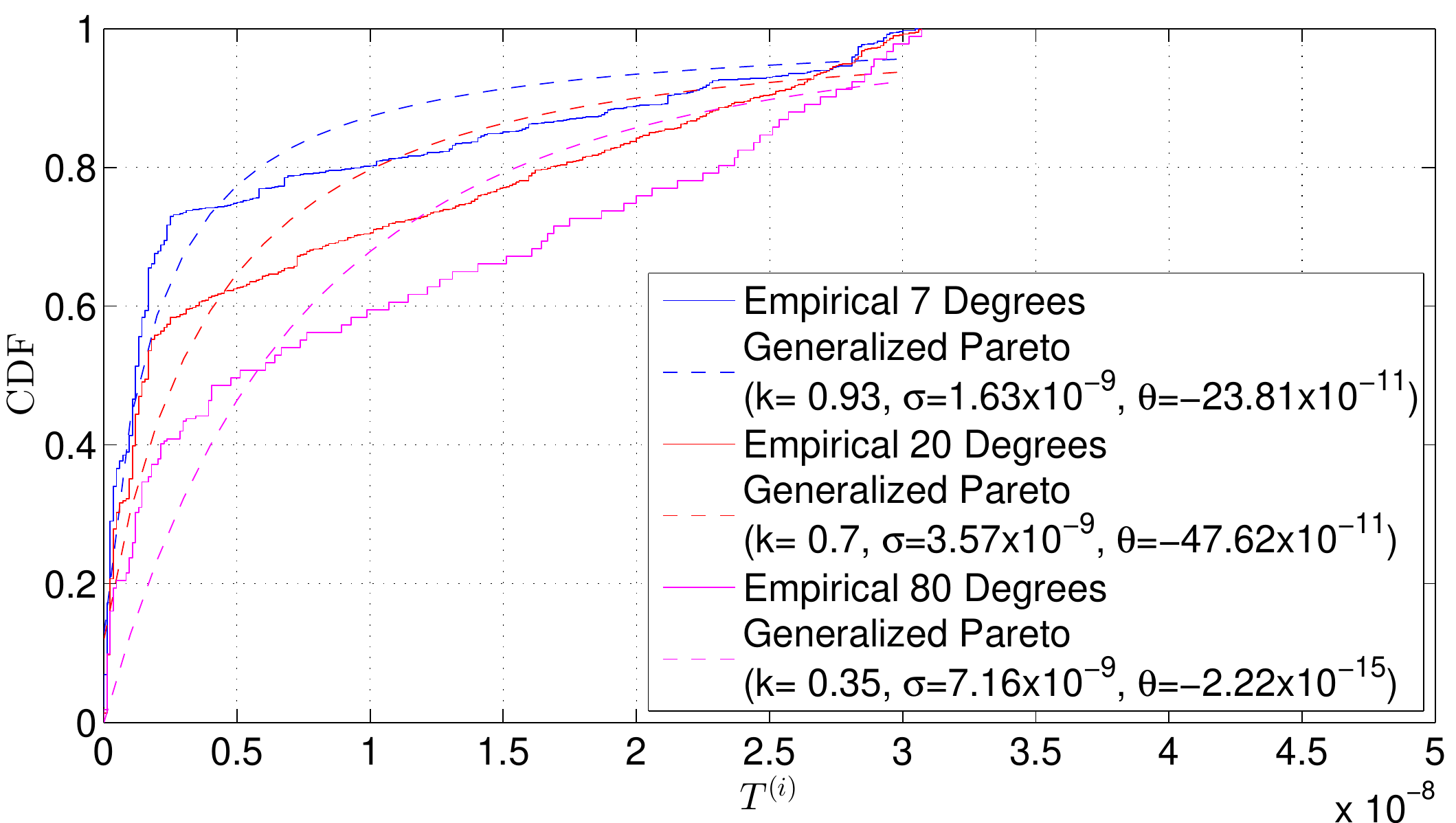}\label{fig:delays_tunnel}}
\subfigure[\hall{}]{\includegraphics[width=0.5\columnwidth]{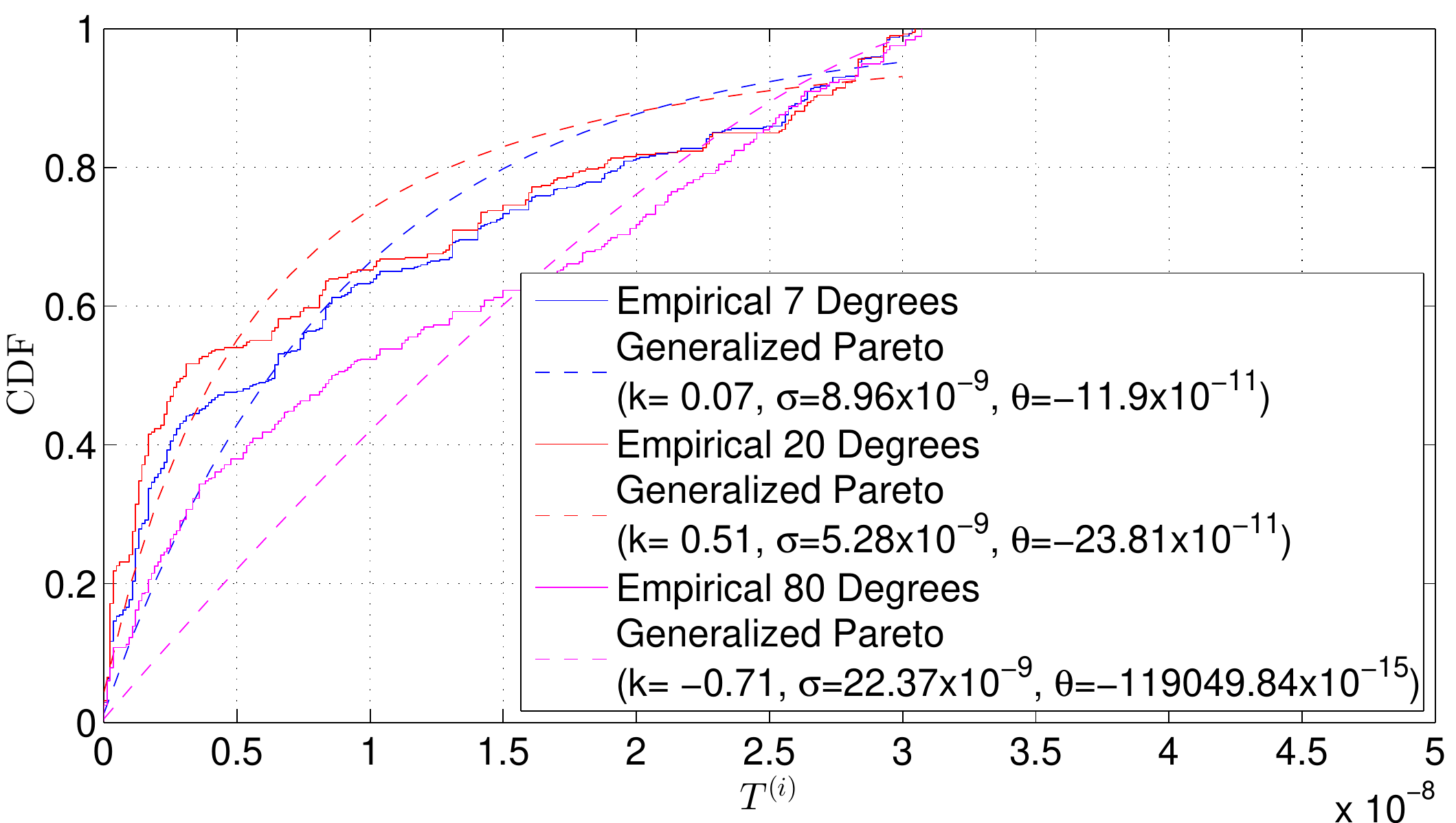}\label{fig:delays_hall}}
\subfigure[\pipe{}]{\includegraphics[width=0.5\columnwidth]{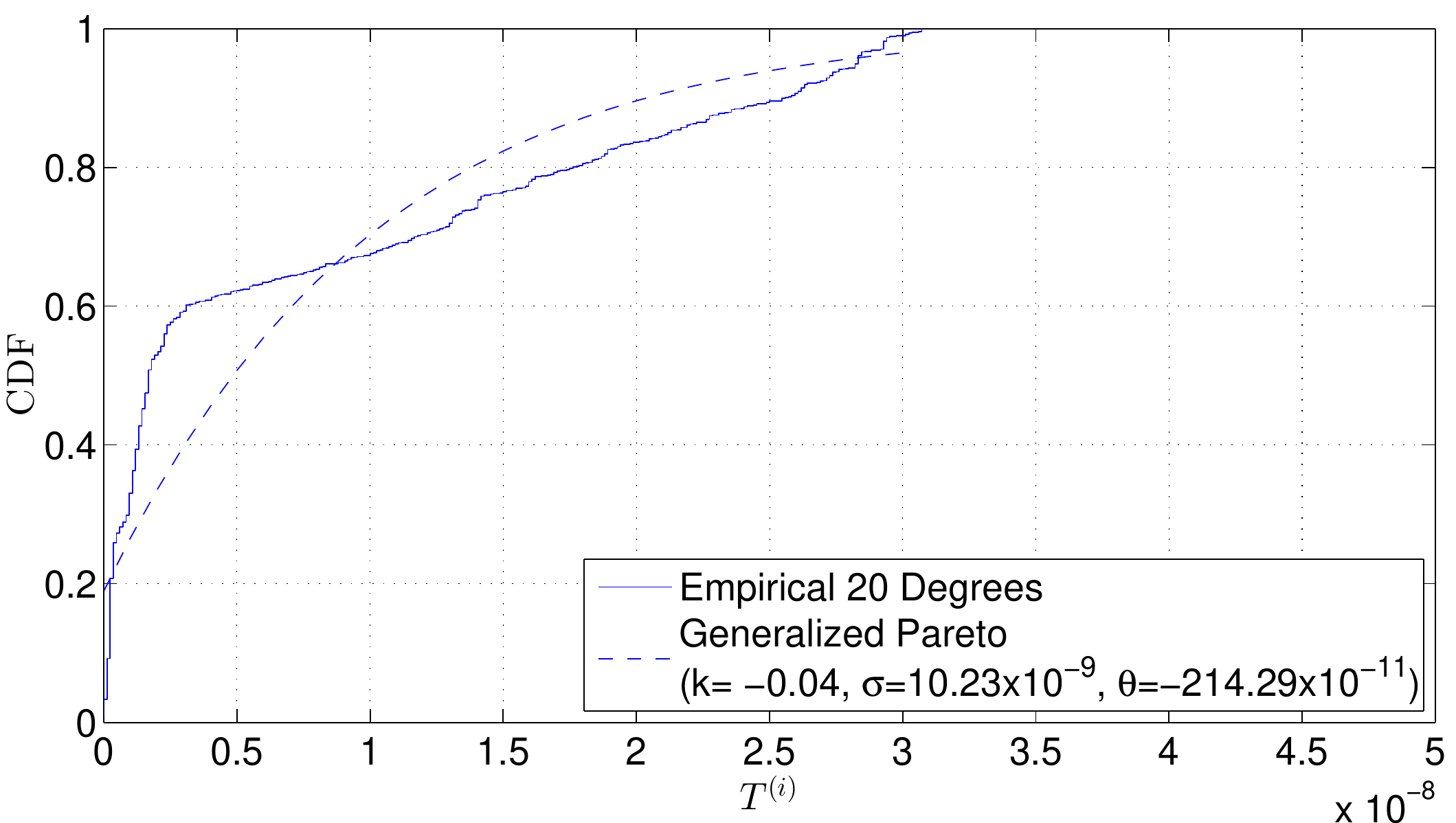}}
\subfigure[\side{}]{\includegraphics[width=0.5\columnwidth]{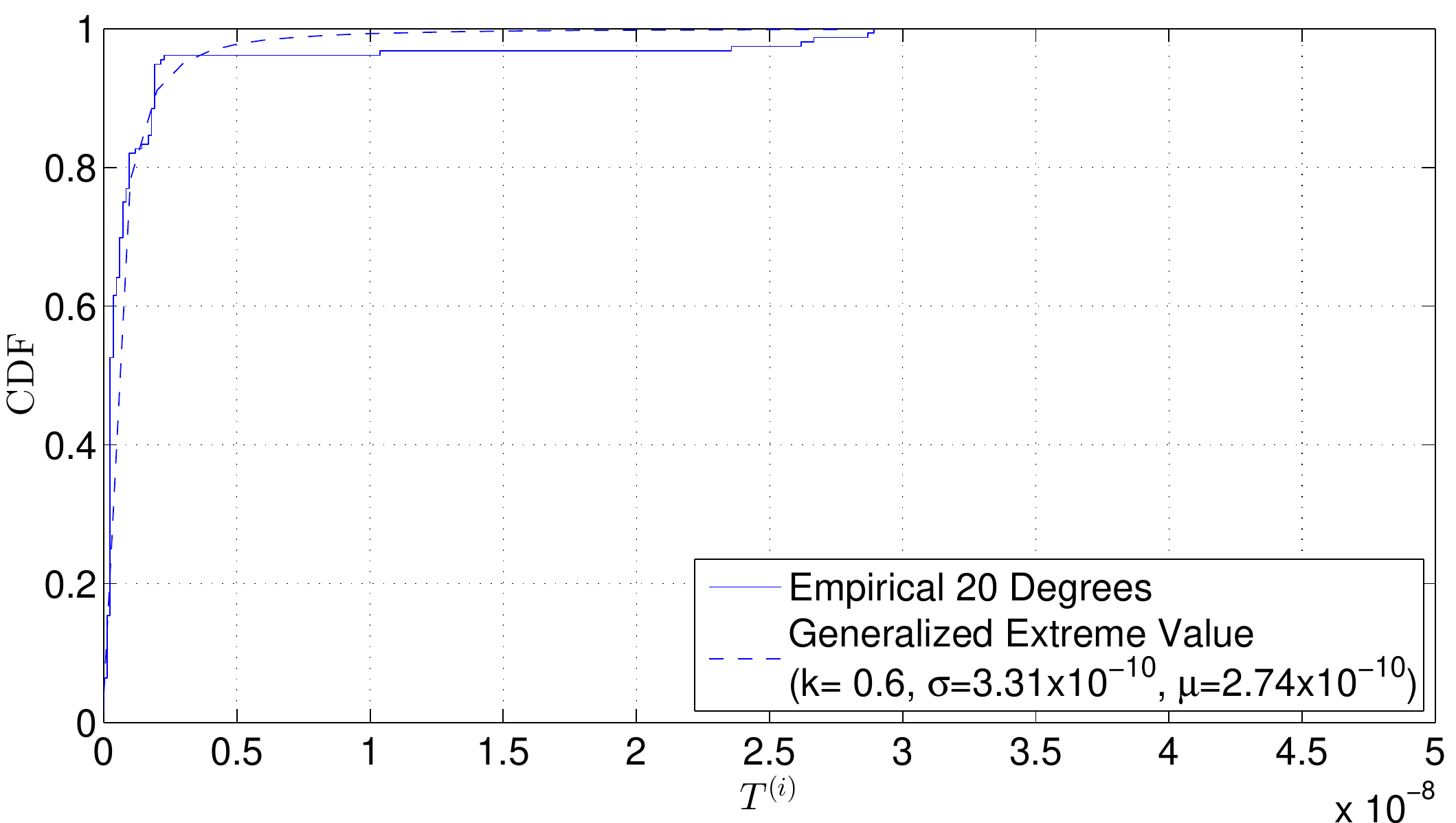}}
\caption{Empirical and best goodness of fit CDFs of the inter-cluster delay ($T^{(i)}$).}
\label{fig:delays}
\end{figure*}

Similar to \fref{fig:npaths}, \fref{fig:delays} shows the empirical and best goodness of fit CDFs of the inter-cluster delays ($T^{(i)}$). The inter-cluster delay is essentially the time between the arrival of two subsequent clusters in the CIR (c.f. \sref{sec:model}).
In this case, we observe rather long inter-cluster delays in the \tunnel{}, \hall{}, and \pipe. Specifically, $20$-$60\%$ of the cases lie in the range of $5$ to $30$~ns. This means that the propagation paths on which the clusters travel differ about 1.5 to 9 meters in terms of length. As expected, \fref{fig:delays_hall} shows that the latter is particularly frequent in the \hall{} when using the 80-degree antenna. The large dimensions of the \hall{} and the wide beam at the receiver allow us to detect clusters that travel along very different paths. This is very beneficial to avoid transient blockage, since the probability that all of the paths are blocked simultaneously is low. Transceivers do not need to use a wide 80-degree antenna for communication to benefit from this resilience to blockage. Instead, they may use narrow beam widths---which result in higher gain and thus higher throughput---and simply re-steer to a different path whenever the current one is blocked. 

For both the \tunnel{} and the \hall{}, we observe that narrower beam-widths tend to result in smaller inter-cluster delays. This is expected, since spatial selectivity increases. However, the effect is less apparent in the \hall{} than in the \tunnel{}. From this, we conclude that the paths are particularly diverse in the \hall{}. As depicted in \fref{fig:ring}, reflectors are located far and at a significant angle from the link. Thus, the difference among the 7-degree and the 20-degree antennas is limited---we need to switch to the 80-degree beam pattern in order to capture some of the reflections. In other words, the angular spread of this particular propagation environment is high. Conversely, the narrow and elongated nature of the \tunnel{} results in a clear difference among the three antennas in \fref{fig:delays_tunnel}. The angular spread is limited and thus switching from 7 to 20 and from 20 to 80 degrees makes a clear difference in terms of the length of the paths that we observe. The length differences are still significant in spite of the \tunnel{} being narrow due to its sheer length. That is, even if paths are roughly parallel due to the limited angular spread, a slight angle difference over up to 115 meters makes a difference.

While we observe an equivalent behavior in the \pipe{}, the CDF reveals that difference in terms of length of the paths is clearly smaller than in the \hall{}, rather resembling the \tunnel{}. This is due to the high number of obstacles---although the dimensions of the room would allow for diverse propagation paths, most of them are blocked. In contrast, the \side{} exhibits a distinctly different behavior. The length of almost all of the paths differs by at most 60 centimeters. This is due to the lack of reflectors in the \side{}. Since barely any reflections exist, we observe a single LOS cluster in most cases. Additional clusters, if at all, travel on similar paths. 

In \tref{tbl:delays}, we provide again numerical values of the 1st quartile, median, and 3rd quartile to allow for computational processing of our data. From \fref{fig:delays} we observe that for all rooms except for the \side{}, the inter-cluster delays can be characterized by a Generalized Pareto distribution with parameters $-0.71 \leq k \leq 0.93$, $1.63\cdot 10^{-9} \leq \sigma \leq 22.37\cdot 10^{-9}$ and $-214.29\cdot 10^{-11} \leq \theta \leq -2.22\cdot 10^{-15}$. Again we observe that the distributions of this parameter in the \tunnel{}, \hall{}, and \pipe{} are quite similar. The distribution of the inter-cluster delay for the \side{} differs considerably and can be better described by a Generalized Extreme Value distribution.

\begin{table}[tb!]
\centering
\caption{Quartiles of the inter-cluster delay ($T^{(i)}$).\label{tbl:delays}\vspace{-3.5mm}}{%
\begin{tabular}{|c|c|c|c|c|} \hline
Scenario & Degree & 1st Quartile & Median & 3rd Quartile\\ \hline
Tunnel & 7 & $2.4\cdot10^{-10}$ & $1.3\cdot10^{-9}$ & $5.8\cdot10^{-9}$ \\ \hline
Tunnel & 20 & $3.6\cdot10^{-10}$ & $1.7\cdot10^{-9}$ & $1.4\cdot10^{-8}$ \\ \hline
Tunnel & 80 & $1.1\cdot10^{-9}$ & $5.1\cdot10^{-9}$ & $2.0\cdot10^{-8}$\\ \hline
Exp. Hall & 7 & $1.3\cdot10^{-9}$ & $6.4\cdot10^{-9}$ & $1.6\cdot10^{-8}$ \\ \hline
Exp. Hall & 20 & $1.1\cdot10^{-9}$ & $3.1\cdot10^{-9}$ & $1.6\cdot10^{-8}$  \\ \hline
Exp. Hall & 80 & $2.4\cdot10^{-9}$ & $8.7\cdot10^{-9}$ & $2.1\cdot10^{-8}$  \\ \hline
Pipe Room & 20 & $3.6\cdot10^{-10}$ & $1.7\cdot10^{-9}$ & $1.4\cdot10^{-8}$ \\ \hline
Side Tunnel & 20 & $2.4\cdot10^{-10}$ & $2.4\cdot10^{-10}$ & $3.6\cdot10^{-10}$  \\ \hline
\end{tabular}}
\end{table}

\subsubsection{Amplitude of Each Cluster}
\label{subsubsec:cluster_amplitude}

\begin{figure*}[hhh!] 
\centering
\subfigure[\tunnel{}]{\includegraphics[width=0.5\columnwidth]{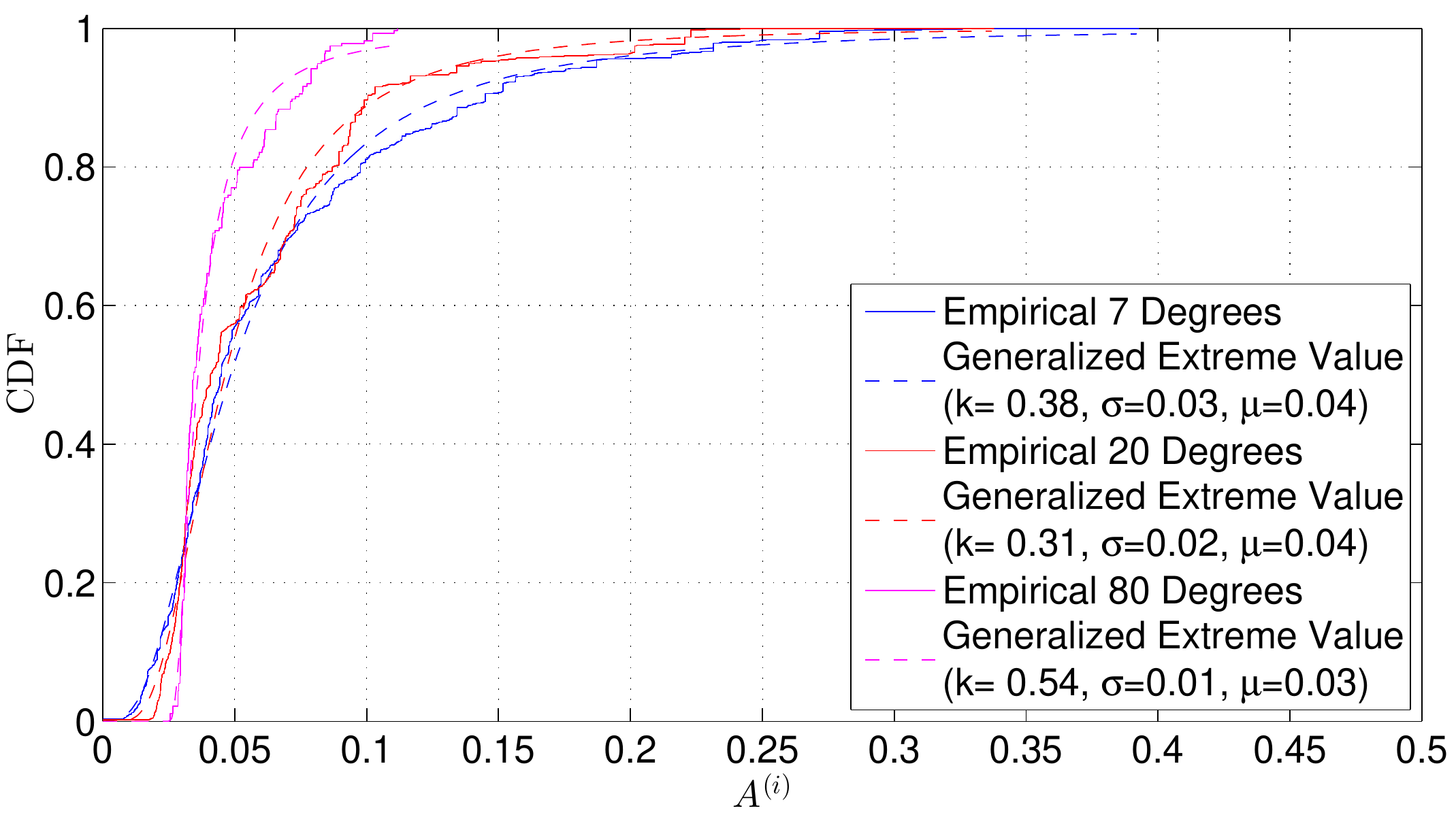}}
\subfigure[\hall{}]{\includegraphics[width=0.5\columnwidth]{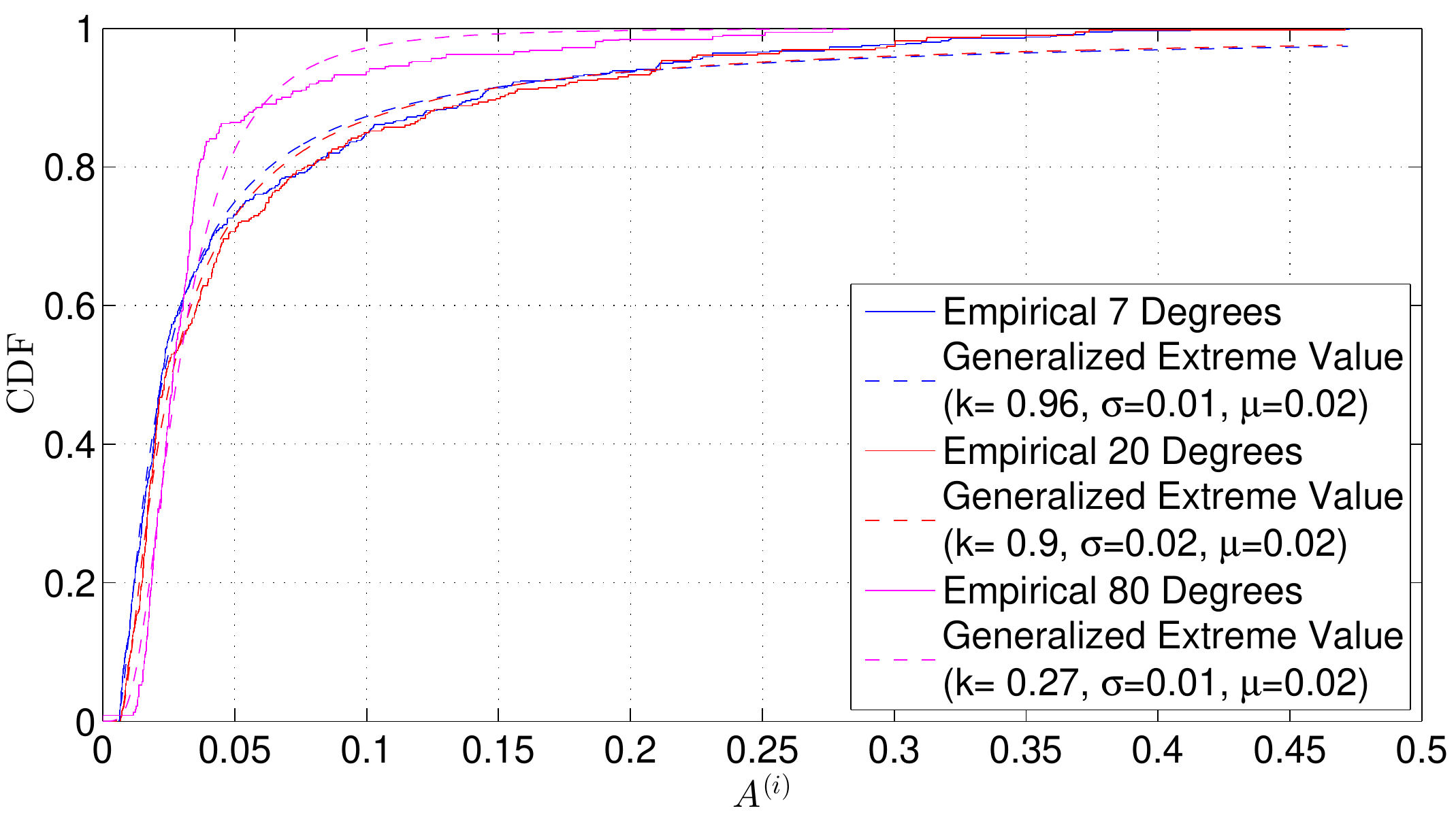}}
\subfigure[\pipe{}]{\includegraphics[width=0.5\columnwidth]{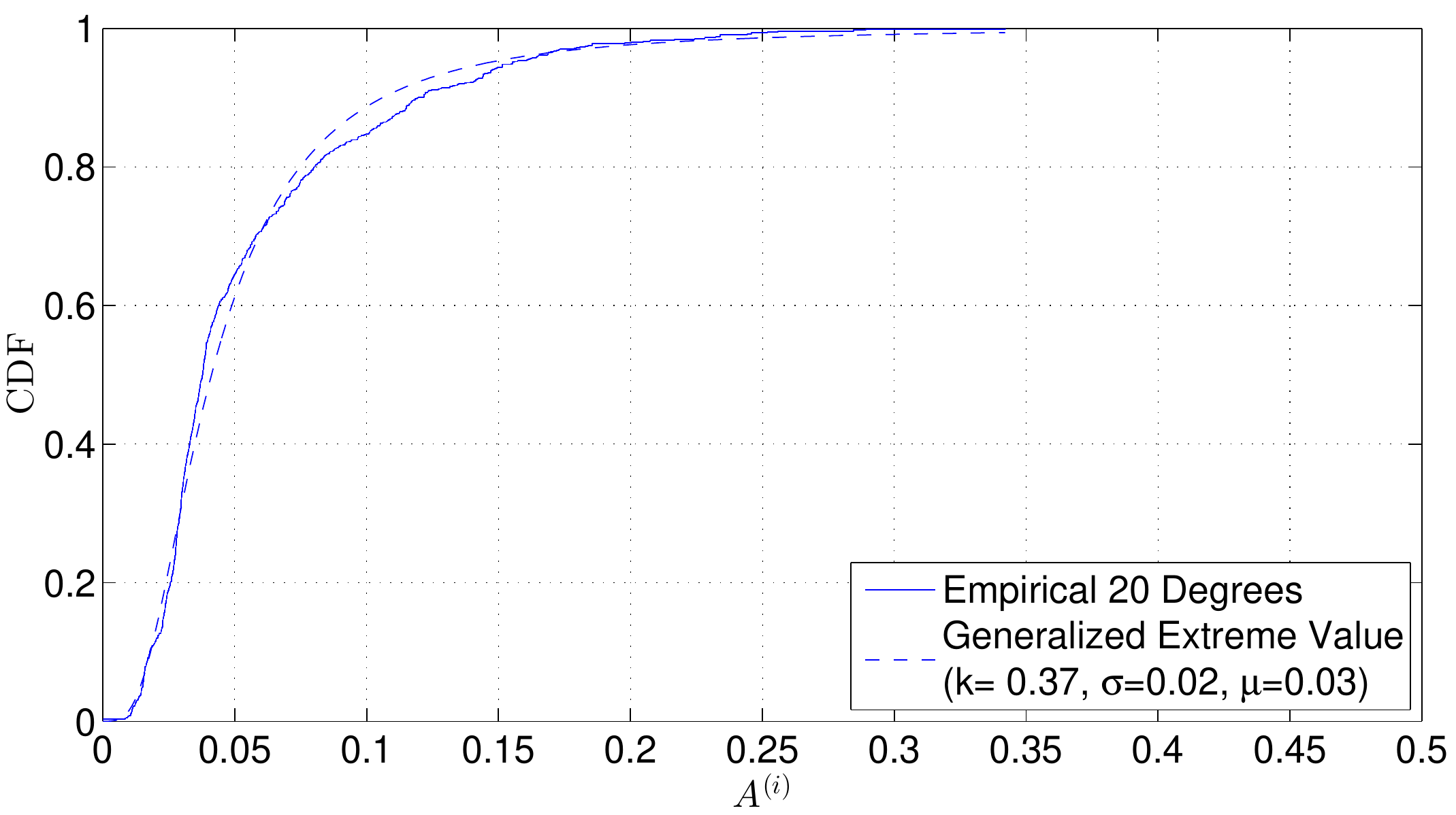}}
\subfigure[\side{}]{\includegraphics[width=0.5\columnwidth]{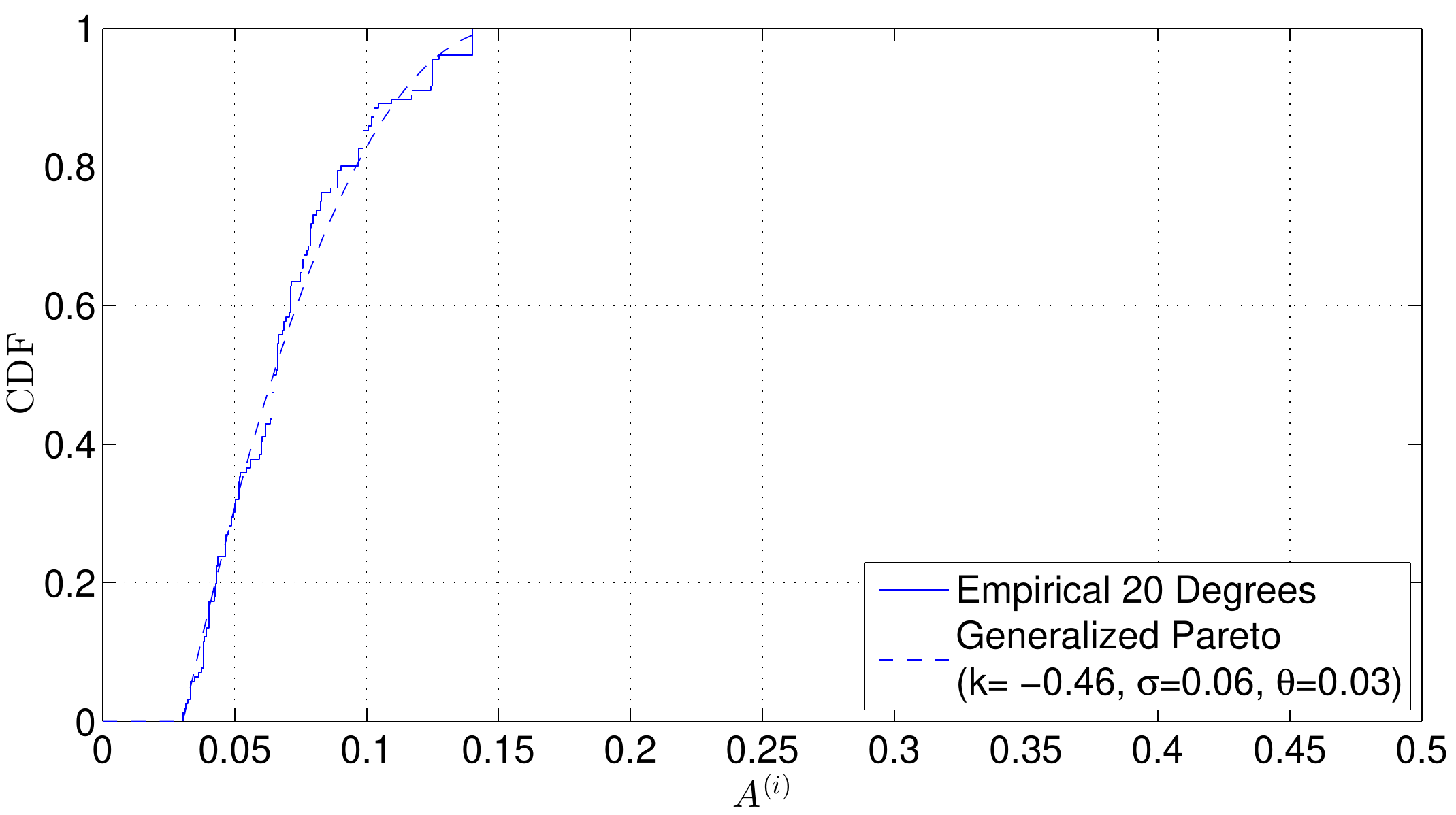}}
\caption{Empirical and best goodness of fit CDFs of the amplitude of each cluster ($A^{(i)}$).}
\label{fig:mpower}
\end{figure*}

\fref{fig:mpower} shows the CDFs of the amplitude of each cluster ($A^{(i)}$). Large amplitude values result from a strong signal at the receiver. We expect the CDF to shift to the left for wide beam patterns since wide patterns capture more reflected paths, which are longer and thus inherently weak due to the high propagation loss. While a wide antenna pattern still captures the comparatively strong LOS path, this strong path plays a minor role in the overall distribution of all the reflected paths. \fref{fig:mpower} confirms this behavior for the two scenarios for which we compare different antennas. In both the \tunnel{} and the \hall{}, the distributions tend to the left as we increase the beam width. For the same reasons as in \sref{subsubsec:inter-cluser-delays}, the 7 and 20 degree antenna behave similarly in the \hall{}. However, in this case both antennas also behave roughly similar in the \tunnel{}. That is, the strength of the paths that we observe with the 7-degree antenna is to some extent similar to the one we observe with the 20-degree antenna. Still, the results in \sref{subsubsec:inter-cluser-delays} show that the paths for both antennas are indeed different, unlike in the \hall{}. While the data does not allow for a clear explanation of this behaviour, the underlying reason is likely related to the number of reflections within each path. Due to the elongated nature of the \tunnel{}, the paths that we observe with the 7- and 20-degree antennas probably reflect only once on the metallic pipes along the tunnel. In contrast, the 80-degree antenna is capable of capturing paths that reflect twice (or more) on the pipes, thus arriving at the receiver with a smaller amplitude.

Overall, \fref{fig:mpower} shows larger amplitudes for the \hall{} than for the \tunnel{}. This is due to the link distance, which is much shorter in the \hall{}. In the \pipe{}, cluster amplitudes are similar to the behavior in the \hall{} for the 20-degree antenna because the length of most of the unobstructed links is in the same order of magnitude. Still, we observe a higher fraction of strong clusters in the \hall{} than in the \pipe{}. This is expected since the \hall{} features a clear and strong LOS path, whereas in the \pipe{} more paths are NLOS. In the \side{}, the cluster amplitudes are distributed similarly to the vast majority of our measurements in the other three scenarios. However, in contrast to the \hall{}, \tunnel{}, and \pipe{}, we do not observe any particularly strong clusters. We conjecture that such strong clusters are the result of constructive interference. Since the \side{} lacks reflective elements, constructive interference does not occur and the maximum amplitude of clusters is limited.

Numerical values of the 1st quartile, median, and 3rd quartile are provided in \tref{tbl:mpower}. Regarding the modelling of the distributions, in \fref{fig:mpower} we see that the amplitude of each cluster can be accurately described by a Generalized Extreme Value Distribution for the \tunnel{}, \hall{}, and \pipe{} with parameters ranging among $0.27 \leq k \leq 0.96$, $0.01 \leq \sigma \leq 0.03$, $0.02 \leq \mu \leq 0.04$.
As before, we observe that these distributions are quite similar for the different rooms.
In contrast, the unique behavior of the \side{} in terms of the cluster amplitude can be described as a Generalized Pareto Distribution.

\begin{table}[tb!]
\centering
\caption{Quartiles of the amplitude of each cluster ($A^{(i)}$).\label{tbl:mpower}\vspace{-3.5mm}}{%
\begin{tabular}{|c|c|c|c|c|} \hline
Scenario & Degree & 1st Quartile & Median & 3rd Quartile\\ \hline
Tunnel & 7 & 0.030 & 0.044 & 0.083 \\ \hline
Tunnel & 20 & 0.031 & 0.041 & 0.075 \\ \hline
Tunnel & 80 & 0.031 & 0.034 & 0.046\\ \hline
Exp. Hall & 7 & 0.014 & 0.022 & 0.054 \\ \hline
Exp. Hall & 20 & 0.016 & 0.024 & 0.063 \\ \hline
Exp. Hall & 80 & 0.020 & 0.027 & 0.035 \\ \hline
Pipe Room & 20 & 0.028 & 0.038 & 0.070 \\ \hline
Side Tunnel & 20 & 0.047 & 0.065 & 0.083 \\ \hline
\end{tabular}}
\end{table}

\subsubsection{Number of Paths Within a Cluster}

\begin{figure*}[hhh!] 
\centering
\subfigure[\tunnel{}]{\includegraphics[width=0.5\columnwidth]{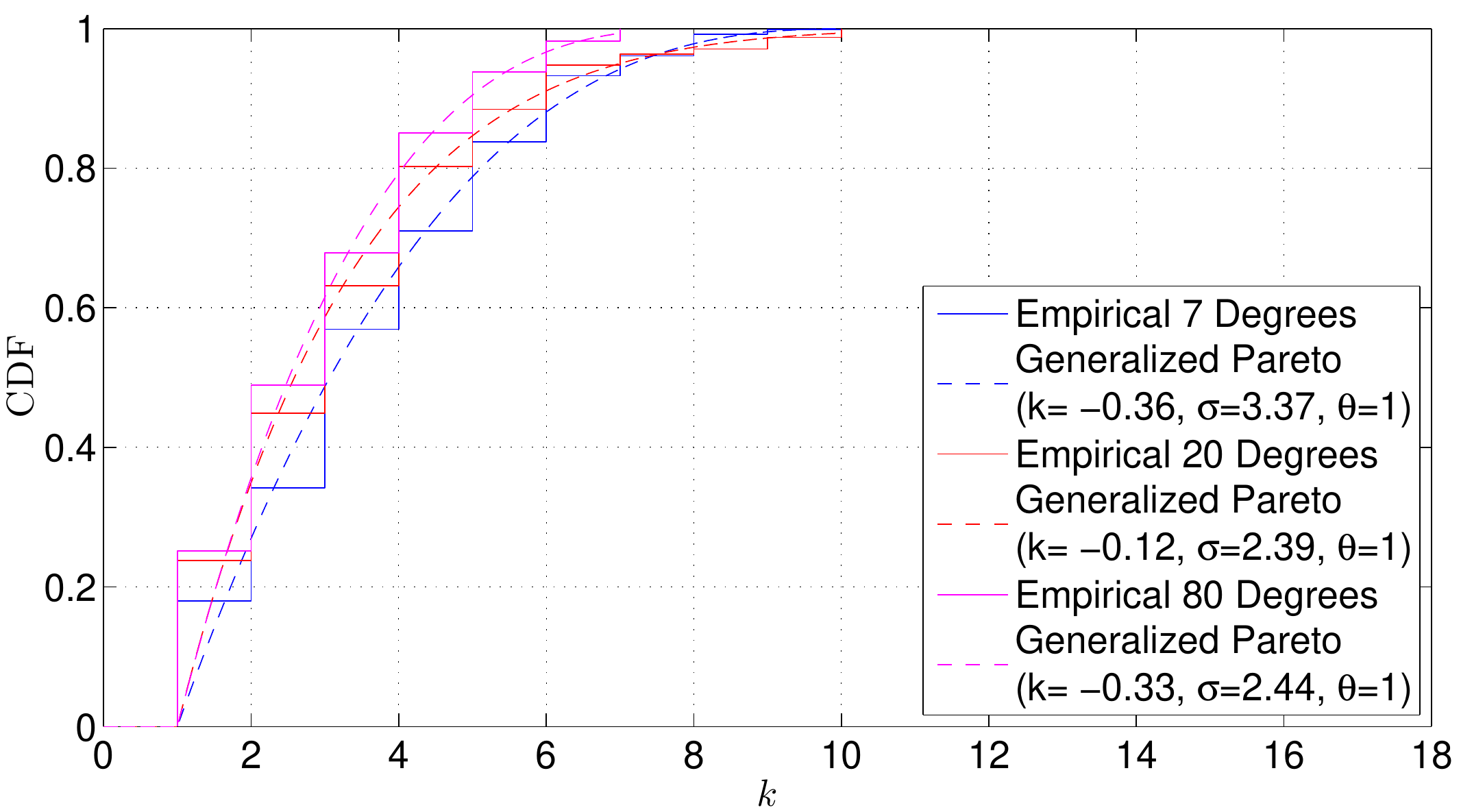}}
\subfigure[\hall{}]{\includegraphics[width=0.5\columnwidth]{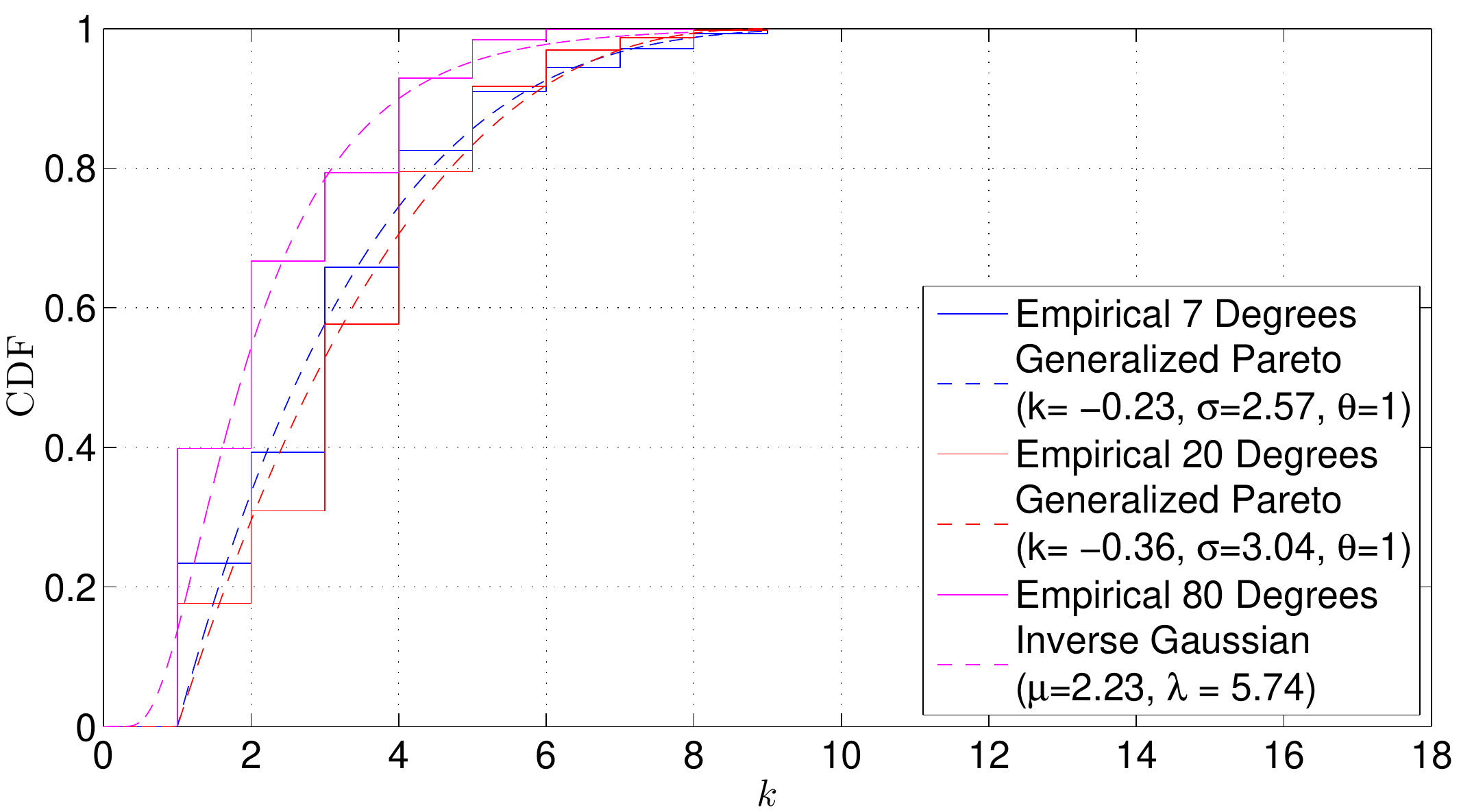}}
\subfigure[\pipe{}]{\includegraphics[width=0.5\columnwidth]{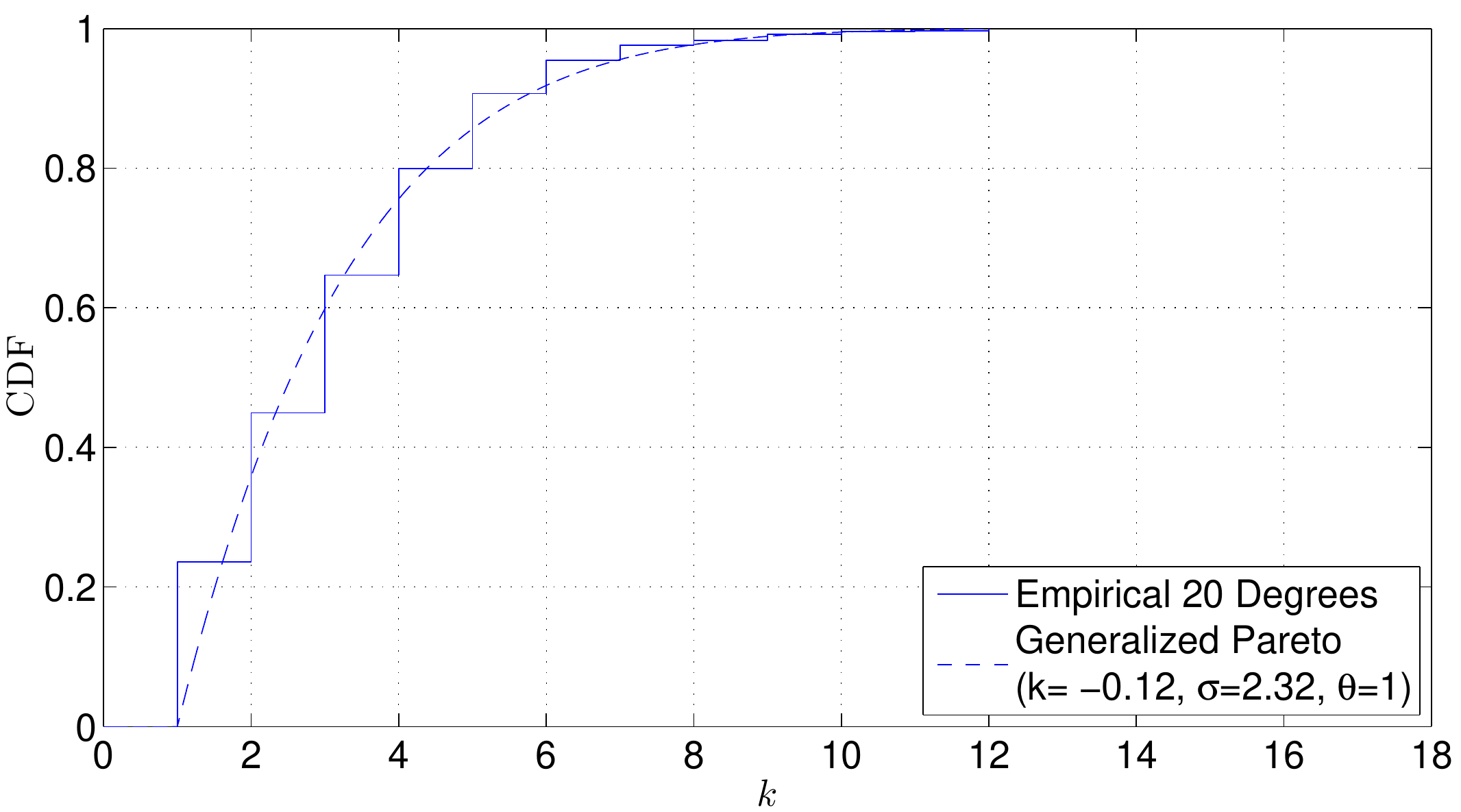}\label{fig:intra_elem_pipe}}
\subfigure[\side{}]{\includegraphics[width=0.5\columnwidth]{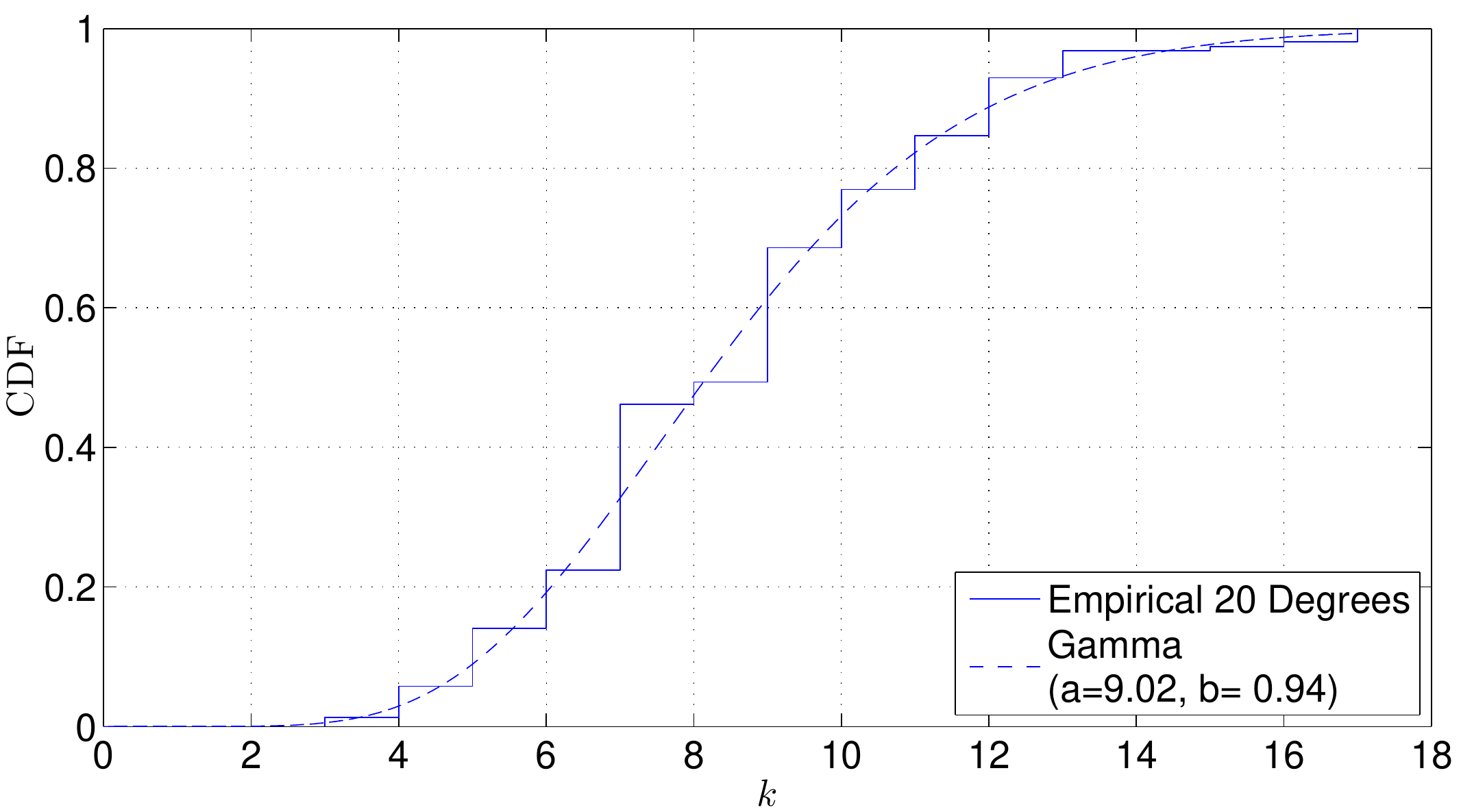}\label{fig:intra_elem_side}}
\caption{Empirical and best goodness of fit CDFs of the number of paths within a cluster ($k$).}
\label{fig:intra_elem}
\end{figure*}

\fref{fig:intra_elem} shows the CDFs of the number of paths within a cluster ($k$). This number is typically related to the type of materials in a certain propagation environment. Specifically, rough surfaces such as raw concrete result in diffuse reflections which spread the incoming ray of a certain path onto a number of approximately parallel rays (c.f. \sref{sec:model}). At the receiver, we observe this as multiple paths in each cluster. In the industrial scenarios that we consider, most of the reflections occur on the metallic surfaces of different types of machinery. These surfaces are typically even and thus result in almost specular reflections of the paths. However, the walls in industrial settings are very often made of rough concrete. Hence, we expect to observe a mixture of both types of reflections. The specific ratio of that mixture should depend on the particular characteristics of each scenario.

\fref{fig:intra_elem} confirms the above discussion. For all of our scenarios in which a significant amount of machinery and other metallic elements are located (i.e., the \tunnel{}, \hall{}, and \pipe{}), the majority of measurements exhibits a limited number of paths within each cluster. Specifically, we observe barely any instances showing more than $8$ paths, and $50\%$ of the cases show between $1-3$ paths. While reflections on rough concrete walls are possible in all of the three scenarios, the fraction of such instances is negligible. Also, diffuse reflections spread the energy in many directions, which means that the strength of paths reflected on rough walls is limited compared to specular reflections on metal. This difference in terms of signal power makes the former hard to observe at the receiver when it occurs along with the latter. For the scenarios for which we compare different beam widths (i.e., the \tunnel{} and \hall{}), we observe only little difference among the 7-, 20-, and 80-degree antennas. This means that all of the paths that we capture with each of the antennas experience the same type of reflections, that is, predominantly specular. Still, wider beam widths tend to result in a slightly smaller number of paths within each cluster. The underlying reason is most probably related to the lengths of the paths on which the clusters travel. Wider antennas capture longer paths (c.f. \sref{subsubsec:inter-cluser-delays}) on which the overall propagation losses are higher. Hence, some of the paths within each cluster may fall below the noise level, which results in a smaller number of observed paths at the receiver.

The \pipe{} exhibits a slightly higher maximum number of paths in each cluster compared to the \hall{} and the \tunnel{}. In particular, \fref{fig:intra_elem_pipe} depicts a fraction of about $2$\% of clusters with more than $8$ paths. This behavior is likely due to the larger number of pipes in the \pipe{}. Since the pipes are made out of curved metal, the angular spread resulting of the limited diffusion on such a surface is higher than in the \hall{} and \tunnel{}. Still, our results show that this effect has a limited impact, and thus our above discussion is also valid for the \pipe{}. In contrast, the \side{} clearly exhibits a different behavior when compared to the \tunnel{}, \hall{}, and \pipe{}. The CDF in \fref{fig:intra_elem_side} reveals barely any instance with less than three paths and $60\%$ of the samples show more than $8$ paths. This is as expected due to the limited number of reflectors in this scenario. Further, the walls of the \side{} are made of rough concrete, which means that the vast majority of the limited number of reflections (c.f. \sref{subsubsec:num_clusters}) is strongly diffuse. This again highlights the impact of metallic surfaces in industrial scenarios on the propagation environment.

\tref{tbl:intra_elem} shows the values of the 1st quartile, median, and 3rd quartile of our results. In \fref{fig:intra_elem} we also observe that the number of paths within a cluster for the \tunnel{}, \hall{}, and \pipe{} can be described as a Generalized Pareto Distribution with parameters $-0.36 \leq k \leq -0.12$, $2.32 \leq \sigma \leq 3.37$ and $\theta = 1$.
Again we note the similarity of the distributions of this parameter for each scenario. As expected from our above analysis, the number of paths within a cluster for the \side{} can be better described as a Gamma distribution instead.

\begin{table}[tb!]
\centering
\caption{Quartiles of the number of paths within a cluster ($k$).\label{tbl:intra_elem}\vspace{-3.5mm}}{%
\begin{tabular}{|c|c|c|c|c|} \hline
Scenario & Degree & 1st Quartile & Median & 3rd Quartile\\ \hline
Tunnel & 7 & 2 & 3 & 4 \\ \hline
Tunnel & 20 & 2 & 3 & 4 \\ \hline
Tunnel & 80 & 2 & 3 & 4 \\ \hline
Exp. Hall & 7 & 2 & 3 & 4 \\ \hline
Exp. Hall & 20 & 2 & 3 & 4  \\ \hline
Exp. Hall & 80 & 1 & 2 & 3 \\ \hline
Pipe Room & 20 & 2 & 3 & 4 \\ \hline
Side Tunnel & 20 & 7 & 9 & 10 \\ \hline
\end{tabular}}
\end{table}

\subsubsection{Amplitude of Each Path in the Cluster}

\begin{figure*}[hhh!] 
\centering
\subfigure[\tunnel{}]{\includegraphics[width=0.5\columnwidth]{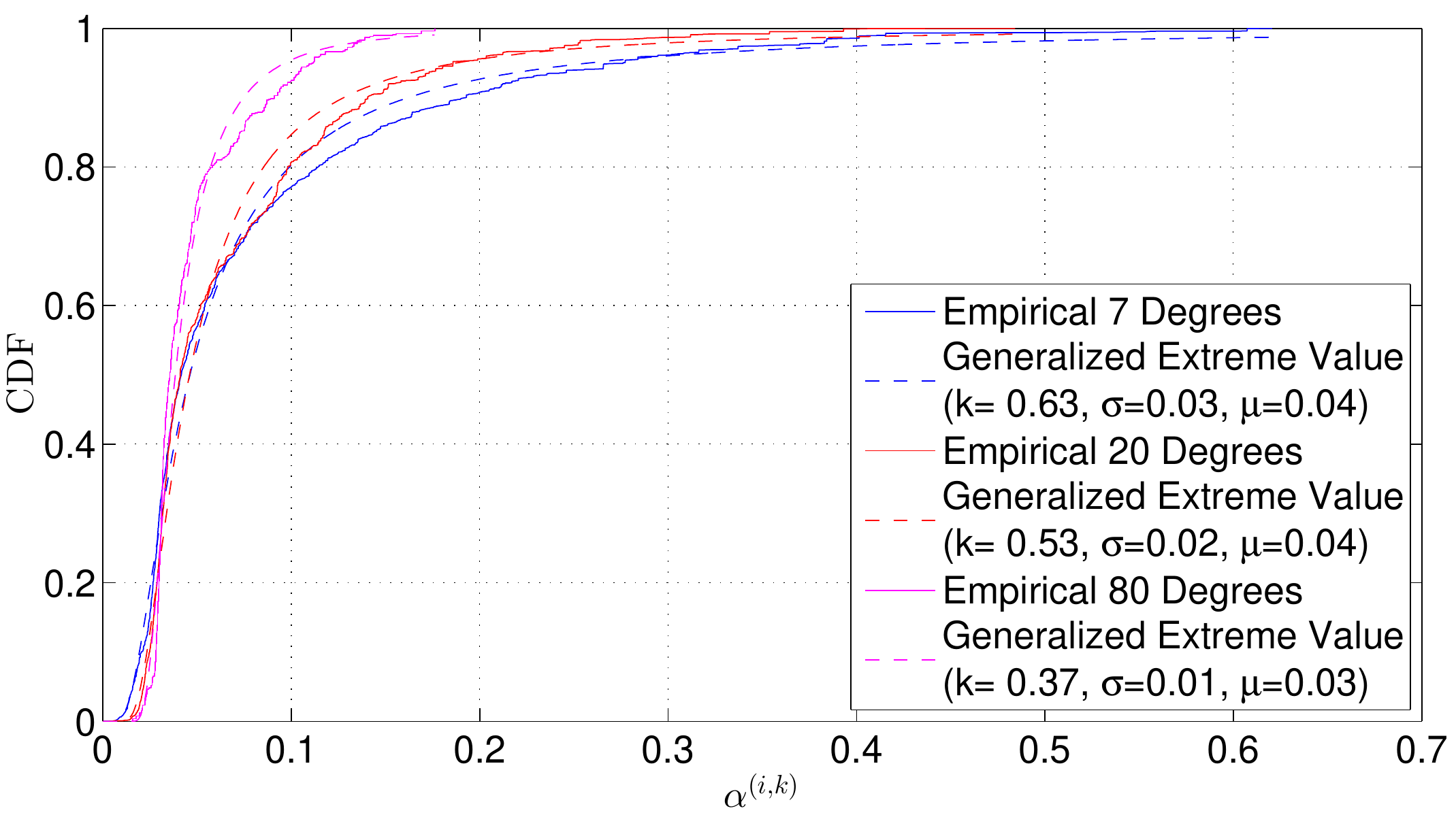}}
\subfigure[\hall{}]{\includegraphics[width=0.5\columnwidth]{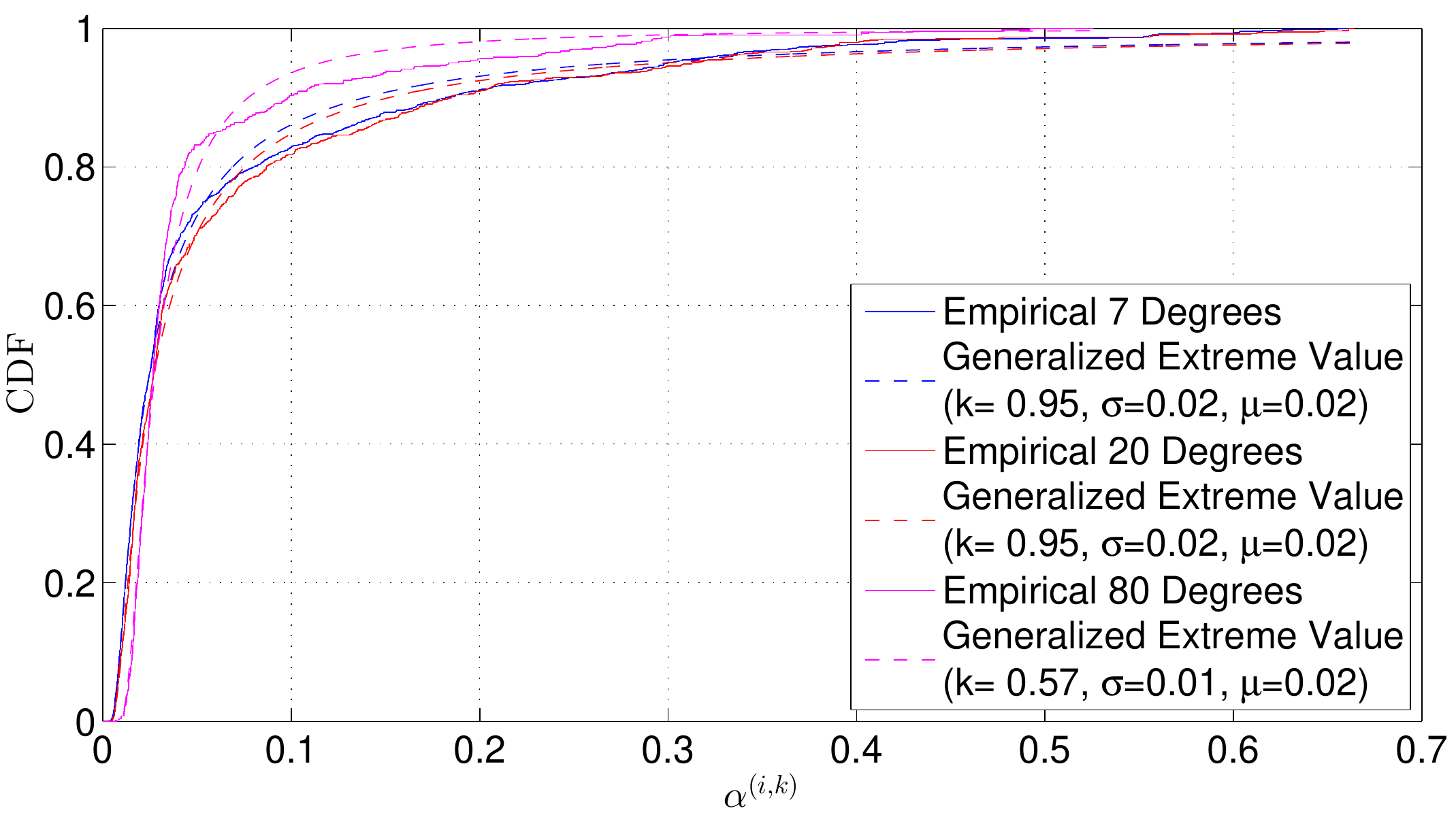}}
\subfigure[\pipe{}]{\includegraphics[width=0.5\columnwidth]{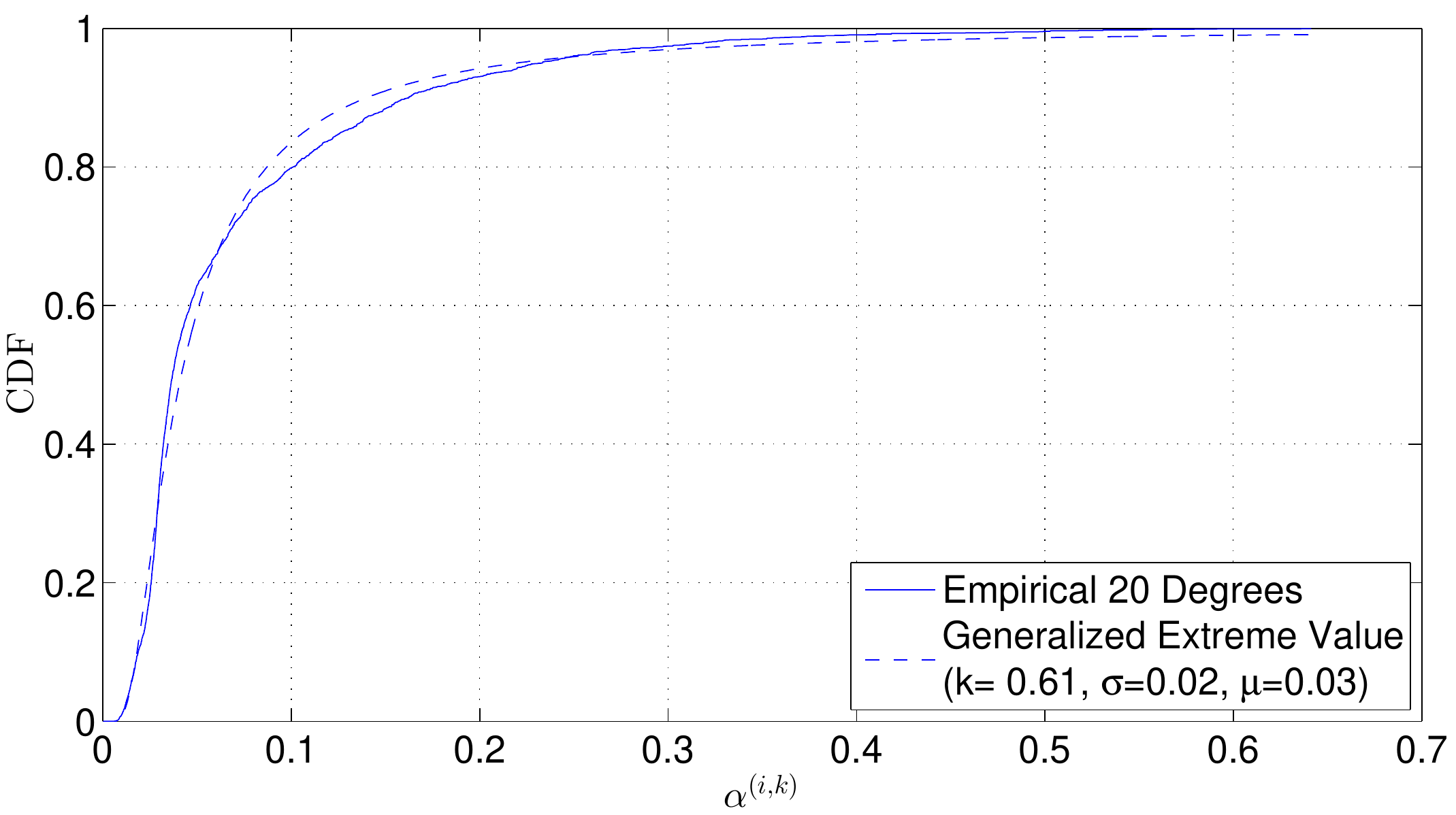}}
\subfigure[\tunnel{}]{\includegraphics[width=0.5\columnwidth]{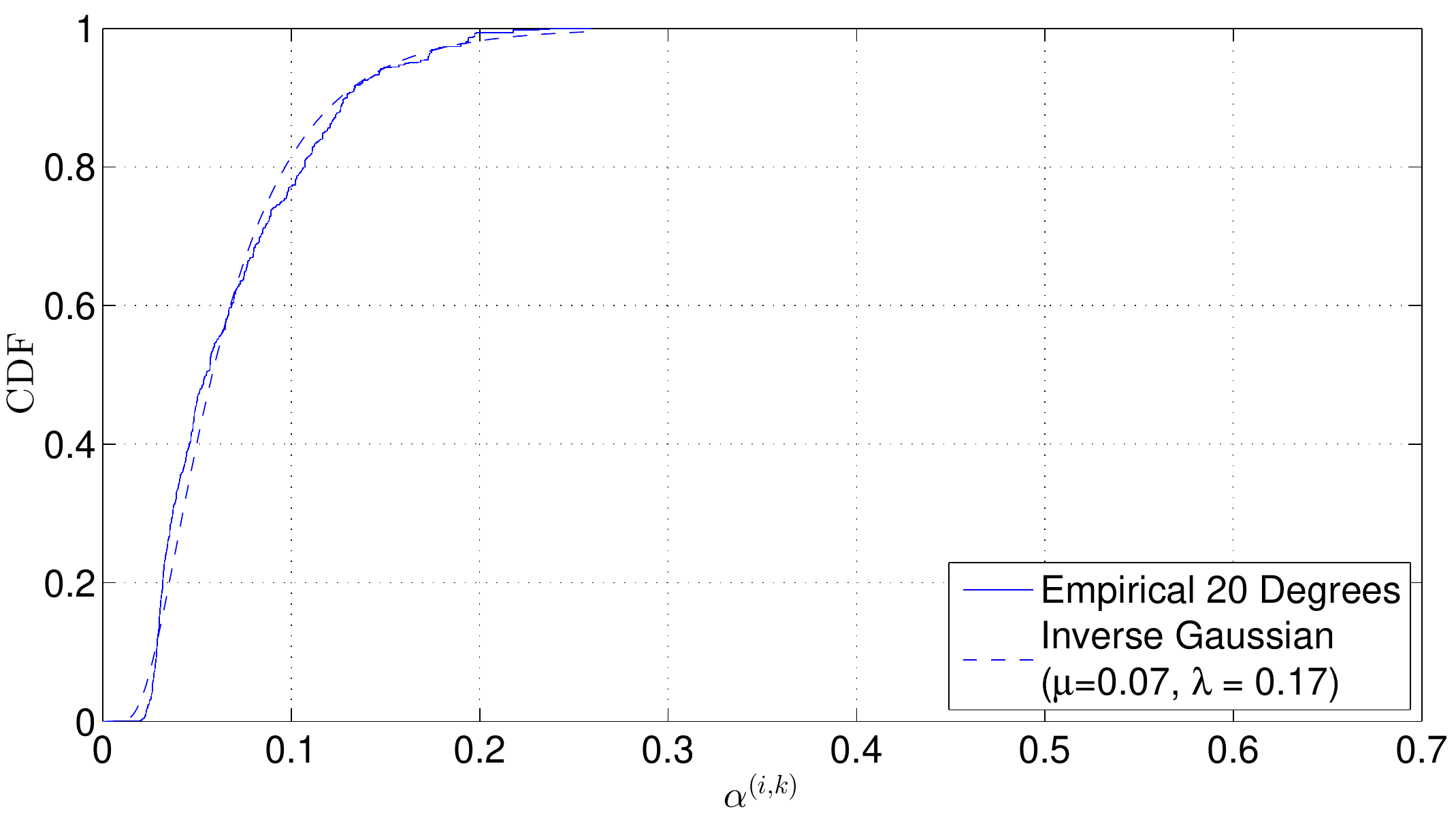}}
\caption{Empirical and best goodness of fit CDFs of the amplitude of each path in the cluster ($\alpha^{(i,k)}$).}
\label{fig:intra_ampl}
\end{figure*}

\fref{fig:intra_ampl} shows the amplitude of each path in the cluster ($\alpha^{(i,k)}$). For all of our scenarios, we observe a very similar behavior than for the average amplitude of the clusters as a whole $A^{(i)}$ in \fref{fig:mpower}. This is expected since the CDFs show the aggregated distribution of all of the paths in all of the observed clusters. However, while in \sref{subsubsec:cluster_amplitude} we averaged the amplitudes of the paths within each cluster, in \fref{fig:intra_ampl} we compute the distributions of the amplitudes of each individual path. As a result, we observe again similar effects, such as a shift to the left of the CDFs for wide beam patterns. As discussed in \sref{subsubsec:cluster_amplitude}, this occurs because such beam patterns capture more weaker paths. \fref{fig:intra_ampl} also shows the maximum amplitude of the main path in each cluster, which is significantly larger than the average value of the cluster depicted in \fref{fig:mpower}. This means that the main path is much stronger than the diffused paths that conform the rest of the cluster. This is expected since the propagation direction of the diffused paths is not exactly the same than the main path, which means that the receiver only captures part of their energy when aligned with the main path.

From the above, we conclude that the comparison of \fref{fig:mpower} and \fref{fig:intra_ampl} essentially reveals the peak to average ratio of the amplitudes of the paths within the clusters. For the \tunnel{}, we observe that this ratio is close to one for the 7- and 20-degree antennas. That is, the average amplitude distribution of the cluster in \fref{fig:mpower} is similar to the one in \fref{fig:intra_ampl}. In contrast, for the 80-degree antenna, we observe that the peak value in \fref{fig:intra_ampl} is about double as large than in \fref{fig:mpower}. The underlying reason is that in the 80-degree case, the antenna captures more diffused paths for each cluster, thus resulting in a smaller average amplitude. For the aforementioned narrower antennas, the main path is still the predominant component of the cluster. We also observe this effect in the \hall{}. However, in this case, the peak to average ratio of the path amplitudes in a cluster is relatively large for all of the antennas. This is probably due to the larger number of reflectors in the \hall{} compared to the \tunnel{}---even when using the narrowest beam width, the receiver still observes a relatively large number of diffused paths. This suggests that the frequency selectivity of the channel may be on average larger in the \hall{} than in the \tunnel{}. This effect is further exacerbated in the \pipe{}, for which the largest observed amplitudes in \fref{fig:intra_ampl} doubles the value in \fref{fig:mpower} even for the 20-degree antenna. As expected, this effect is limited for the \side{} due to the fewer reflections.

The 1st quartile, median, and 3rd quartile of our data are shown in \tref{tbl:intra_ampl}. The distributions of the amplitude of each path in a cluster can be well described by a Generalized Extreme Value Distribution for the \tunnel{}, \hall{}, and \pipe{} with parameters $0.37 \leq k \leq 0.95$, $0.01 \leq \sigma \leq 0.03$ and $0.02 \leq \mu \leq 0.04$. 
We observe again similar distributions of the parameter for the different environments. On the contrary, in the \side{} this parameter is better described by an Inverse Gaussian distribution due to the limited number of reflections.

\begin{table}[tb!]
\centering
\caption{Quartiles of the the amplitude of each path in the cluster ($\alpha^{(i,k)}$).\label{tbl:intra_ampl}\vspace{-3.5mm}}{%
\begin{tabular}{|c|c|c|c|c|} \hline
Scenario & Degree & 1st Quartile & Median & 3rd Quartile\\ \hline
Tunnel & 7 & 0.029 & 0.041 & 0.090 \\ \hline
Tunnel & 20 & 0.030 & 0.041 & 0.090 \\ \hline
Tunnel & 80 & 0.031 & 0.036 & 0.050\\ \hline
Exp. Hall & 7 & 0.014 & 0.025 & 0.054 \\ \hline
Exp. Hall & 20 & 0.015 & 0.027 & 0.065 \\ \hline
Exp. Hall & 80 & 0.020 & 0.027 & 0.038 \\ \hline
Pipe Room & 20 & 0.027 & 0.037 & 0.079 \\ \hline
Side Tunnel & 20 & 0.034 & 0.055 & 0.095 \\ \hline
\end{tabular}}
\end{table}

\subsection{Throughput Measurements}
\label{subsec:throughput}

In this section we evaluate how our previous insights on the specific propagation environments of industrial settings translate into practical throughput values. In addition to the four scenarios discussed in \sref{subsec:param_fitting}, we also provide results for the \ring{} and \ups{}. To measure the aforementioned practical throughput, we use the COTS setup described in \sref{subsec:cots_setup}. As discussed in \sref{sec:setup}, we place our COTS transceivers at a large number of locations in each scenario, establish a connection, and exchange data in both directions.

\subsubsection{\pipe{}}

\begin{figure*}[hhht!] 
\centering
\subfigure[South.]{\includegraphics[width=0.5\columnwidth]{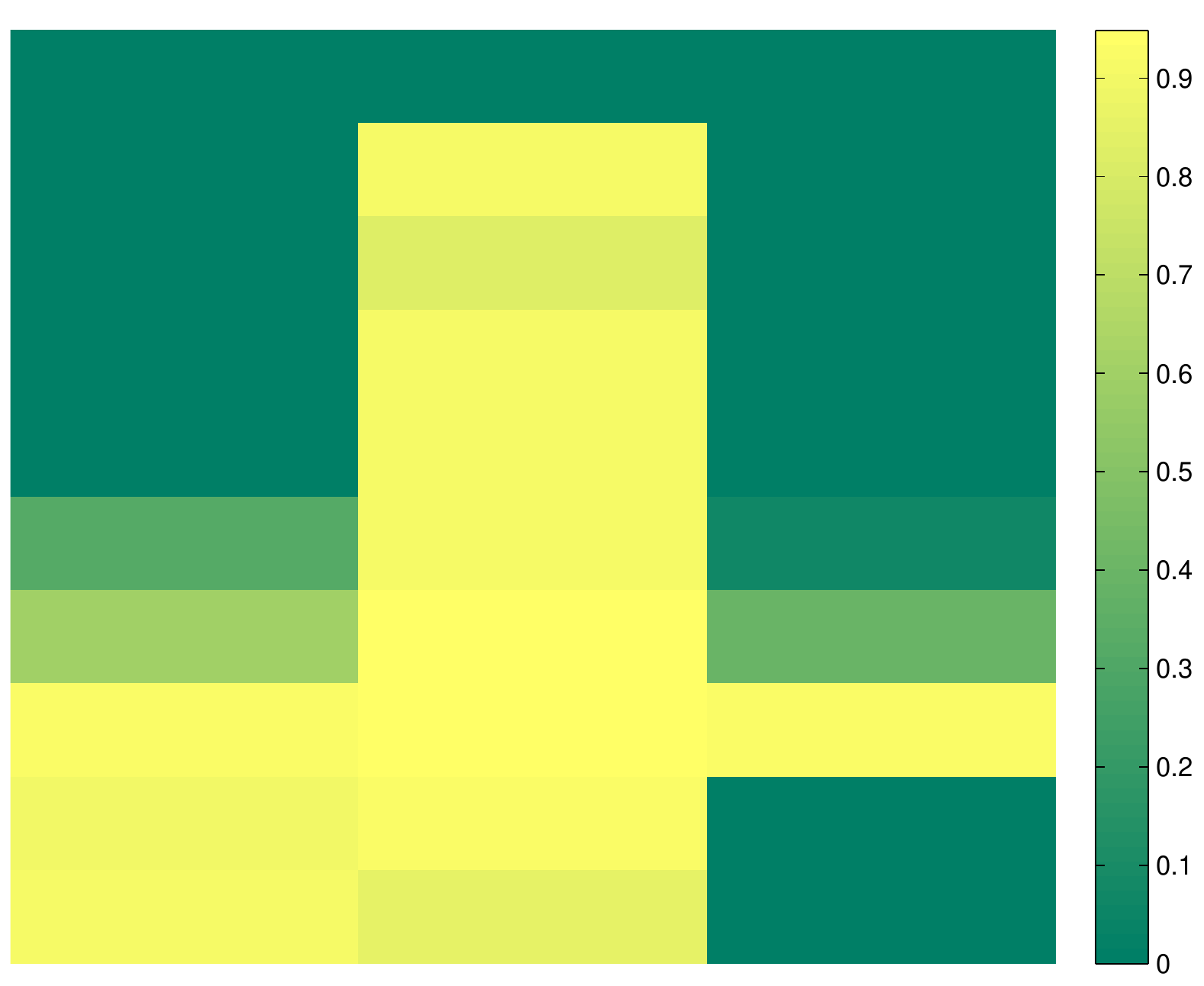}\label{fig:tp_pipe_s}}
\subfigure[East.]{\includegraphics[width=0.5\columnwidth]{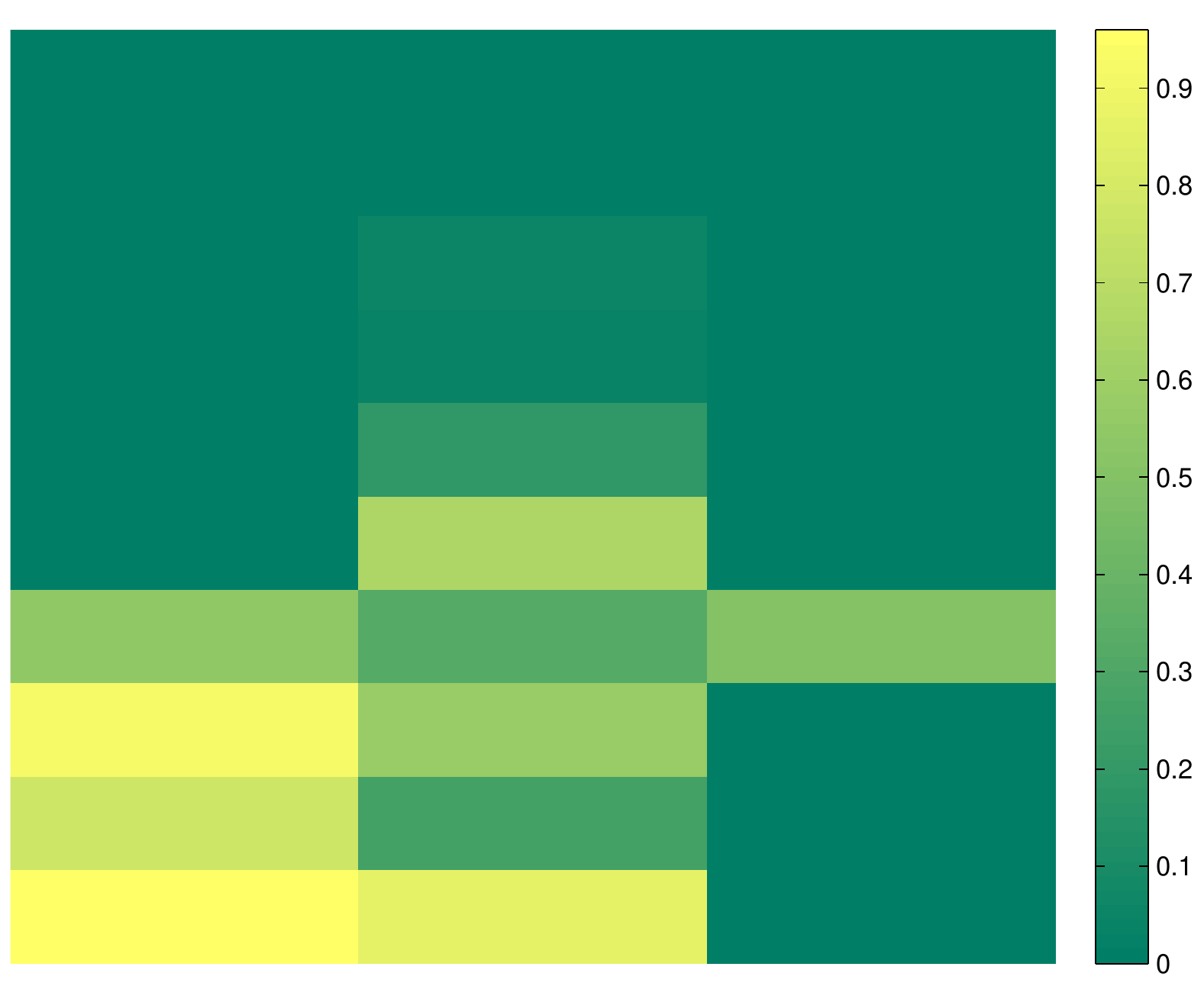}\label{fig:tp_pipe_e}}
\subfigure[North.]{\includegraphics[width=0.5\columnwidth]{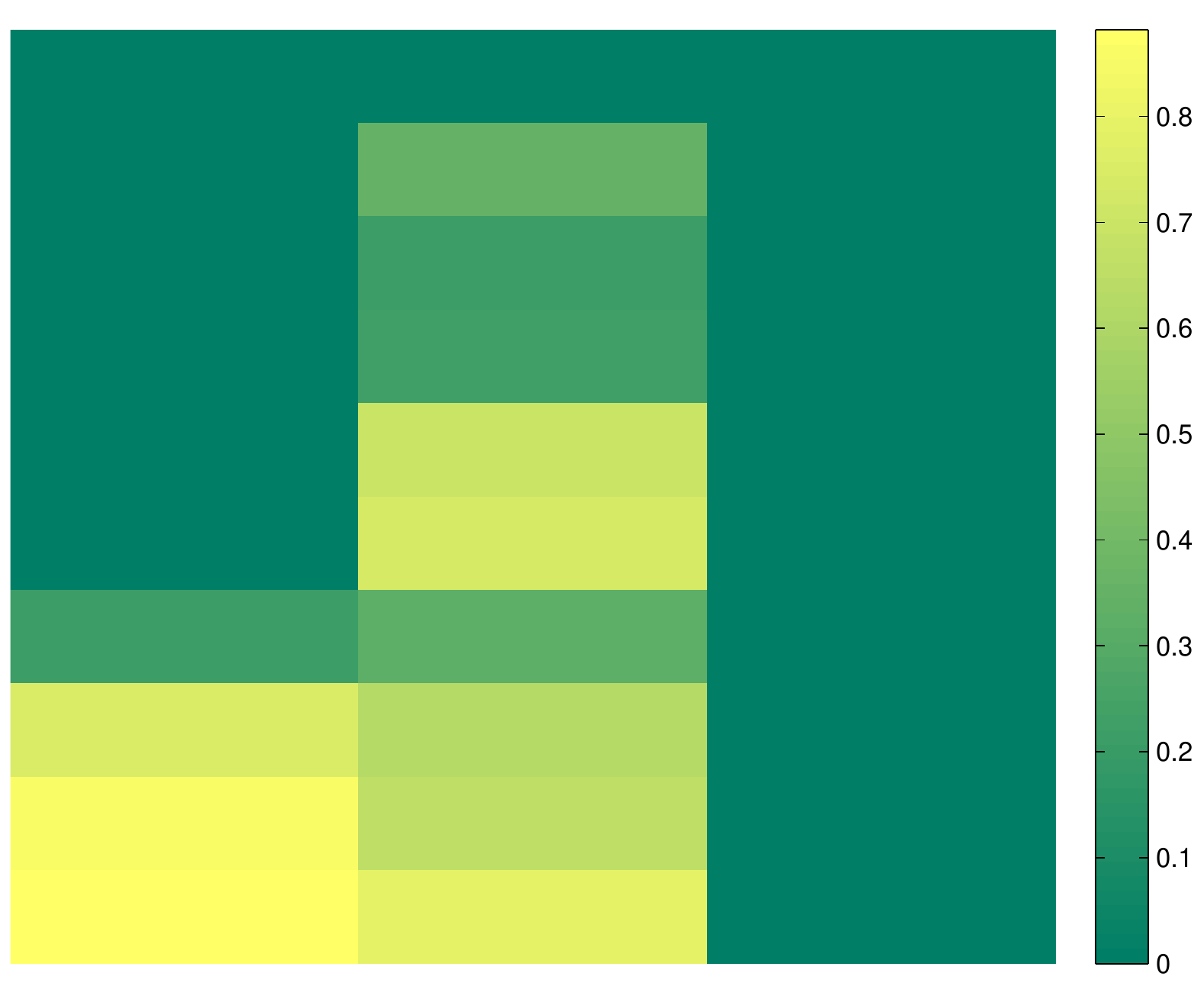}\label{fig:tp_pipe_n}}
\subfigure[West.]{\includegraphics[width=0.5\columnwidth]{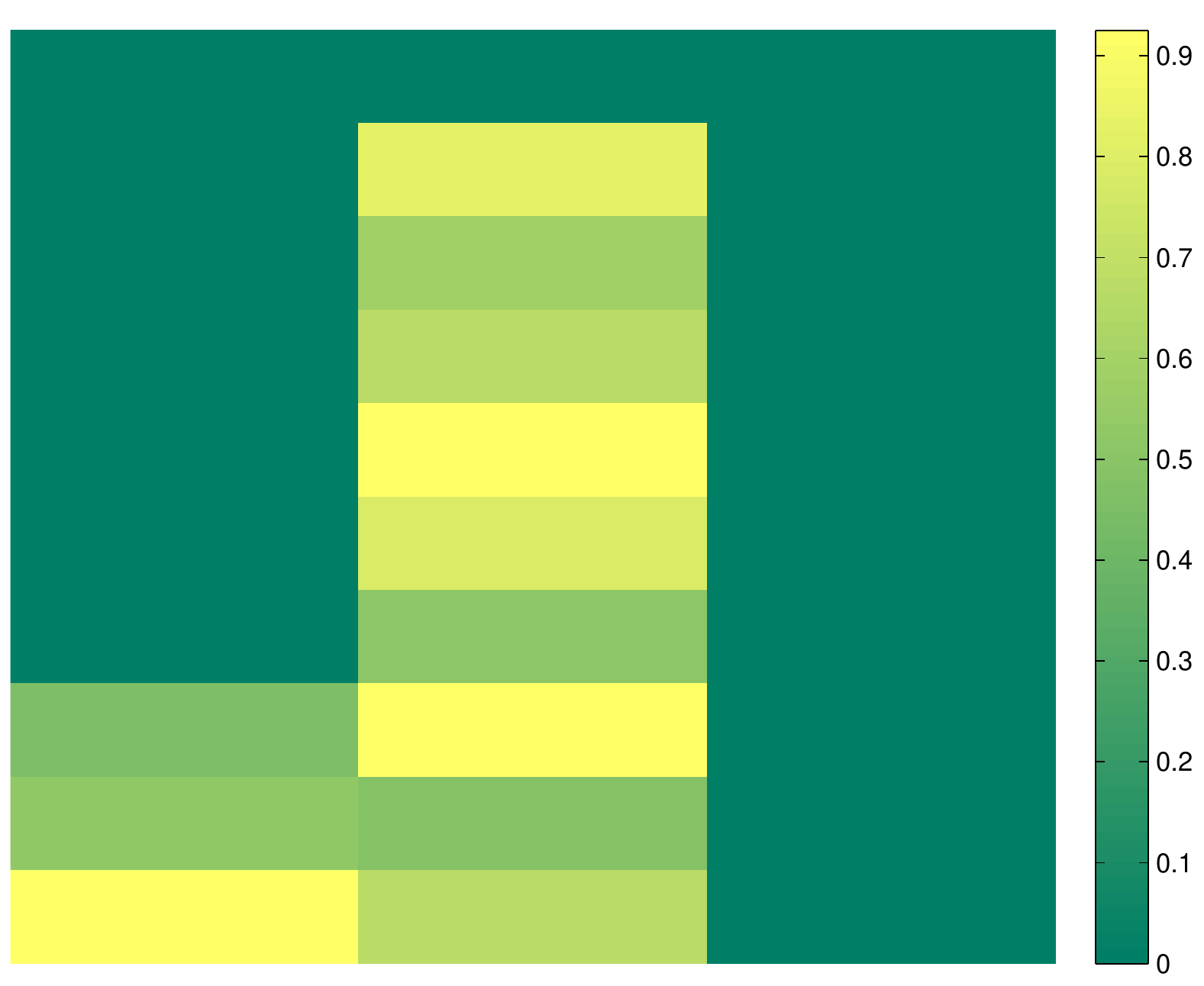}\label{fig:tp_pipe_w}}\\
\caption{Throughput results in the \pipe{}.}
\label{fig:tp_pipe}
\end{figure*}

\fref{fig:tp_pipe} shows a heat-map of the throughput (in Gbps) in the \pipe{} measured according to the methodology described in \sref{subsec:cots_setup}. That is, we place a transmitter at a fixed location in the room and measure the throughput for different positions of the receiver. Further, at each position we consider four different orientations for the receiving node: \emph{i)} south, pointing at the transmitter, \emph{ii)} east, \emph{iii)} north, pointing in the opposite direction of the transmitter, and \emph{iv)} west. Note that the antenna module of the TP-Link Talon AD7200 router that we use in this setup does support beam steering. Thus, even if the receiver device itself is not pointing towards the transmitter, it can adjust its beam in order to receive data. However, beam steering on practical devices typically works best in the direction towards which the device is physically pointing, while steering to the side---or backwards, if possible at all---results in strong side-lobes and low antenna gain. This allows us to capture the limitations of millimeter-wave antennas in industrial scenarios.

The heat-map in \fref{fig:tp_pipe} depicts the throughput at each of the positions showed in the room layout in \fref{fig:pipe}. As expected, we see that the throughput is highest when the receiver is pointing south (\fref{fig:tp_pipe_s}). Additionally, we observe that even in the cases where the receiver is pointing towards other directions (\fref{fig:tp_pipe_e}-\fref{fig:tp_pipe_w}), the received throughput is significant for the positions in the middle of the room. This is also the case for the positions close to the transmitter at its left side. These outcomes show that, while practical beam-steering is sub-optimal in terms of antenna gain, current commercial devices are able to successfully establish a link even if strongly misaligned. Most interestingly, \fref{fig:tp_pipe_n} shows that communication is possible even when the receiver points in the opposite direction of the transmitter. This is significant because the antenna module at the receiver is not designed to receive in that direction, and actually the antenna circuitry partially blocks the signal. Beam pattern measurements of the device reveal that partial backwards reception is feasible, but with significantly lower antenna gain. Further, the throughput in \fref{fig:tp_pipe_n} fluctuates significantly as we move away from the transmitter on the positions $20$ to $11$ in \fref{fig:pipe}. In contrast, for the case in \fref{fig:tp_pipe_s} facing towards the transmitter, throughput is much more stable. Along with the limited backwards reception of our device, this suggests that in \fref{fig:tp_pipe_n} the receiver is receiving the signal not only from the LOS path but also from NLOS paths which are available at a subset of the aforementioned positions $20$ to $11$.

For the cases in which the receiver points to the side (\fref{fig:tp_pipe_e} and \fref{fig:tp_pipe_w}), we conclude that reception takes place both via strong side-lobes and reflections. Indeed, a close inspection of the beam patterns that our devices use in this experiment confirms that the transceivers tend to use beams with significant side-lobes that point in each other's direction. Further, we observe that the positions on the right side of the room are hardly reachable for any of the four device orientations that we measure. This is due to the very high degree of blockage in that area, as depicted in \fref{fig:pipe}.

\subsubsection{\hall{}}

\begin{figure}[hhht!] 
\centering
\subfigure[]{\includegraphics[width=0.49\columnwidth]{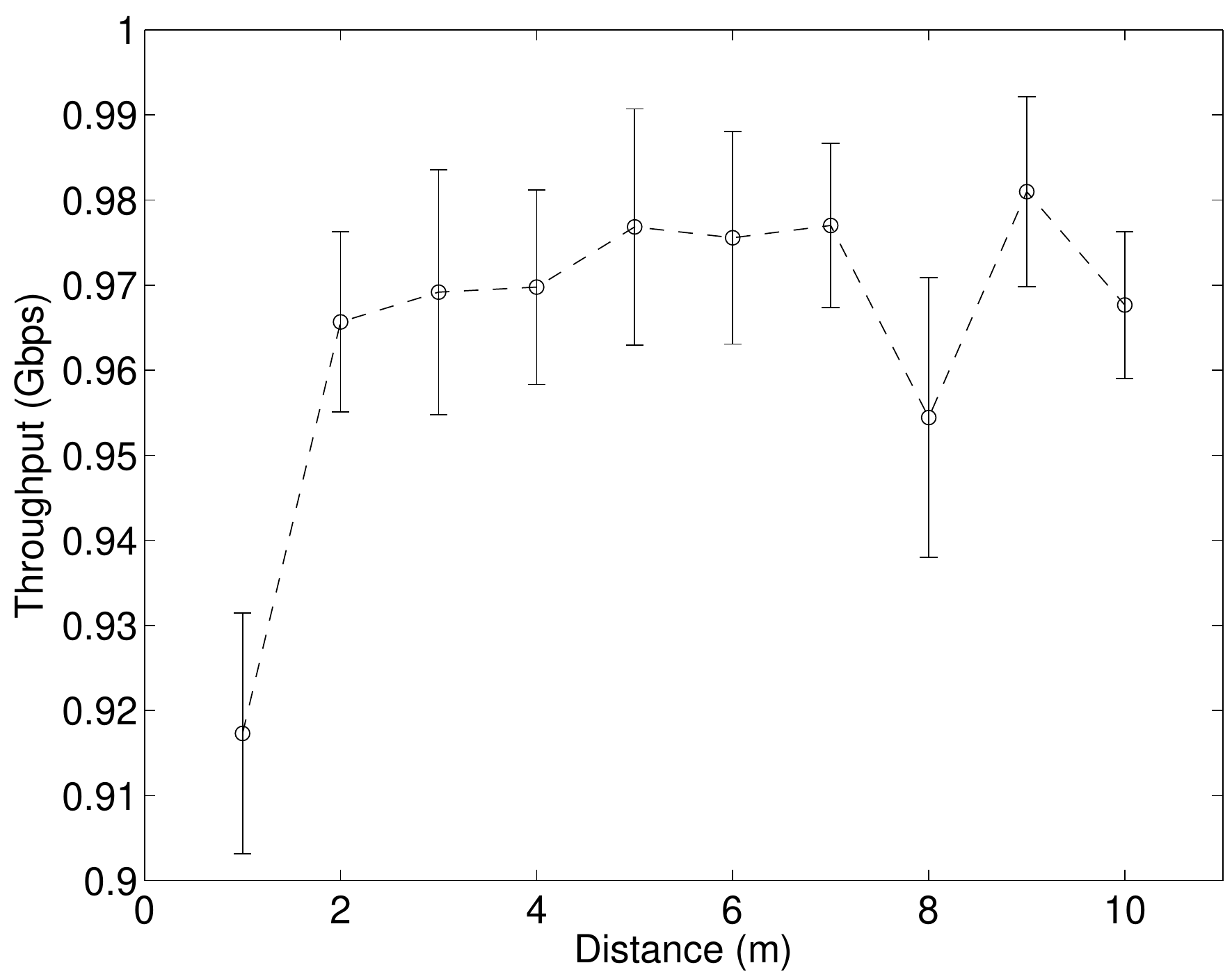}\label{fig:tp_exphall}}
\subfigure[]{\includegraphics[width=0.49\columnwidth]{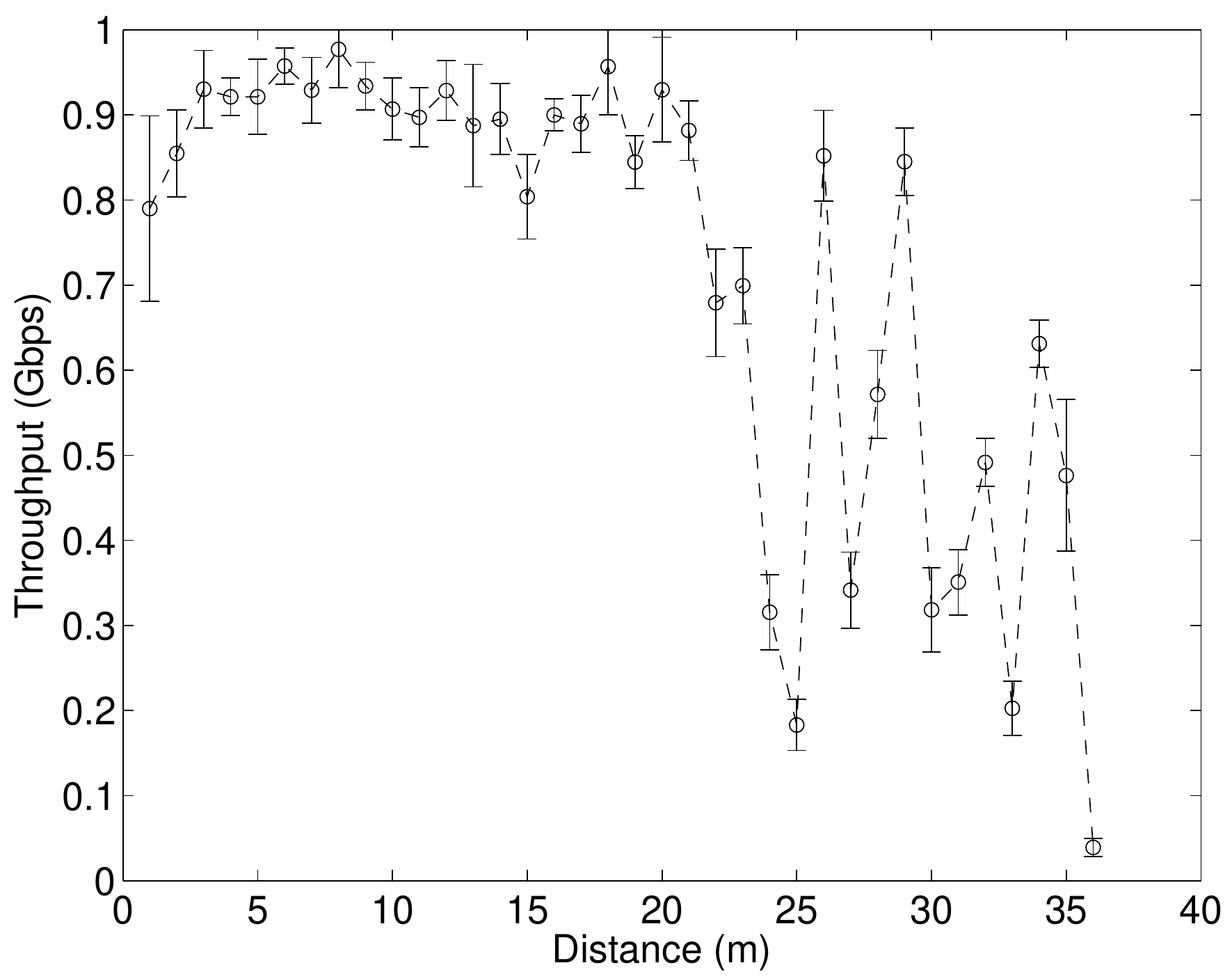}\label{fig:tp_ring}}
\caption{Throughput results in (a) the \hall{} (floor level) and in (b) the \ring{}.}
\label{fig:tp_exphall_ring}
\end{figure}

\fref{fig:tp_exphall} shows the throughput results in the \hall{} on the floor level at the different distances shown in \fref{fig:ring}. In this case, we performed the experiments with the receiver pointing towards the transmitter. Given this alignment, our devices are known to use rather directional beam patterns with a strong main lobe and negligible side lobes \cite{conextDaniel}. Further, as discussed in \sref{subsec:param_fitting}, the \hall{} does feature a significant number of reflectors but many of them only play a role for wide antenna beam patterns due to the size and openness of the area. Thus, in this experiment we expect the Talon router to receive primarily the LOS path and only limited reflections, if at all. This would result in a channel with very low frequency selectivity and thus benevolent for the single carrier modulation used in IEEE 802.11ad. While we cannot measure the channel at the Talon device itself, our results in \fref{fig:tp_exphall} show a relatively stable behaviour with throughput close to 1 Gbps at all locations. The standard allows for higher throughput values but given the hardware constraints of the Talon devices, 1 Gbps is close to the highest achievable throughput in practice at the application layer. As discussed in \sref{subsec:hall_setup}, we also measure throughput for links connecting the lower and the upper level of the experimental hall. If we align the receiver with the transmitter but place the former at the upper floor, we achieve similar throughput ($\sim0.93$~Gbps). However, if we move the receiver at an angle as shown in \fref{fig:ring}, we observe a drop to $\sim0.13$~Gbps. This suggests that the impact of elevation itself is limited but that misalignment---and the resulting signal strength loss along with the frequency selectivity due to reflections captured by side-lobes---plays a significant role. This observation matches the design of the Talon antenna array, which offers limited vertical beam forming capabilities \cite{conextDaniel}. That is, the antenna gain is similar for different elevation angles.


\subsubsection{\ring{}}

The \ring{} is a unique environment due to its curvature and the large amount of reflectors that it contains. While we cannot carry out our full channel measurements (c.f. \sref{subsec:param_fitting}) in this scenario due to access restrictions, we are able to perform measurements as depicted in \fref{fig:tp_ring}. Essentially, we place the transmitter at a fixed location in the ring and move the receiver along the ring to increase the link distance. Our goal is to analyze whether the elongated shape of the ring allows us to exploit the many reflectors in order to set up a long link along the curve of the ring even if the transceivers are not in LOS. In other words, we study whether the effects that we observed for the \tunnel{} in \sref{subsec:param_fitting} hold for curved environments. In \fref{fig:tp_ring}, we clearly observe the distance at which the link becomes NLOS---at $20$ meters, the throughput drops from about $0.9$~Gbps to values that fluctuate between $0.2$~Gbps to $0.8$~Gbps for a link distance of up to $36$ meters. This high throughput values in the NLOS half of the graph are highly encouraging, since the peaks at $\sim0.8$~Gbps suggest that strong reflections allow communication along curved narrow structures such as the \ring{}, thus increasing the coverage of each individual AP beyond its LOS range. The high fluctuations in \fref{fig:tp_ring} are most likely because reflectors are scattered along the \ring{} (c.f. \fref{fig:ring}) and thus reflections may or may not be available. If homogeneous reflectors are available along the curved environment, such as the pipes in the \tunnel{}, we expect a more stable throughput.


\subsubsection{\tunnel{} and \side{}}

\begin{figure}[hhht!] 
\centering
\subfigure[]{\includegraphics[width=0.49\columnwidth]{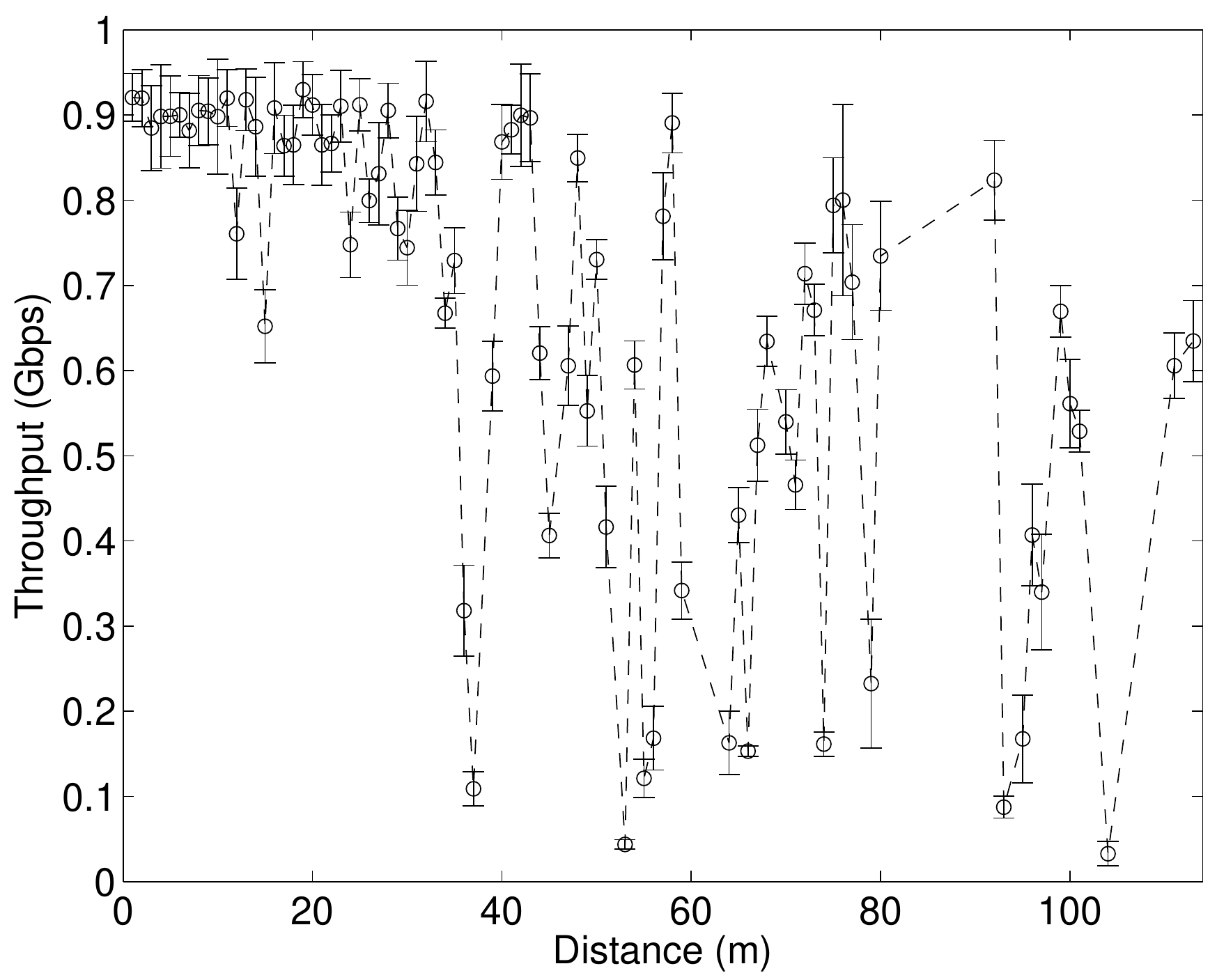}\label{fig:tp_tunnel}}
\subfigure[]{\includegraphics[width=0.49\columnwidth]{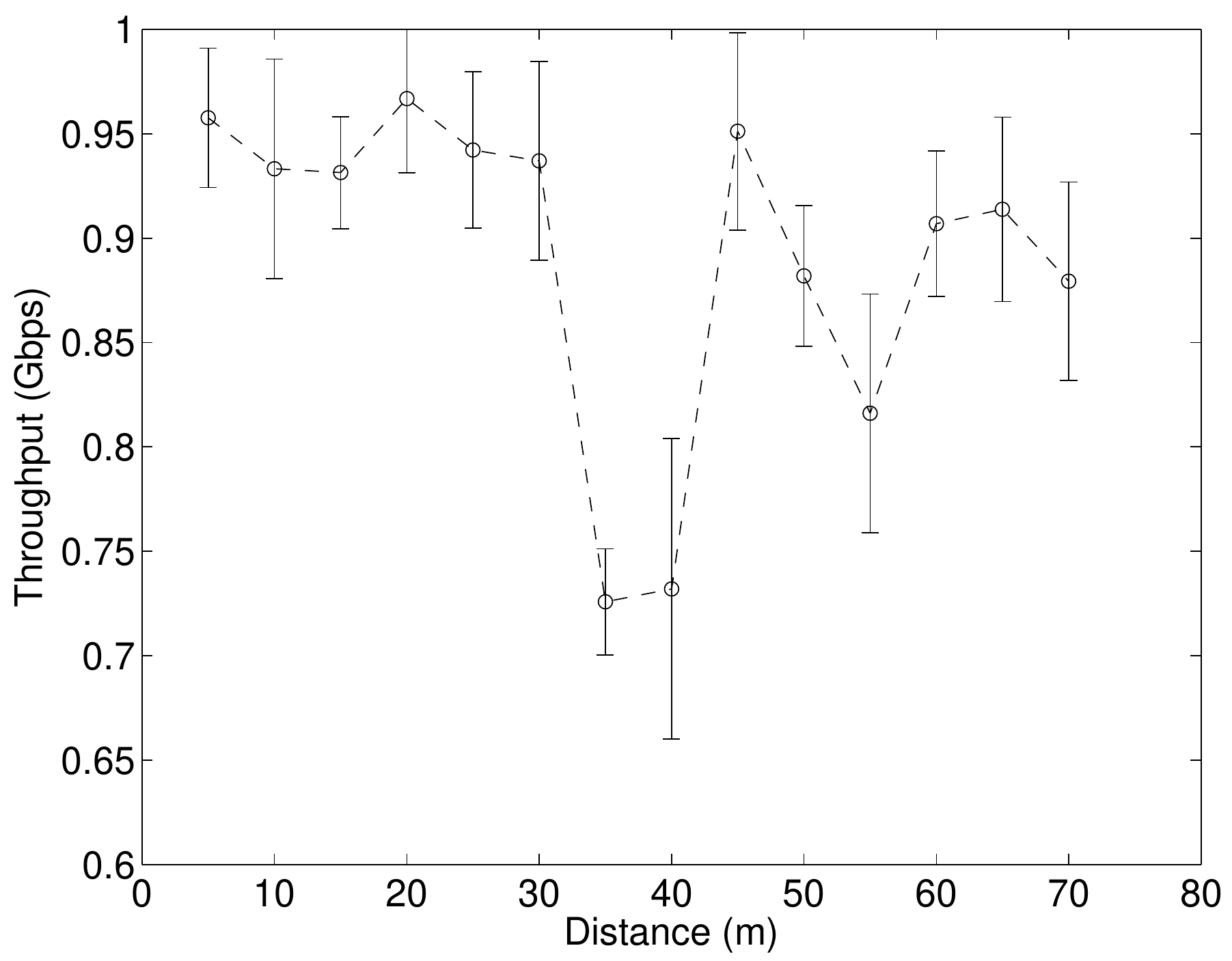}\label{fig:tp_side_tunnel}}
\caption{Throughput results in (a) the \tunnel{} and in (b) the \side{}.}
\label{fig:tp_tunnel_side_tunnel}
\end{figure}

\fref{fig:tp_tunnel} shows the results of the \tunnel{} for increasing link lengths. We observe that the throughput remains high and reasonably stable until the transmitter and receiver are $30$~m apart. Beyond that point, throughput becomes sensitive to the specific link length, which suggests a high impact of constructive and destructive reflections. While our COTS hardware only provides limited insights regarding the underlying physical layer effects, our full channel measurements in this environment (c.f. \sref{subsec:param_fitting}) suggest a number of reasons for this behavior. First, in \sref{subsubsec:inter-cluser-delays} we observed a particularly high inter-cluster delay. Due to the length of the tunnel, we expect this delay to increase with the link distance as the difference in terms of length between the LOS and the reflected paths increases. This results in a large delay spread at the receiver, which makes channel equalization more challenging for long links. Second, in \sref{subsubsec:cluster_amplitude} we concluded that most reflected paths in the \tunnel{} were likely to reflect only once before reaching the receiver. Thus, we expect reflections to be relatively strong at the receiver when compared to the LOS, which results in higher frequency diversity and hence challenging equalization due to the single-carrier modulation used in IEEE 802.11ad. While \fref{fig:tp_tunnel} shows that the above physical layer effects have a strong impact, existing mechanisms such as Orthogonal Frequency-Division Multiplexing (OFDM) with a long cyclic prefix can easily tackle them. Hence, our results in \fref{fig:tp_tunnel} are encouraging. Remarkably, we achieve $>0.6$~Gbps at $110$~m despite the use of single-carrier modulation. To the best of our knowledge, this is the largest 802.11ad link distance reported in literature \cite{Nitsche:2015vq, fastninfuriating, 7179366, 7568477, SahaVGK15}.

In \fref{fig:tp_side_tunnel} we depict the equivalent results for the \side{}. In this case, we observe that the throughput is much more stable than in \fref{fig:tp_tunnel} even for distances beyond $30$~m. This matches our above reasoning. As we concluded from our full channel measurements in \sref{subsec:param_fitting}, the \side{} exhibits a very different behavior at the physical layer compared to all of our other scenarios due to the lack of reflectors. This translates into shorter delay spreads and less frequency selectivity, which in turn makes channel equalization easier. As a result, the fluctuations in \fref{fig:tp_side_tunnel} are small compared to \fref{fig:tp_tunnel}.



\subsubsection{\ups{}}

\begin{figure*}[hhht!] 
\centering
\subfigure[South.]{\includegraphics[width=0.5\columnwidth]{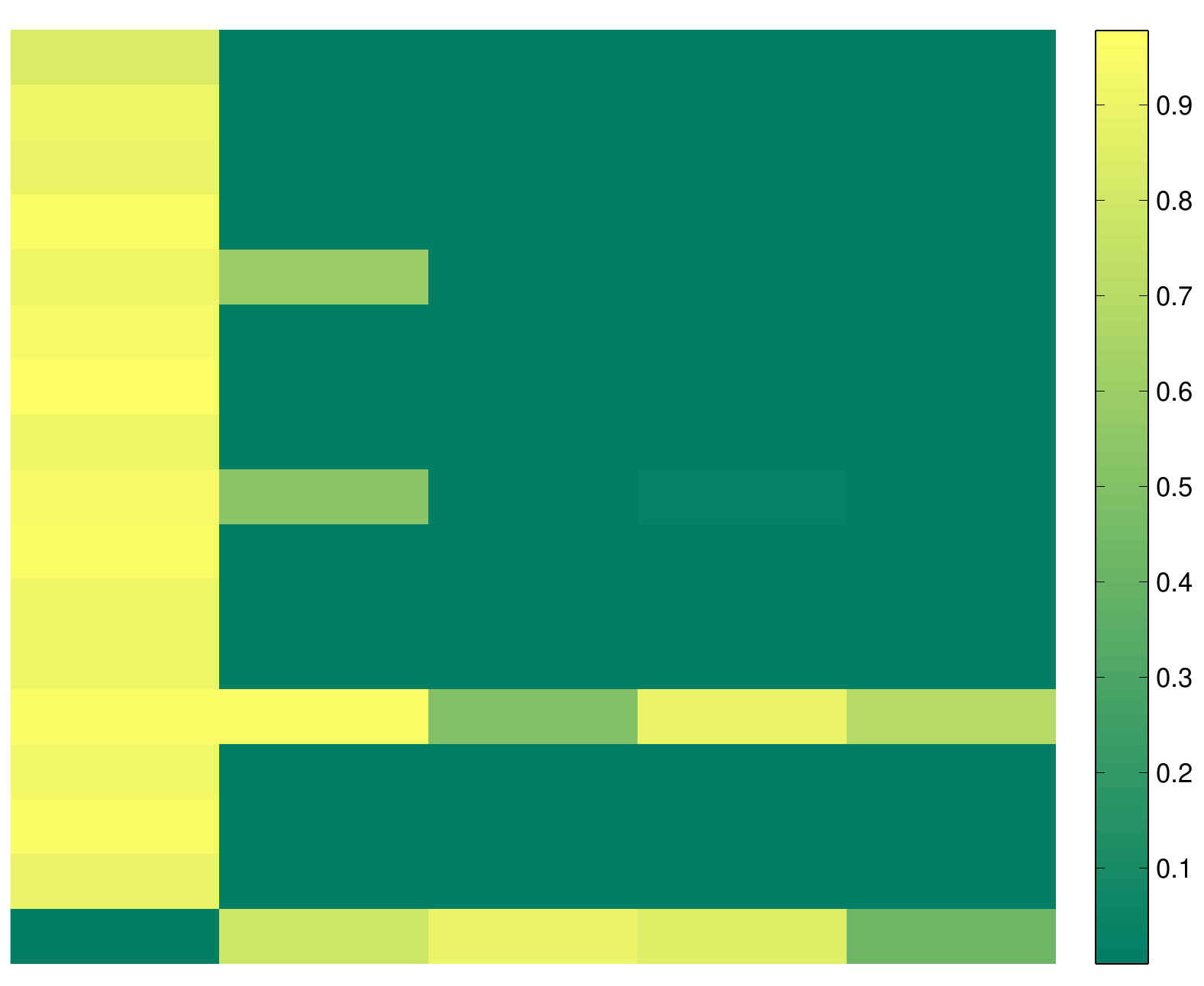}\label{fig:tp_ups_s}}
\subfigure[East.]{\includegraphics[width=0.5\columnwidth]{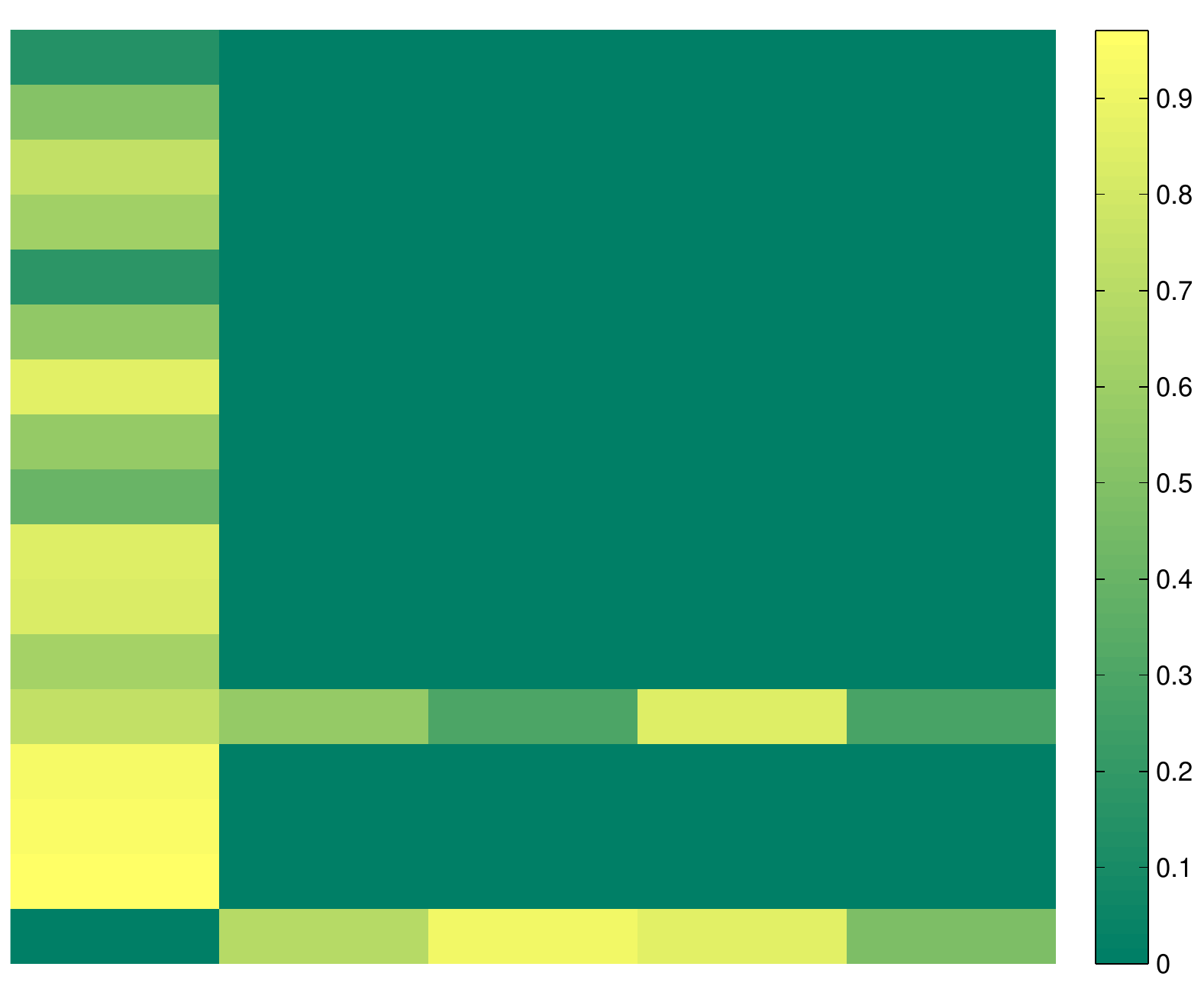}\label{fig:tp_ups_e}}
\subfigure[North.]{\includegraphics[width=0.5\columnwidth]{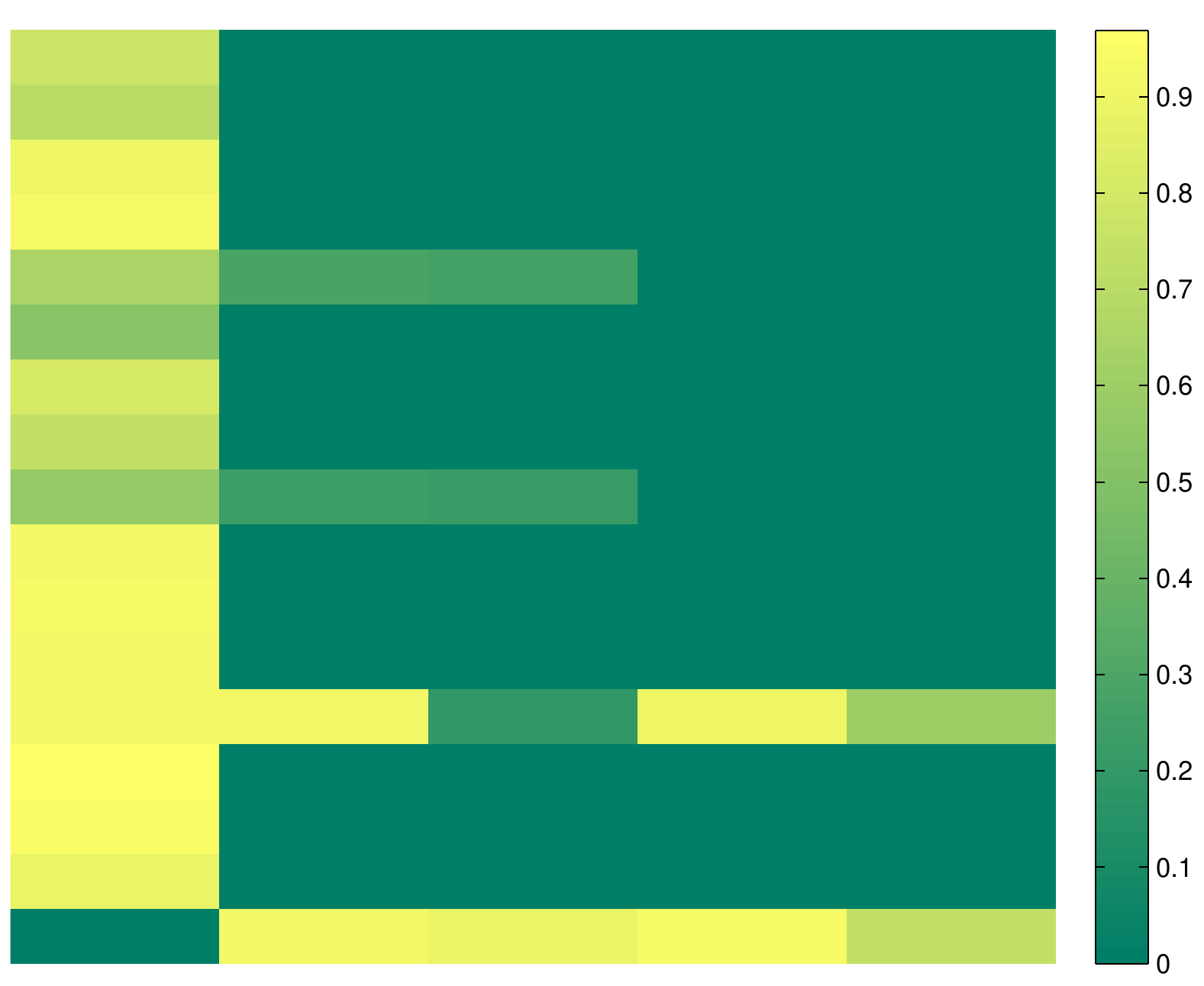}\label{fig:tp_ups_n}}
\subfigure[West.]{\includegraphics[width=0.5\columnwidth]{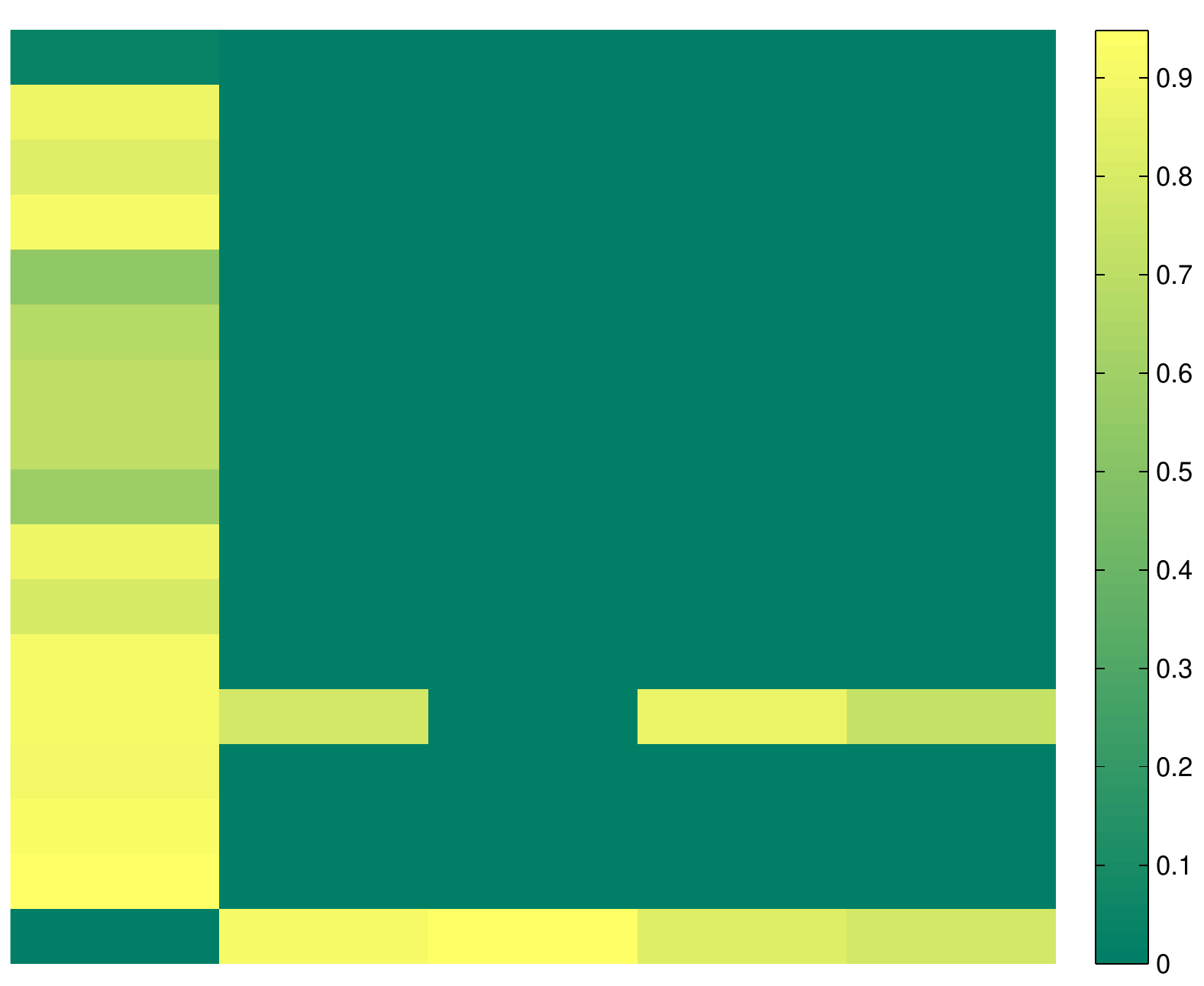}\label{fig:tp_ups_w}}\\
\caption{Throughput results in the \ups{}.}
\label{fig:tp_ups}
\end{figure*}

Similarly to the \pipe{}, in \fref{fig:tp_ups} we show the heatmap of the throughput in the \ups{} for different orientations of the receiver. As depicted in \fref{fig:ups}, the \ups{} contains racks of hardware enclosed in massive metal cabinets which are impossible to traverse for the IEEE 802.11ad signal. As a result, we essentially obtain coverage in the LOS locations only. At those locations, the impact of the receiver orientation is as expected---when the receiver points south (\fref{fig:tp_ups_s}) and thus towards the transmitter, we obtain the best results. Other orientations result in degradation of the throughput but still achieve reasonable coverage (e.g., \fref{fig:tp_ups_w}). In the bulk of the NLOS locations, our devices are not able to establish a connection. While the metal cabinets are potentially good reflectors, the intricate geometry of the \ups{} requires devices to be located at very specific locations to benefit from such reflections. This is in clear contrast to the \tunnel{} and the \ring{}. Further, most locations would only be reachable via second or third order reflections, which inherently results in high losses and thus a weak signal at the receiver. Still, for the first corridor---measurement locations 7 to 10---in \fref{fig:ups}, \fref{fig:tp_ups} reveals some coverage. Most interestingly, the throughput is highest for the north (\fref{fig:tp_ups_n}) and south (\fref{fig:tp_ups_s}) orientations, although the orientation of the corridor is east to west. This indicates that the connection is established via one of the above reflections. We conclude that for those locations path loss is limited due to being close to the transmitter and thus the receiver can still decode the signal despite the above high reflection losses. This shows that intricate environments and blockage are important challenges for millimeter-wave in industrial scenarios.

\section{Conclusions}
\label{sec:conc}

We carry out an extensive set of channel measurements in the millimeter-wave band for the particular case of industrial scenarios such as factories, production lines, and server rooms. To this end, we obtain link characteristics both at the physical layer using high-precision hardware to record the raw channel impulse response, as well as at the application layer using COTS devices to record TCP throughput. We collect about 70 gigabytes of raw traces and analyze them in depth to obtain a statistical channel model. Our model along with our data enables the research community to generate arbitrary channels that represent the typical propagation characteristics of the above industrial scenarios. In addition, we discuss in detail our measurement observations for each scenario. We conclude that millimeter-wave communication in industrial environments is not only feasible but occasionally even easier than in typical home or office settings. Thus, our measurement campaign is highly encouraging regarding the use of such very high frequency wireless networks as part of the upcoming Industry 4.0.

\ifCLASSOPTIONcaptionsoff
  \newpage
\fi

\section*{Acknowledgment}

We would like to thank ALBA Synchrotron for allowing us performing these experiments in their facilities and for the kind support of their computing and management sections. This work has been partially supported by the European Regional Development Fund under grant TEC2015-71303-R (MINECO/FEDER).

\bibliographystyle{IEEEtran}
\bibliography{biblio}

\begin{thebibliography}{10}
\providecommand{\url}[1]{#1}
\csname url@rmstyle\endcsname
\providecommand{\newblock}{\relax}
\providecommand{\bibinfo}[2]{#2}
\providecommand\BIBentrySTDinterwordspacing{\spaceskip=0pt\relax}
\providecommand\BIBentryALTinterwordstretchfactor{4}
\providecommand\BIBentryALTinterwordspacing{\spaceskip=\fontdimen2\font plus
\BIBentryALTinterwordstretchfactor\fontdimen3\font minus
  \fontdimen4\font\relax}
\providecommand\BIBforeignlanguage[2]{{%
\expandafter\ifx\csname l@#1\endcsname\relax
\typeout{** WARNING: IEEEtran.bst: No hyphenation pattern has been}%
\typeout{** loaded for the language `#1'. Using the pattern for}%
\typeout{** the default language instead.}%
\else
\language=\csname l@#1\endcsname
\fi
#2}}

\bibitem{martinez2018square}
B.~Martinez, C.~Cano, and X.~Vilajosana, ``A square peg in a round hole: The
  complex path for wireless in the manufacturing industry,'' \emph{IEEE
  Communications Magazine (Pending publication)}, 2019.

\bibitem{Engelhard2018sigcom}
P.~Engelhard, B.~Holfeld, J.~Schulz-Zander, and M.~Oberle, ``Software-defined
  networking in an industrial multi-radio access technology environment,'' in
  \emph{Proceedings of the Symposium on SDN Research}, ser. SOSR '18, 2018, pp.
  23:1--23:2.

\bibitem{Cheffena2016commag}
M.~Cheffena, ``Industrial wireless communications over the millimeter wave
  spectrum: opportunities and challenges,'' \emph{IEEE Communications
  Magazine}, vol.~54, no.~9, pp. 66--72, September 2016.

\bibitem{geng2009millimeter}
S.~Geng, J.~Kivinen, X.~Zhao, and P.~Vainikainen, ``Millimeter-wave propagation
  channel characterization for short-range wireless communications,''
  \emph{IEEE Transactions on Vehicular Technology}, vol.~58, no.~1, pp. 3--13,
  2009.

\bibitem{soma2000modeling}
P.~Soma, Y.-M. Chia, and L.~Ong, ``Modeling and analysis of time varying radio
  propagation channel for lmds,'' in \emph{Radio and Wireless
  Conference}.\hskip 1em plus 0.5em minus 0.4em\relax IEEE, 2000.

\bibitem{elrefaie1997propagation}
A.~F. Elrefaie and M.~Shakouri, ``{Propagation measurements at 28 GHz for
  coverage evaluation of local multipoint distribution service},'' in
  \emph{Wireless Communications Conference, 1997., Proceedings}.\hskip 1em plus
  0.5em minus 0.4em\relax IEEE, 1997, pp. 12--17.

\bibitem{samimi20163}
M.~K. Samimi and T.~S. Rappaport, ``{3-D millimeter-wave statistical channel
  model for 5G wireless system design},'' \emph{IEEE Transactions on Microwave
  Theory and Techniques}, vol.~64, no.~7, 2016.

\bibitem{maccartney2016millimeter}
G.~R. MacCartney~Jr, S.~Sun, T.~S. Rappaport, Y.~Xing, H.~Yan, J.~Koka,
  R.~Wang, and D.~Yu, ``Millimeter wave wireless communications: New results
  for rural connectivity,'' in \emph{Proceedings of the 5th Workshop on All
  Things Cellular: Operations, Applications and Challenges}.\hskip 1em plus
  0.5em minus 0.4em\relax ACM, 2016, pp. 31--36.

\bibitem{maccartney2017rural}
G.~R. MacCartney and T.~S. Rappaport, ``Rural macrocell path loss models for
  millimeter wave wireless communications,'' \emph{IEEE journal on selected
  areas in communications}, vol.~35, no.~7, 2017.

\bibitem{meinila2009winner}
J.~Meinil{\"a}, P.~Ky{\"o}sti, T.~J{\"a}ms{\"a}, and L.~Hentil{\"a}, ``{WINNER
  II channel models},'' \emph{Radio Technologies and Concepts for
  IMT-Advanced}, 2009.

\bibitem{Hur:jv}
S.~Hur, S.~Baek, B.~Kim, Y.~Chang, A.~F. Molisch, T.~S. Rappaport, K.~Haneda,
  and J.~Park, ``{Proposal on Millimeter-Wave Channel Modeling for 5G Cellular
  System},'' \emph{IEEE Journal of Selected Topics in Signal Processing},
  vol.~10, no.~3, pp. 454--469, 2016.

\bibitem{Fan:2016dg}
W.~Fan, I.~Carton, J.~{\O}. Nielsen, K.~Olesen, and G.~F. Pedersen, ``{Measured
  wideband characteristics of indoor channels at centimetric and millimetric
  bands.}'' \emph{EURASIP J. Wireless Comm. and Networking}, vol. 2016, no.~1,
  p. 335, 2016.

\bibitem{zhang2017millimeter}
P.~Zhang, H.~Wang, H.~Wang, and R.~Bai, ``Millimeter-wave channel measurement
  and spatial characteristics for indoor environments,'' in \emph{Applied
  Computational Electromagnetics Society Symposium (ACES), 2017
  International}.\hskip 1em plus 0.5em minus 0.4em\relax IEEE, 2017, pp. 1--2.

\bibitem{zhao2017channel}
X.~Zhao, S.~Li, Q.~Wang, M.~Wang, S.~Sun, and W.~Hong, ``{Channel measurements,
  modeling, simulation and validation at 32 GHz in outdoor microcells for 5G
  radio systems},'' \emph{IEEE Access}, vol.~5, 2017.

\bibitem{rappaport1991statistical}
T.~S. Rappaport, S.~Y. Seidel, and K.~Takamizawa, ``Statistical channel impulse
  response models for factory and open plan building radio communicate system
  design,'' \emph{IEEE Transactions on Communications}, vol.~39, no.~5, pp.
  794--807, 1991.

\bibitem{zaaimia201660}
M.~Zaaimia, R.~Touhami, L.~Talbi, M.~Nedil, and M.~Yagoub, ``{60-GHz
  statistical channel characterization for wireless data centers},'' \emph{IEEE
  antennas and wireless propagation letters}, vol.~15, 2016.

\bibitem{nyuwireless}
\BIBentryALTinterwordspacing
{NYU WIRELESS - 5G mmWave Research}. [Online]. Available:
  \url{http://wireless.engineering.nyu.edu/}
\BIBentrySTDinterwordspacing

\bibitem{3GPP25.996}
{3GPP}, ``{Spatial Channel Model for Multiple Input Multiple Output (MIMO)
  Simulations},'' \emph{TR 25.996 V12.0.0}, 2014.

\bibitem{mmmagic}
\BIBentryALTinterwordspacing
mmmagic project. [Online]. Available: \url{https://5g-mmmagic.eu/}
\BIBentrySTDinterwordspacing

\bibitem{shuangchannel}
Y.-j.~L. Shuang-de Li, L.-k. Lin, Z.~Sheng, X.-c. Sun, Z.-p. Chen, and X.-j.
  Zhang, ``Channel measurements and modeling at {6 GHz} in the tunnel
  environments for {5G} wireless systems,'' \emph{International Journal of
  Antennas and Propagation}, 2017.

\bibitem{hawbaker1990indoor}
D.~Hawbaker and T.~Rappaport, ``Indoor wideband radiowave propagation
  measurements at 1.3 ghz and 4.0 {GHz},'' \emph{Electronics Letters}, vol.~26,
  no.~21, pp. 1800--1802, 1990.

\bibitem{lede}
\BIBentryALTinterwordspacing
Openwrt/lede project. [Online]. Available: \url{https://openwrt.org/}
\BIBentrySTDinterwordspacing

\bibitem{talon-tools:project}
\BIBentryALTinterwordspacing
D.~Steinmetzer, D.~Wegemer, and M.~Hollick. (2017) {Talon Tools: The Framework
  for Practical IEEE 802.11ad Research}. [Online]. Available:
  \url{https://seemoo.de/talon-tools/}
\BIBentrySTDinterwordspacing

\bibitem{Nitsche:2015vq}
T.~Nitsche, G.~Bielsa, I.~Tejado, A.~Loch, and J.~Widmer, ``{Boon and Bane of
  60 GHz Networks: Practical Insights Into Beamforming, Interference, and Frame
  Level Operation},'' in \emph{CoNEXT 2015}, 2015.

\bibitem{intel_model}
A.~M. et~al., ``{Channel Models for 60 GHz WLAN Systems},'' Tech. Rep., 2010.

\bibitem{conextDaniel}
D.~Steinmetzer, D.~Wegemer, M.~Schulz, J.~Widmer, and M.~Hollick, ``Compressive
  millimeter-wave sector selection in off-the-shelf ieee 802.11ad devices,'' in
  \emph{CoNEXT'17}.\hskip 1em plus 0.5em minus 0.4em\relax ACM, 2017.

\bibitem{fastninfuriating}
S.~K. Saha, H.~Assasa, A.~Loch, N.~M. Prakash, R.~Shyamsunder, S.~Aggarwal,
  D.~Steinmetzer, D.~Koutsonikolas, J.~Widmer, and M.~Hollick, ``Fast and
  infuriating: Performance and pitfalls of 60 ghz wlans based on consumer-grade
  hardware,'' in \emph{SECON 2018}, 2018.

\bibitem{7179366}
S.~K. Saha, V.~V. Vira, A.~Garg, A.~Tennenbaum, and D.~Koutsonikolas, ``On the
  feasibility of indoor ieee 802.11ad wlans,'' in \emph{IEEE INFOCOM WKSHPS},
  2015.

\bibitem{7568477}
S.~K. Saha, V.~V. Vira, A.~Garg, and D.~Koutsonikolas, ``A feasibility study of
  60 ghz indoor wlans,'' in \emph{ICCCN}, 2016.

\bibitem{SahaVGK15}
------, ``60 ghz multi-gigabit indoor wlans: Dream or reality?'' \emph{CoRR},
  vol. abs/1509.04274, 2015.

\end{thebibliography}

\end{document}